\documentclass[twocolumn,table,twocolappendix]{aastex63}

\usepackage{amsmath,amssymb}
\usepackage{gensymb}

\usepackage{epsfig}
\usepackage{comment}
\usepackage{color}
\usepackage[normalem]{ulem}
\usepackage{txfonts}
\usepackage{microtype}
\usepackage{booktabs}
\usepackage{color}
\usepackage{relsize}

\newcommand{\cxc}{{\em Chandra}}
\newcommand{\xmm}{XMM-{\em Newton}}

\def\Mg{M_{\rm g}}

\def\DA{D_{\rm A}}
\def\c{c^\star}

\def\Mv {M_{\rm 500}}
\def \Rv {R_{500}}

\def\TX{T_{\rm X}}
\def\TXc{T_{\rm X,CXO}}
\def\TXx{T_{\rm X,XMM}}
\def\YX{Y_{\rm X}}
\def\YXarc{\DA^{-2}\YX}

\def\YSZ{Y_{\rm SZ}}

\def\YSZPSZ2{Y_{\rm SZ}^{\rm PSZ2}}
\def\YSZmpc{\DA^{2}\YSZ}

\def\MX{M_{500}^{\rm Y_X}}
\def\MSZ{M_{500}^{\rm SZ}}
\def\MPSZ{M_{500}^{\rm SZ,PSZ2}}
\def\MPSX{M_{500}^{\rm SZ,PSX}}

\def\YXc {Y_{\rm X,CXO}}
\def\YXx {Y_{\rm X,XMM}}

\def\YSZc{Y_{\rm SZ, CXO}}
\def\YSZx{Y_{\rm SZ,XMM}}

\def\MXc{M_{\rm 500, CXO}^{\rm Y_X}}
\def\MXx{M_{\rm 500, XMM}^{\rm Y_X}}
\def\MSZc{M_{\rm 500, CXO}^{\rm SZ}}
\def\MSZx{M_{\rm 500, XMM}^{\rm SZ}}

\def\YM{Y_{\rm M}}
\def\YMarc{\DA^{-2}\YM}

\def\MYX {$\Mv$--$\YX$}

\def\MYSZ {$\Mv$--$\DA^2\YSZ$}
\def\YSZM {$\YSZmpc$--$\Mv$}
\def\YXM {$\YX$--$\Mv$}

\def\YSZYXarc{$\YSZ$--$\DA^{-2}\YX$}
\def\YSZYXmpc{$D_{\rm A}^2\YSZ$--$\YX$}
\def\YSZYX {$\YSZ$--$\YX$}

\def\YSZYM {$\YSZ$--$\YMarc$}

\def\MSZMX {$\MSZ$--$\MX$}

\def\sigsz {\sigma_{\log,\YSZ}}
\def\sigx {\sigma_{\log,\YX}}

\def\msol {{\rm M_{\odot}}}
\def\Msun{$M_{\odot}$}

\def\lesssim{\mathrel{\hbox{\rlap{\hbox{\lower4pt\hbox{$\sim$}}}\hbox{$<$}}}}
\def\gtrsim{\mathrel{\hbox{\rlap{\hbox{\lower4pt\hbox{$\sim$}}}\hbox{$>$}}}}

\newfont{\gwpfont}{cmssq8 scaled 1000}

\def\xmm{{\it XMM-Newton}}
\def\chandra{{\it Chandra}}
\def\planck{{\it Planck}}

\def\lira{\texttt{LIRA}}
\def\bces{\texttt{BCES}}

\def\linmix{\texttt{LINMIX}}

\shorttitle{{\em P{\scriptsize LANCK}}  ESZ with {\em C{\scriptsize HANDRA}}: A Re-Examination of Mass Proxies}
\shortauthors{Andrade-Santos et al.}

\begin{document}

\title{\textit{Chandra} Observations of the \textit{Planck} ESZ Sample: A Re-Examination of Masses and Mass Proxies}

\author[0000-0002-8144-9285]{Felipe Andrade-Santos}
\affiliation{Clay Center Observatory, Dexter Southfield, 20 Newton Street, Brookline, MA 02445, USA}
\affiliation{Center for Astrophysics \text{\textbar} Harvard \& Smithsonian, 60 Garden Street, Cambridge, MA 02138, USA}

\author{Gabriel W. Pratt}
\affiliation{AIM, CEA, CNRS, Universit\'e Paris-Saclay, Universit\'e Paris Diderot, Sorbonne Paris Cit\'e, F-91191 Gif-sur-Yvette, France}

\author{Jean-Baptiste Melin}
\affiliation{IRFU, CEA, Universit{\'e} Paris-Saclay, F-91191 Gif-sur-Yvette, France}

\author{Monique Arnaud}
\affiliation{AIM, CEA, CNRS, Universit\'e Paris-Saclay, Universit\'e Paris Diderot, Sorbonne Paris Cit\'e, F-91191 Gif-sur-Yvette, France}

\author{Christine Jones}
\affiliation{Center for Astrophysics \text{\textbar} Harvard \& Smithsonian, 60 Garden Street, Cambridge, MA 02138, USA}

\author[0000-0002-9478-1682]{William R. Forman}
\affiliation{Center for Astrophysics \text{\textbar} Harvard \& Smithsonian, 60 Garden Street, Cambridge, MA 02138, USA}

\author[0000-0001-8488-3645]{Etienne Pointecouteau}
\affiliation{CNRS; IRAP; 9 Av. Colonel Roche, BP 44346, F-31028 Toulouse Cedex 4, France}
\affiliation{Universit\'e de Toulouse; UPS-OMP; IRAP; Toulouse, France}

\author[0000-0001-7703-9040]{Iacopo Bartalucci}
\affiliation{AIM, CEA, CNRS, Universit\'e Paris-Saclay, Universit\'e Paris Diderot, Sorbonne Paris Cit\'e, F-91191 Gif-sur-Yvette, France}

\author[0000-0001-8121-0234]{Alexey Vikhlinin}
\affiliation{Center for Astrophysics \text{\textbar} Harvard \& Smithsonian, 60 Garden Street, Cambridge, MA 02138, USA}

\author{Stephen S. Murray}
\altaffiliation{Steve Murray passed away on 2015 August 10. Completing the {\em Chandra}\\ observations would not have been possible without his invaluable contri-\\butions.}
\affiliation{Department of Physics and Astronomy, The Johns Hopkins University, 3400 N. Charles St., Baltimore, MD 21218, USA}

\author[0000-0002-5411-1748]{Pasquale Mazzotta}
\affiliation{Dipartimento di Fisica, Universit\`a di Roma Tor Vergata, Via della Ricerca Scientifica 1, 00133 Roma, Italy}
\affiliation{Center for Astrophysics \text{\textbar} Harvard \& Smithsonian, 60 Garden Street, Cambridge, MA 02138, USA}

\author[0000-0001-6151-6439]{Stefano Borgani}
\affiliation{Dipartimento di Fisica dell' Universit\`a di Trieste, Sezione di Astronomia, Via Tiepolo 11, I-34131 Trieste, Italy}
\affiliation{INAF, Osservatorio Astronomico di Trieste, via Tiepolo 11, I-34131, Trieste, Italy}

\author[0000-0002-3754-2415]{Lorenzo Lovisari}
\affiliation{INAF - Osservatorio di Astrofisica e Scienza dello Spazio di Bologna, via Piero Gobetti 93/3, I-40129 Bologna, Italy}
\affiliation{Center for Astrophysics \text{\textbar} Harvard \& Smithsonian, 60 Garden Street, Cambridge, MA 02138, USA}

\author[0000-0002-0587-1660]{Reinout J. van Weeren}
\affiliation{Leiden Observatory, Leiden University, PO Box 9513, 2300 RA Leiden, The Netherlands}

\author[0000-0002-0765-0511]{Ralph P. Kraft}
\affiliation{Center for Astrophysics \text{\textbar} Harvard \& Smithsonian, 60 Garden Street, Cambridge, MA 02138, USA}

\author{Laurence P. David}
\affiliation{Center for Astrophysics \text{\textbar} Harvard \& Smithsonian, 60 Garden Street, Cambridge, MA 02138, USA}

\author{Simona Giacintucci}
\affiliation{Naval Research Laboratory, 4555 Overlook Avenue SW, Code 7213, Washington, DC 20375, USA}


\begin{abstract}
Using \chandra\ observations, we derive the $\YX$ proxy and associated total mass measurement, $\MX$, 
for 147 clusters with $z < 0.35$ from the \planck\ Early Sunyaev-Zel'dovich catalog, and for 80 clusters with $z < 0.22$ from an X-ray flux-limited sample. 
We re-extract the \planck\ $\YSZ$ measurements and obtain the corresponding mass proxy, $\MSZ$, from the full \planck\ mission maps, minimizing the Malmquist bias due to observational scatter.
The masses re-extracted using the more precise X-ray position and characteristic size agree with the  published PSZ2 values, but yield a significant reduction in the scatter (by a factor of two) in the \MSZMX\ relation. The slope is $0.93\pm0.03$, and  
the median ratio,  $\MSZ/\MX= 0.91\pm0.01$, is  within the expectations from known X-ray calibration systematics.
The  $\YSZ/\YX$ ratio is $0.88 \pm 0.02$, in good agreement with predictions from cluster structure, and implying a low level of clumpiness.
In agreement with the findings of the \planck\ Collaboration,  the slope of the \YSZYXarc\ flux relation is  significantly less than unity ($0.89\pm0.01$). Using extensive simulations, we  show that this result is not due to selection effects, intrinsic scatter, or covariance between quantities.
We demonstrate analytically that  changing the \YSZYX\ relation from apparent flux to intrinsic properties  results in a best-fit slope that is closer to unity
and increases the dispersion about the relation. The redistribution resulting from  this transformation implies that 
the best fit parameters of the \MSZMX\ relation will be sample-dependent.
\end{abstract}



\section{Introduction}

Galaxy clusters reside in the highest ranges of the halo mass function. 
In the standard $\Lambda$CDM cosmology,
massive halos form by the accretion of smaller sub-clumps \citep[e.g.,][]{1979Jones,1982Forman,2011Allen,2012Kravtsov}. 
Under the influence of gravity, uncollapsed and collapsed sub-clumps fall into
larger halos and, occasionally, objects of comparable mass merge with
one another. X-ray observations of substructures in galaxy clusters 
\citep[see, for instance,][]{1984Jones,1999Jones,1995Mohr,1996Buote,2005Jeltema,2010Bohringer,2010Lagana,
2012Andrade-Santos,2013Andrade-Santos} and measurements of the growth of structure
\citep{2009bVik,2010Mantz,2011Allen,2013Benson,2014Planck,2016Planck}
show that these objects are still in the process of formation.

A well established relation between the cluster mass and its observables
(such as X-ray luminosity, gas temperature, etc)
is crucial to any work that explores the theoretical relation between
the number density of collapsed halos (the mass function) 
and the underlying cosmological parameters \citep{2009bVik,2014Planck,2016Planck,2019Pratt}.
The relations between a cluster's observables and its mass are a direct
consequence of gravity, which is the main force driving cluster
evolution \citep{1986Kaiser}. Departures from the
purely gravitational self-similar expectation are usually attributed to non-gravitational
processes, such as radiative cooling, AGN feedback, etc. \citep[e.g.][]{2005Voit,2010Pratt}.

\begin{figure*}[!t]
\centerline{
\includegraphics[width=1.2\textwidth]{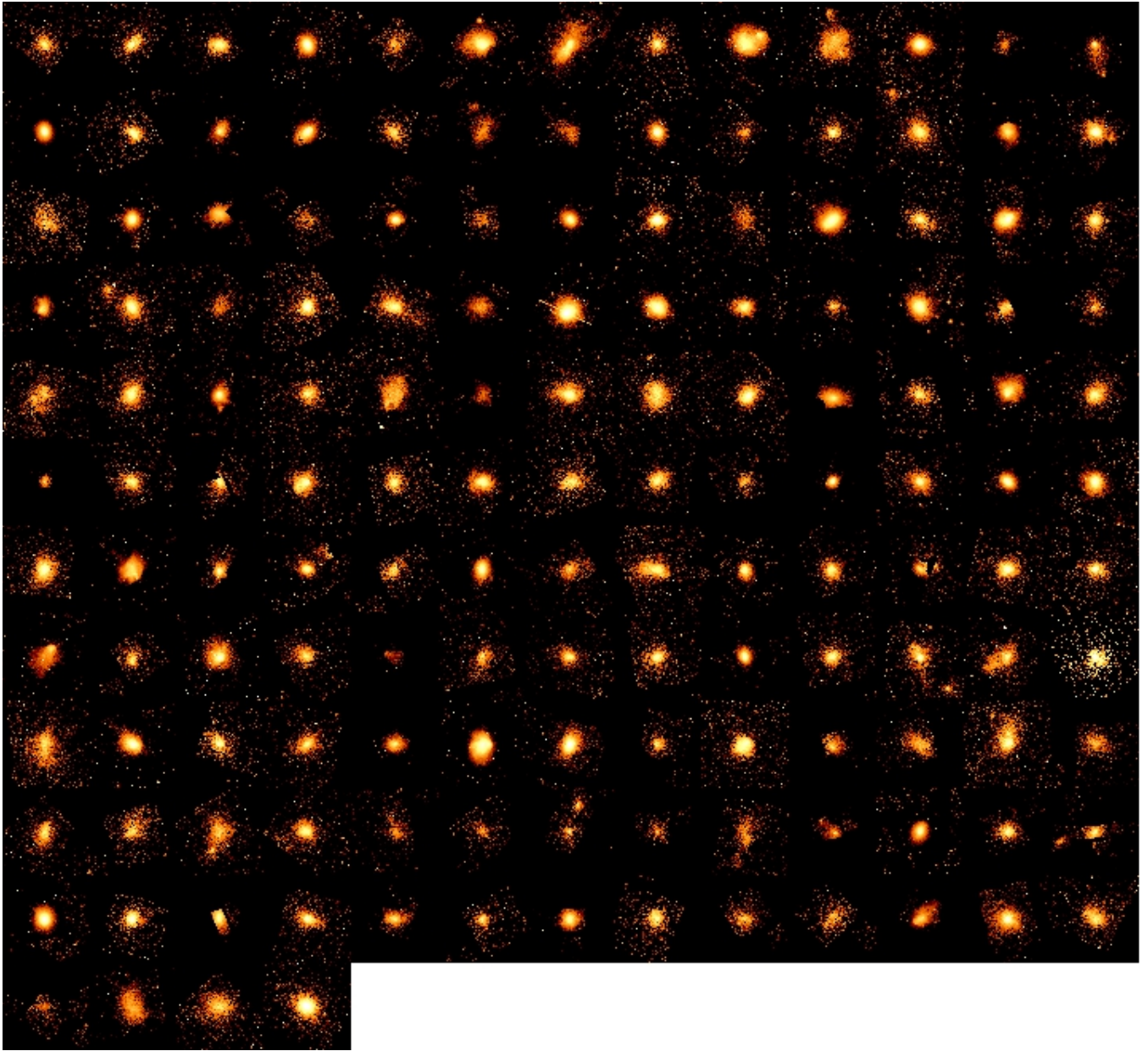}
}
\caption{\small{
Mosaic of the 147 clusters from the ESZ sample that were used in this work. Each cluster image was filtered to 0.5-2.0 keV, background-subtracted, exposure map corrected, binned to 4$\times$4 pixels, and smoothed to 3 pixels. Image sizes are $3 \times \Rv$ on a side, where $\Rv$ is obtained from the $\MX$. Images are corrected for surface brightness cosmological dimming with redshift, and are displayed in logarithmic scale.  The colour table and scale range is the same for all clusters, so that the images would be identical if clusters obeyed strict self-similarity. Order of clusters follows that from the machine-readable table available online, here in row-major order. Using standard matrix notation, examples are: the Bullet cluster in position $c_{96}$, Abell 3411 in position $c_{81}$, and Coma in position $c_{33}$.
}}\label{fig:ESZmosaic}
\end{figure*} 

The {\em Planck} satellite surveyed the entire sky across nine
microwave frequency bands. The resulting data set allowed the detection
of galaxy clusters through the thermal Sunyaev-Zel'dovich (SZ) effect
\citep{1972SZ}. The thermal SZ effect is the shift
of the CMB spectrum towards higher frequencies caused by the inverse
Compton scattering of the CMB photons by the hot electrons in the ICM.
Being proportional to the line-of-sight integral of the ICM pressure, 
the SZ signal is expected to be closely related to
the cluster total mass  \citep[e.g.][]{2004dasilva,2014Pike}, and 
the amplitude of its signal is independent of
the
cluster redshift, since the CMB photons were all emitted at a constant
redshift of $z \sim 1000$. As a consequence, SZ surveys can
potentially build unbiased cluster samples, covering higher redshifts
than X-ray samples, and are expected to be close to mass-limited. 

The second {\em Planck} catalogue \citep[PSZ2,][]{2016PSZ2}, derived from the full High Frequency Instrument (HFI) survey of 29 months, detected 1653 cluster candidates. The vast majority ($>$ 1200) of these candidates have been confirmed, making the PSZ2 catalog a reference for cluster studies. Among the many quantities provided by the PSZ2 catalog for each cluster, the mass estimate is arguably the most important. Since the {\em Planck} masses were determined using {\em XMM-Newton} data to calibrate the \YSZYX\ relation, comparing {\em Chandra} X-ray 
derived masses to SZ derived masses
provides an independent and invaluable test for the {\em Planck} mass estimates.  

Here we use {\em Chandra} observations to compute the masses for 147 {\em Planck} ESZ clusters at $z < 0.35$. All of the clusters have {\em Planck} SZ masses, and we compare these to the X-ray derived masses from our {\em Chandra} analysis. We compare the results with those obtained from observations of an X-ray selected sample of the 80 of the 100 highest flux X-ray clusters at Galactic latitudes $|b| > 20^\circ$ and $0.025 < z < 0.22$. We also study the \YSZYX\ relation. Our aim is to examine the \YSZYX\ relation and to compare X-ray and SZ masses for these clusters. We use the SZ-selected clusters as our baseline sample, and the X-ray-selected objects only for comparison. Examination of the impact of our findings on the \planck\ SZ cosmological constraints is beyond the scope of the present work.

Throughout this paper, we assume a standard $\Lambda$CDM cosmology
with $H_0= ~ 70 \rm ~km~s^{-1}~Mpc^{-1}$, $\Omega_{\rm M}=0.3$, and
$\Omega_{\rm \Lambda}=0.7$. The variables $\Mv$ and $\Rv$ are the total mass within $\Rv$ and radius corresponding to a total density contrast  
$\Delta=500$ as compared to $\rho_{\rm c}(z)$, the critical density of the universe at the cluster redshift. The SZ flux is characterised by $Y_{500}^{\rm SZ}$ (in arcmin$^2$), or simply $\YSZ$. The quantity
$D_{\rm A}^2 \YSZ$ is then the spherically integrated Compton parameter within $\Rv$ (in Mpc$^2$), where $D_{\rm A}$ is the angular-diameter distance of the cluster. All uncertainties are quoted at the $1\sigma$ (68\% confidence) level. The natural logarithm is denoted ln, while log denotes the decimal logarithm. 

\begin{figure*}[!t]
\centerline{
\includegraphics[width=1.0\textwidth]{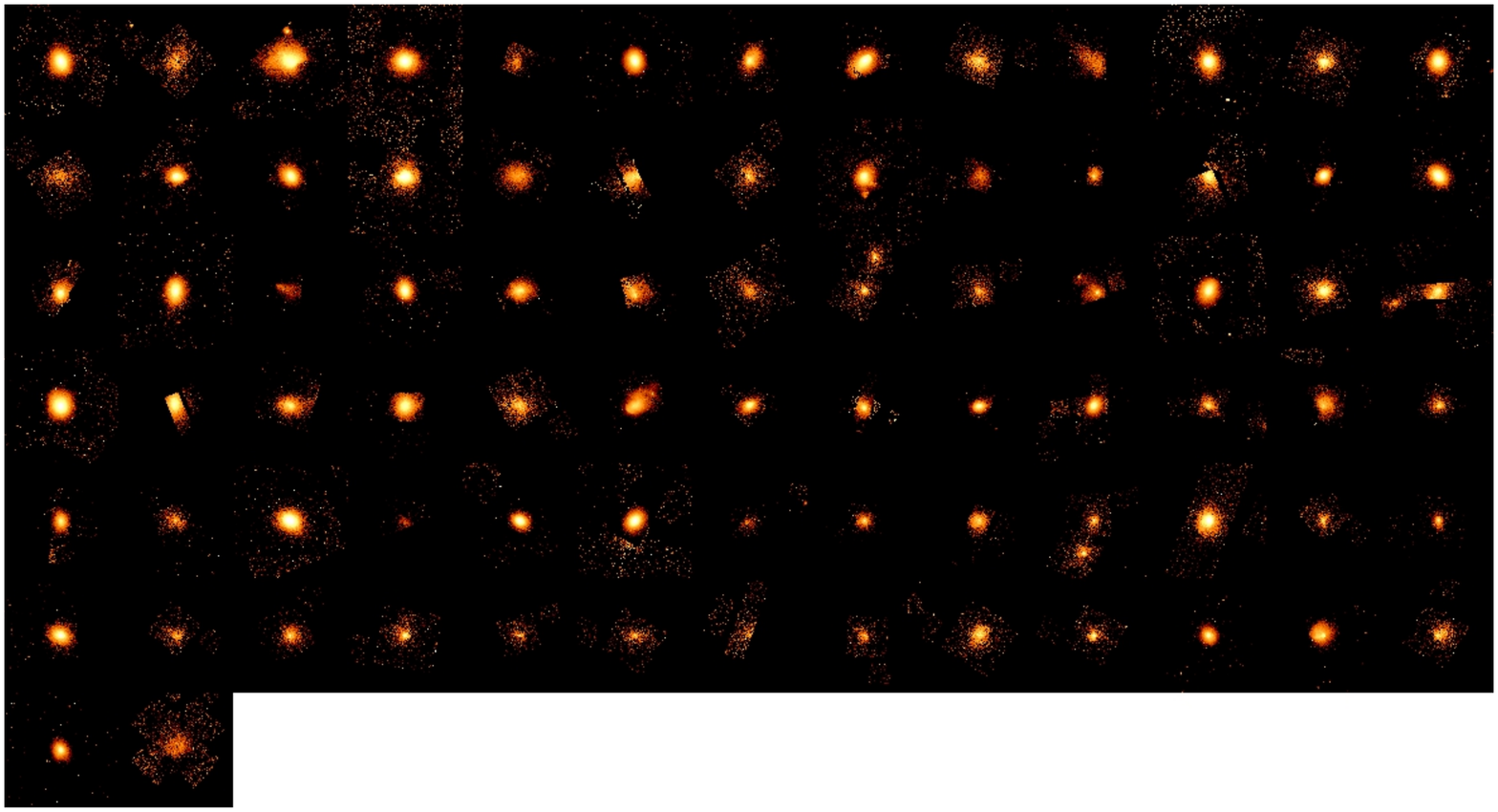}
}
\vspace{-2cm}
\caption{\small{
Same as Figure \ref{fig:ESZmosaic}, except for the 80 clusters from the X-ray sample that were used in this work. Clusters that are also part of the ESZ sample used in this work are presented from the beginning (top left) until (and including) Abell 3667, which is displayed in position $c_{46}$.
}}\label{fig:Xraymosaic}
\end{figure*} 


\section{SZ and X-ray selected clusters}

For our investigation, we make use of the following cluster samples:

\begin{itemize}

\item An SZ-selected sample of clusters derived from the first catalog of 189 objects detected in the first 10 months of the {\em Planck} survey, and released in early 2011 \citep{2011PlanckCol}.
A {\em Chandra} XVP (X-ray Visionary Program -- PI: Jones) and HRC Guaranteed Time Observations (PI: Murray) were combined to form the {\em
Chandra-Planck} Legacy Program for Massive Clusters of Galaxies.
For each of the 163 ESZ {\em Planck} clusters at $z < 0.35$, we obtained \cxc ~exposures sufficient to collect at least 10,000 source counts\footnote{\scriptsize \href{http://hea-www.cfa.harvard.edu/CHANDRA_PLANCK_CLUSTERS/}{hea-www.cfa.harvard.edu/CHANDRA\_PLANCK\_CLUSTERS/}}. 

The second {\em Planck} catalogue \citep[PSZ2,][]{2016PSZ2}, derived from the full mission maps, provided masses for 
157 of the original set of 163 ESZ clusters at $z < 0.35$. The remaining six clusters are not in the PSZ2 catalog, simply because they fall outside the mask used to construct that catalog (i.e. they are close to point sources). We can thus discard those without changing the sample selection function. We use the PSZ2 mass estimates of the resulting systems as they are largely independent of the original detection (discussed in more detail below).  
Of the resulting 157 clusters, nine are classed as multiple, defined as having more than one object (visually identified in the X-ray images) at a distance less than 10' from the
main system. These were removed from the following as the presence of more than one system may lead to a boosting of the SZ signal due to confusion in the \planck\ beam. We also discarded the very nearby cluster PLCKESZ\,G234.59+73.01 (A 1367  at $z=0.02$) as it is too large to allow a reliable background estimate in the \chandra\ pointing. 
Figure \ref{fig:ESZmosaic} displays a mosaic of the 147 clusters from the ESZ sample that were used in this work. 

\item An X-ray selected sample of high-flux clusters. \citet{2004Voevodkin} compiled a sample of the 52 X-ray brightest
clusters in the local universe by selecting the highest flux objects detected in the 
ROSAT All-Sky survey at $|b| > 20^\circ$ and 
$z > 0.025$ -- using the 
HIFLUGCS\footnote{HIFLUGCS -- The HIghest X-ray FLUx Galaxy Cluster Sample \citep{2002Reiprich,2017Schellenberger}} catalog as
reference. The sample used here is an extension of \citet{2004Voevodkin}'s approach, where the flux limit in the 0.5 -- 2.0 keV band was lowered to $f_{\rm X} > 7.5 \times 10^{-12} \rm~erg ~ s^{-1} ~ cm^{-2}$.  The full sample contains 100 clusters; the highest-redshift object is at $z = 0.215$. All have {\em Chandra} observations. 

Eighty-three of the clusters from this sample have {\em Planck} SZ masses. Three of these were classified as multiple, according to the definition applied to the ESZ sample, so after removing these clusters we obtained a sample of 80 clusters. Of these, 45 are also in the ESZ sample, and 42 are in the HIFLUGCS sample. Figure \ref{fig:Xraymosaic} displays a mosaic of the 80 clusters from the X-ray sample that were used in this work. 

\end{itemize}

\begin{figure*}[!t]
\centerline{
\includegraphics[width=0.5\textwidth]{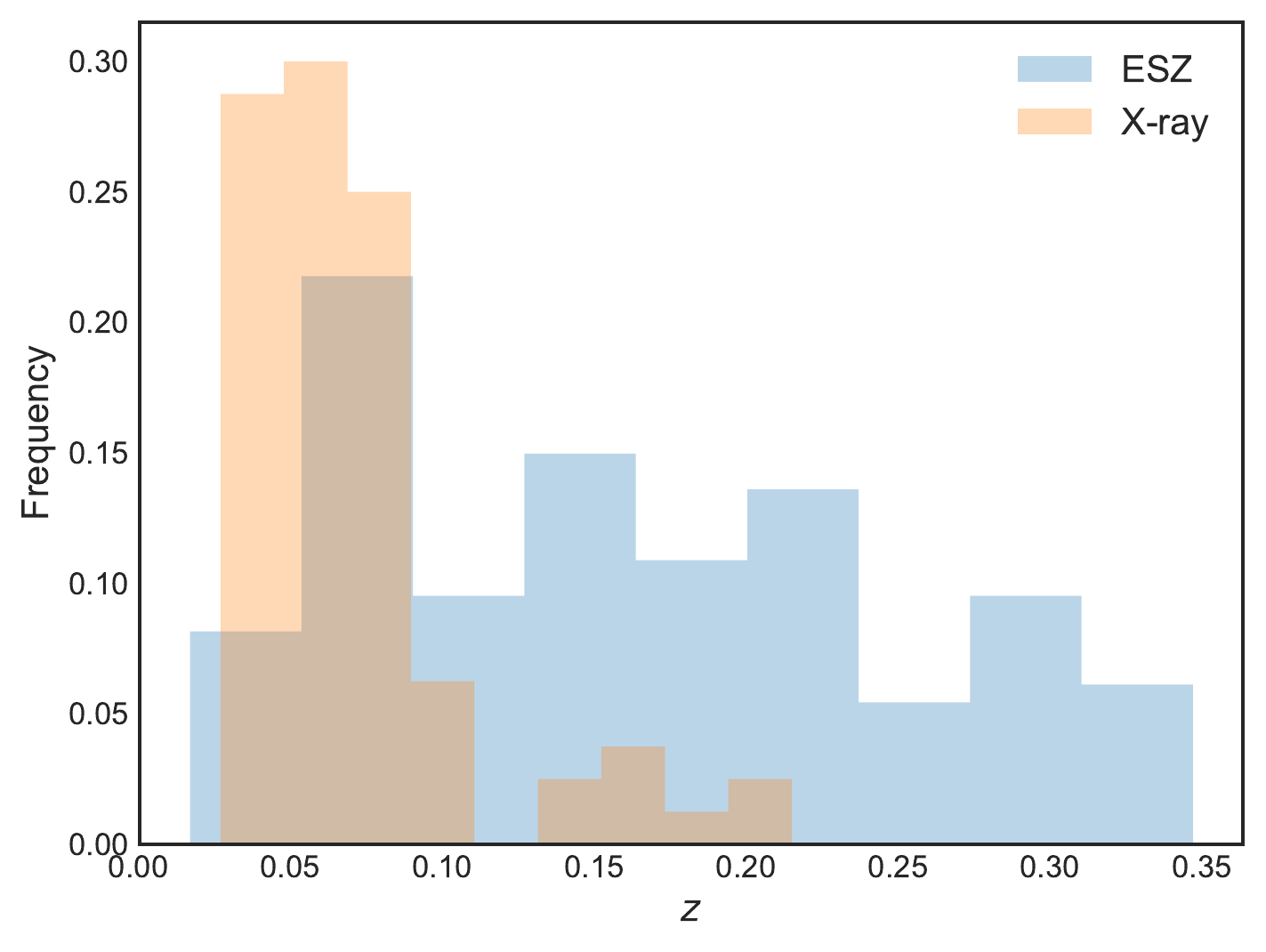}
\includegraphics[width=0.5\textwidth]{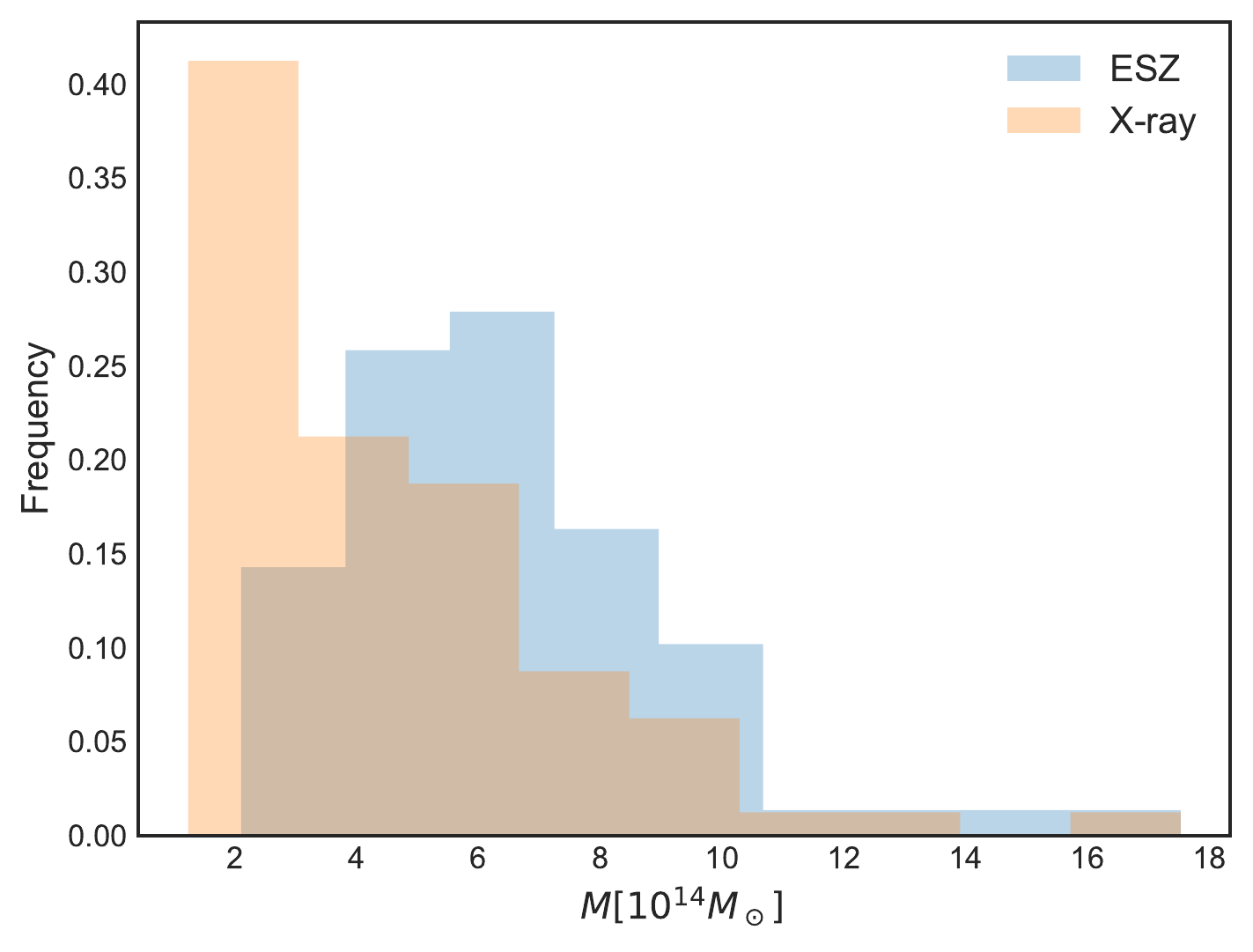}
}
\caption{\small{
The orange histograms show the X-ray flux limited sample, while the blue ones
correspond to the ESZ cluster sample. Left: Distribution of cluster redshifts in the ESZ and the X-ray selected
samples. The ESZ sample extends to higher redshifts than the X-ray
flux-limited sample.
Right: Distribution of cluster masses in the ESZ and X-ray selected
samples. The X-ray sample extends to lower masses than
does the ESZ sample.
}}\label{fig:histz}
\end{figure*}

The left panel of Figure \ref{fig:histz} presents the redshift distribution of
both cluster samples. The {\em Planck} detected clusters are clearly more
broadly distributed in redshift than the X-ray clusters. This is due to the nature of the selection:
for resolved clusters the Sunyaev-Zel'dovich signal is independent of the redshift of
the cluster because it is the CMB that is distorted (the CMB photons
originate  at the epoch of recombination -- from a constant redshift
of $z \sim 1000$), while the X-ray selected clusters constitute
a flux-limited sample, which strongly favors the X-ray brighter, lower redshift clusters. 
The right panel of Figure \ref{fig:histz} presents the mass (derived using the $\YX$ proxy)
distribution of both samples, showing that the X-ray sample has on average 
lower mass than the {\em Planck} ESZ sample.


\section{Data Reduction}
\label{sec:chandra} 

\subsection{\cxc\ data}

\subsubsection{Event file processing}

Our {\em Chandra} data reduction followed the process described in
\citet{2005Vik}. We applied the calibration files \texttt{CALDB 4.7.2}. 
The data reduction included corrections for the time dependence of the charge
transfer inefficiency and gain, and also a check for periods
of high background, which were then omitted. Standard blank sky background files
and readout artifacts were subtracted. We also detected compact X-ray
sources in the 0.7--2.0 keV and 2.0--7.0 keV
bands using \texttt{CIAO}'s \texttt{wavdetect} and then masked these sources, as well as extended substructures, before performing the spectral and spatial
analyses of the cluster emission. For each cluster, we used all available {\em
Chandra} ACIS (ACIS-I and ACIS-S) observations within 2 Mpc from the cluster center. The median maximum radius of the observations is $1.53 \, \Rv$ (with a maximum radius greater than $\Rv$ for $95\%$ of the sample)  and the median azimuthal coverage is $0.92\,\pi \Rv^2$. We have checked that our results do not depend either on the maximum radius or on the coverage.

\subsubsection{Emission measure profiles}

We refer to \citet{2006Vik} for a detailed description of
the procedures we used to compute the emission measure profile for
each cluster.
We outline here only the main aspects of the method. 

We measured the surface brightness profiles in the 0.7--2.0 keV energy band, 
which maximizes the signal to noise ratio in
{\em Chandra} observations for typical cluster gas temperatures. We used the X-ray peak 
as the cluster center.
The readout artifacts and blank-field background \citep[see section 2.3.3 of][]{2006Vik} 
were subtracted from the X-ray images, and the results were then exposure-corrected, using exposure maps
computed assuming an absorbed optically-thin thermal plasma with $kT = 5.0$
keV, abundance = 0.3 solar, with the Galactic column density
and including corrections for bad pixels and CCD gaps, which do not 
take into account spatial variations of the effective area. 
We subtracted a small uniform component, corresponding to 
soft X-ray foreground adjustments, if required (determined by fitting 
a thermal model in a region of the detector field most distant from the
cluster center, properly taking into account the expected thermal contribution from the cluster).

Following these steps, we extracted the surface brightness in narrow 
concentric annuli ($r_{\rm out}/r_{\rm in} = 1.05$) centered on the
X-ray peak and computed the {\em Chandra} area-averaged effective area for each
annulus \citep[see][for details on calculating the effective area]{2005Vik}. 
To compute the emission measure and temperature profiles, we assumed spherical 
symmetry. The spherical assumption is expected to introduce only small 
deviations  in the emission measure profile \citep{2003Piffaretti}.
Using the modeled de-projected temperature \citep[see][]{2017Andrade-Santos}, effective area, and 
metallicity as a function of radius, we converted the {\em Chandra} count 
rate in the 0.7--2.0 keV band into the emission integral, 
${\rm EI} =  \int n_{\rm e} n_{\rm p} dV$, within each cylindrical
shell. 
Online tables in machine-readable format 
list the maximum cluster radius where the emission integral is computed ($r_{\rm max}$)
for each cluster. 142 ($97\%$) clusters in the ESZ sample have 
$r_{\rm max} > \Rv$
\footnote{$\Rv$ defines the radius at the over-density of
500 times the critical density of the Universe at the cluster
redshift. This quantity was obtained from the X-ray derived masses.}, and in the X-ray sample, 72 ($90\%$) clusters
satisfy this condition (four of the eight clusters that do not satisfy
this condition are also in the ESZ sample). 
We have an average azimuthal coverage within $\Rv$ of $(0.8 \pm 0.2) \times \pi \Rv^2$ for the ESZ clusters and $(0.6 \pm 0.2) \times \pi \Rv^2$ for the X-ray clusters.

\begin{figure*}[!t]
\centerline{
\includegraphics[width=0.5\textwidth]{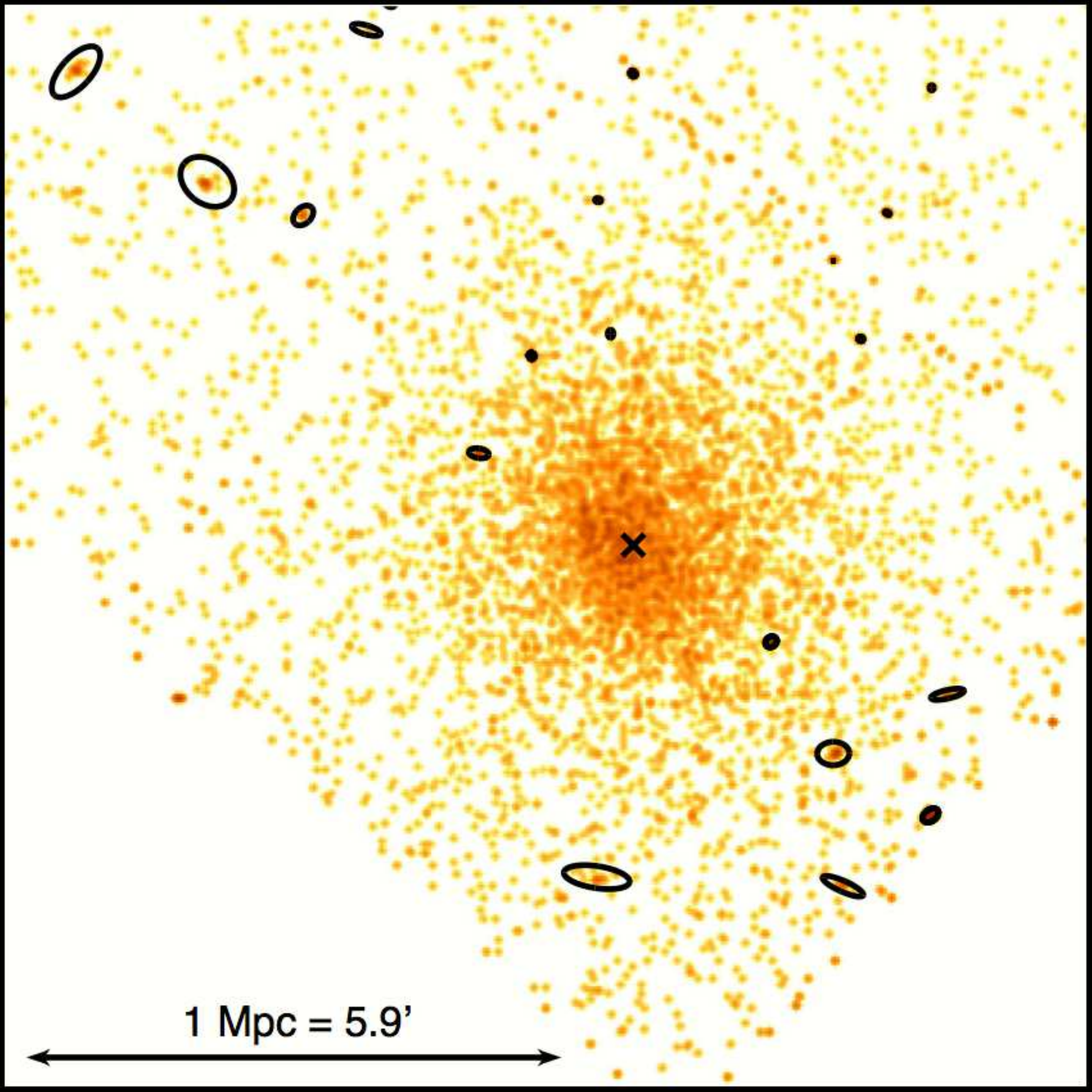}
} 
\centerline{
\includegraphics[width=0.33\textwidth]{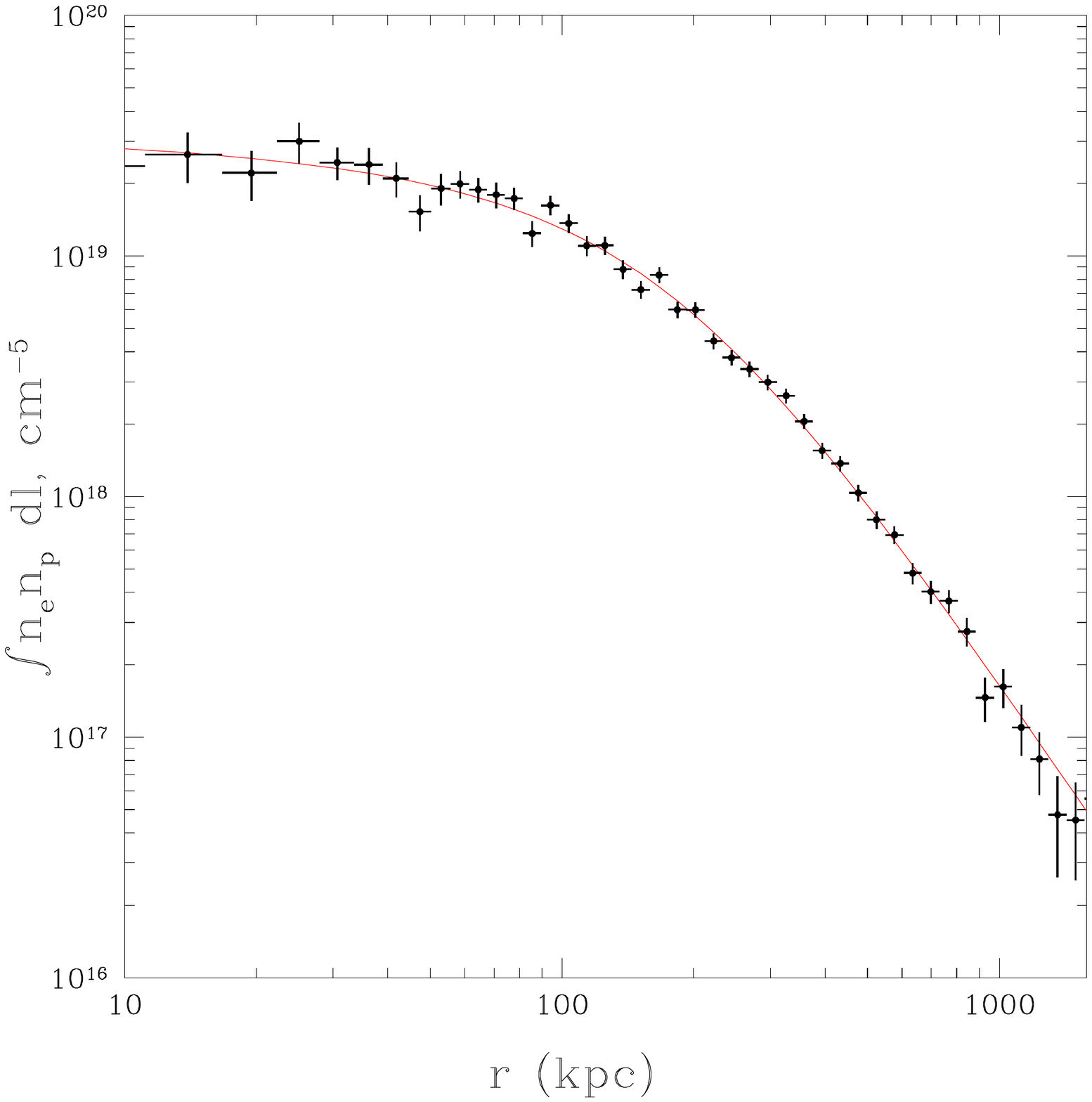}
\includegraphics[width=0.33\textwidth]{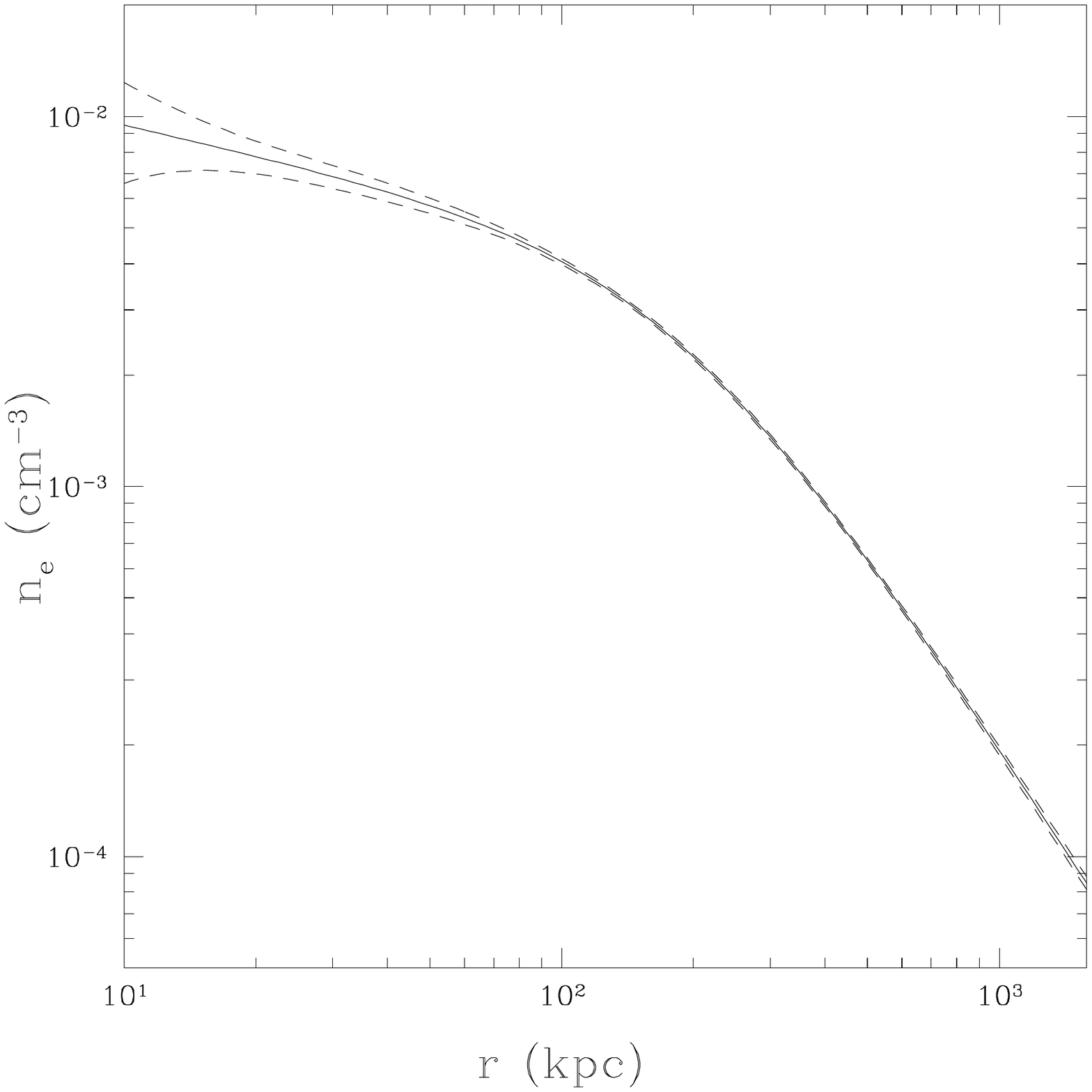}
\includegraphics[width=0.33\textwidth]{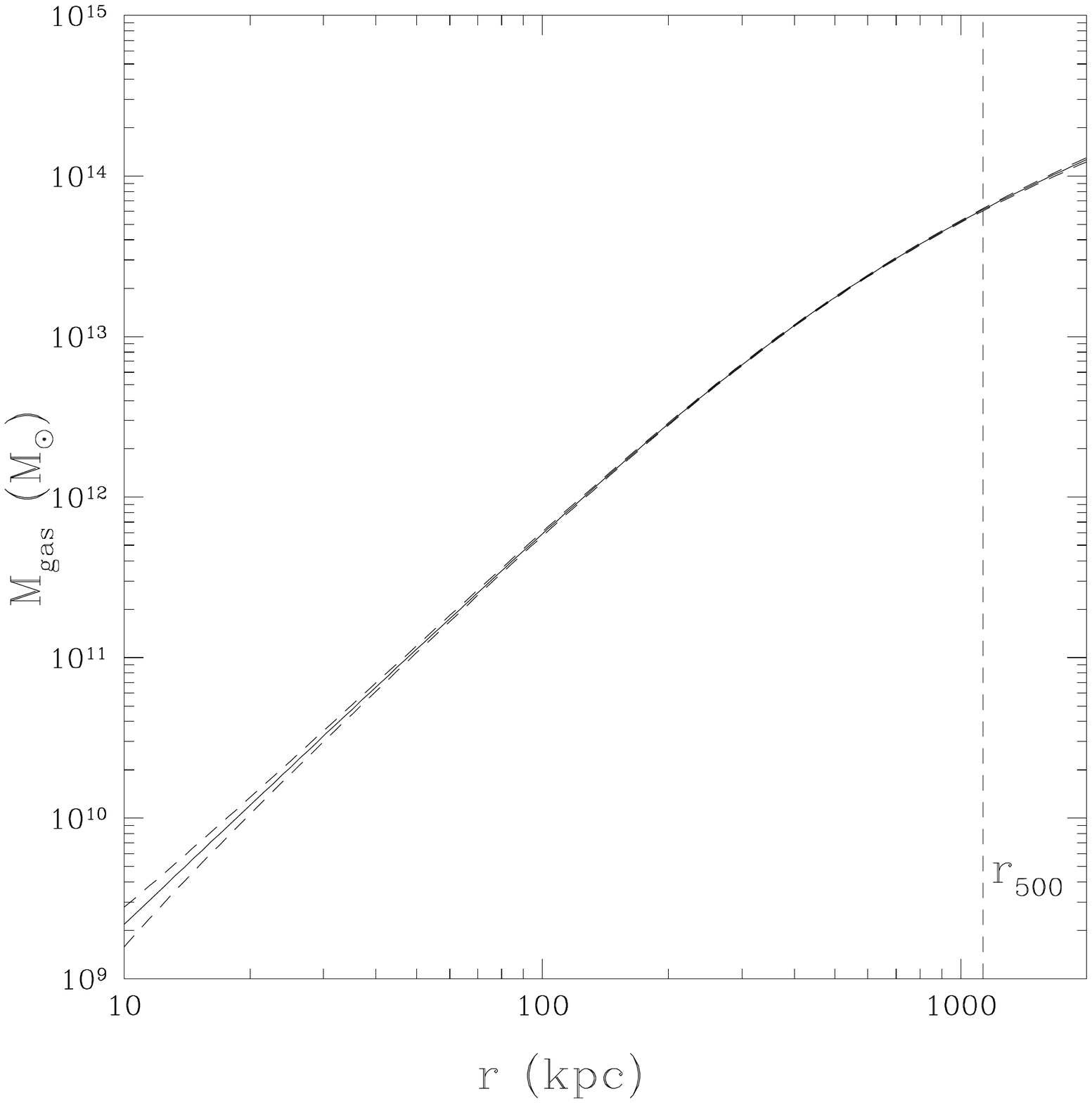}
}
\vspace{-1cm}
\caption{\small{Example of the X-ray image (top panel), projected
emissivity (lower left), gas density (lower center),
and gas mass (lower right) profiles for the cluster  PLCKESZ G000.44-41.83.
Top panel shows the 0.5-2.0 keV, background-subtracted, exposure map corrected
ACIS-I image.  Black ellipses correspond to the masked X-ray point
sources and the cross corresponds to the cluster center. 
Lower left panel shows the projected emissivity profile. The
solid line shows the emission measure integral of the best fit
to the emissivity profile given by Equation (\ref{eq:nenp}). 
Lower center panel shows the gas density profile. The
solid line shows the density profile obtained from the emissivity
profile given by Equation
(\ref{eq:nenp}).
Lower right panel shows the integrated gas mass profile $M_{\rm gas}(< R)$, with the dashed
vertical line indicating $\Rv$. The dashed lines in the
gas density and mass profiles show the 68\%
confidence range.  This is an example of a cluster with 
moderate data quality, which illustrates the average data quality for
the clusters in our samples.
The total filtered {\em Chandra} exposure is 14 ks.
}}\label{fig:emm_dens}
\end{figure*}

We fit the emission measure profile assuming the gas density
profile follows that given by \citet{2006Vik}: 

\begin{eqnarray}
n_{\rm e}n_{\rm p} &=& n_0^2
\frac{ (r/r_{\rm c})^{-\alpha}}{(1+r^2/r_{\rm c}^2)^{3\beta-\alpha/2}}
\frac{1}{(1+r^\gamma/r_{\rm s}^\gamma)^{\epsilon/\gamma}} \nonumber \\
&+& \frac{n^2_{02}}{(1+r^2/r_{\rm c2}^2)^{3\beta_2}},\label{eq:nenp}
\end{eqnarray}
where the parameters $n_0$ and $n_{02}$ determine the normalizations of
both additive components. $\alpha$, $\beta$, $\beta_2$, and $\epsilon$ are indices controlling the slope of the
curve at characteristic radii given by the parameters $r_{\rm c}$,
$r_{\rm c2}$, and $r_{\rm s}$. $\gamma$ controls the width of the
transition region given by $r_{\rm s}$. Although the relation given
by Equation \ref{eq:nenp} is based on a
classic  $\beta$-model \citep{1976Cavaliere}, it is modified to account
for a central power-law type cusp and a steeper emission measure slope at large radii. In addition,
a second $\beta$-model is included, to better characterize the cluster core.
For further details on this equation, we refer the reader to
\citet{2006Vik}. In
the fit to the emissivity profile, all parameters are free to vary.
For a typical metallicity of
0.3 $Z_\odot$, the reference values from \citet{1989AndersGrevesse}
yield $n_{\rm e}/n_{\rm p} = 1.1995$. Examples of projected emissivity
and gas density profiles are presented in Figure \ref{fig:emm_dens}.


\subsubsection{Cluster Mass Estimates}
\label{sec:yxmass}

Using the gas mass and temperature, we estimated
the total cluster mass from the $\Mv$--$Y_{500}^{\rm X}$ scaling relation of \citet{2009Vik},
\begin{eqnarray}
\MX = E^{-2/5}(z)\,A_{\rm YM}\left[\frac{\YX}{3\times10^{14}M_\odot
  {\rm keV}}\right]^{B_{\rm YM}}, \nonumber \\
\label{e_yx_m}
\end{eqnarray}
where $\YX= M_{\rm gas,500} \times kT_{\rm X}$, $M_{\rm
  gas,500}$ is computed using the best fit parameters of Equation (\ref{eq:nenp}), and $T_{\rm X}$ is the measured
temperature in the (0.15--1) $\times ~
\Rv$ range. 
We fit the spectrum in the 0.7--7.0 keV energy range with an absorbed \texttt{APEC} model, fixing NH to the total Galactic value \citep[molecular plus atomic hydrogen contributions --][]{2013Willingale}.
The parameters $A_{\rm YM}=(5.77 \pm 0.20)\times10^{14}h^{1/2}  M_\odot$, and $B_{\rm YM}=0.57
\pm 0.03$ \citep{2009Vik}. Here,
$\MX$ is the total mass within $\Rv$,
and $E(z)=[\Omega_{\rm M}(1+z)^3 + (1-\Omega_{\rm M}-\Omega_\Lambda)(1+z)^2 +
\Omega_\Lambda]^{1/2}$ is the function describing the evolution of the Hubble
parameter with redshift. Using Equation (\ref{e_yx_m}), $\Rv$  is computed by
solving 
\begin{eqnarray}
M_{500} \equiv 500 \rho_c (4\pi/3) \Rv^{3},\label{m500_def}
\end{eqnarray}
where $\rho_c$ is the critical density of the Universe at the cluster
redshift. In practice, Equation (\ref{e_yx_m}) is evaluated at 
a given radius, whose result is compared
to the evaluation of Equation (\ref{m500_def}) at the same radius. This
process is repeated in an iterative procedure, until the fractional mass difference
is less than 1\%. We estimated 1$\sigma$ statistic uncertainties in the $Y_{500}^{\rm X}$ derived masses 
using Monte Carlo simulations. We also added to the Monte Carlo
procedure a 1$\sigma$ systematic uncertainty of 9\% in the mass determination, as discussed by \citet{2009Vik}.


\subsection{\planck\ data}
\label{sec:planckdata}

\subsubsection{Published masses}

Derivation of the published \planck\ masses is extensively described in \citet{2016PSZ2}. The mass was estimated from the relation between $Y_{500}^{\rm SZ}$ and $\Mv$ \citep{2014Planck}:
\begin{equation}
E(z)^{-2/3}\left[\frac{D_{\rm A}^2\,Y_{500}^{\rm SZ}}{\rm 10^{-4}\,Mpc^2}\right] =  10^{-0.186}\left[\frac{h}{0.7}\right]^{-0.211} \left[\frac{M^{\rm SZ}_{500}}{6\times10^{14}\,\msol}\right]^{1.789}.
\label{eq:y_m}
\end{equation}

The SZ size-flux degeneracy was broken through the use of the scaling relation linking the SZ flux $Y_{500}^{\rm SZ}$ to the angular size, $\theta_{500}$, that is then intersected with the two-dimensional posterior probability contours in the $(Y_{500}^{\rm SZ}, \theta_{500}$) plane. For the PSZ2 masses, $\MPSZ$, the $Y_{500}^{\rm SZ}$--$\theta_{500}$ prior was obtained from the intrinsic relation between $\Mv$ and $\theta_{500}$ for a given redshift $z$:
\begin{equation}
{\theta_{500}}=\theta_\ast
\left[\frac{h}{0.7}\right]^{-2/3}\left[{M^{\rm SZ}_{500}\over 3\times
    10^{14} \msol}\right]^{1/3} \,E^{-2/3}(z)\,\left[{D_{\mathrm
      A}(z)\over 500\,{\mathrm{Mpc}}}\right]^{-1}\!\!, 
\label{eq:angle}
\end{equation}
where $\theta_\ast=6.997 \, \mathrm{arcmin}$.

We recall that Equation (\ref{eq:y_m}) was calibrated on a baseline mass proxy relation linking the hydrostatic mass and the  X-ray analog of the SZ Compton
parameter, $Y_{500}^{\rm X}$. As described in \citet{2016PSZ2}, the resulting SZ mass $\MPSZ$ can be viewed as the hydrostatic mass expected for a cluster consistent with the assumed scaling relation, at a given redshift, and given the measured $(Y_{500}^{\rm SZ}, \theta_{500}$) posterior information. Hereafter we will refer to $Y_{500}^{\rm SZ}$ and $Y_{500}^{\rm X}$ simply as $\YSZ$ and $\YX$.

\subsubsection{Re-extracted masses}

As discussed in Appendix A of \citet{planck2011-5.2b}, a refined measurement of the SZ flux can be obtained by using higher-quality prior information on the position and size of the cluster. Such information is readily provided by our \cxc\ observations. In the following, we refer to the re-extracted masses as $\MPSX$.
The SZ signal was thus re-extracted from the six HFI temperature channel maps corresponding to the full \planck\ mission survey. We used full resolution maps of HEALPix \citep{gor05}\footnote{\url{http://healpix.jpl.nasa.gov}} $N_{\rm side}=2048$ and assumed that the  beams were described by circular Gaussians. We adopted beam FWHM values of 9.66, 7.22, 4.90, 4.92, 4.68, and 4.22 arcmin for channel frequencies 100, 143, 217, 353, 545, and 857 GHz, respectively. 
Bandpass uncertainties were taken into account in the flux measurement. Uncertainties  due to beam corrections and map calibrations are expected to be small, as discussed extensively in \citet{2011PlanckCol}, \citet{2011PlanckXMMvalid}, \citet{planck2011-5.2a} and \citet{planck2011-5.2b}.


\begin{deluxetable*}{lcccccccccc}[t!]
\tablecaption{Cluster masses using SZ and X-ray proxies for the clusters in
  the {\em Planck} ESZ sample (this table is available online in its entirety
  in machine-readable format).} 
\tablewidth{0pt} 
\tablehead{ 
\colhead{Cluster} & 
\colhead{RA} &
\colhead{DEC} &
\colhead{$z$} & 
\colhead{$\MX$} &
\colhead{$\sigma_{\MX}$} &
\colhead{$M_{\rm 500}^{\rm PSX}$} &
\colhead{$\sigma_{M_{500}^{\rm PSX}}$} &
\colhead{$M_{500}^{\rm PSZ2}$} &
\colhead{$\sigma_{M_{500}^{\rm PSZ2}}$} &
\colhead{$r_{\rm max}$} \\ 
\colhead{} & 
\colhead{} &
\colhead{} &
\colhead{} & 
\colhead{($10^{14}~M_\odot$)} &
\colhead{($10^{14}~M_\odot$)} &
\colhead{($10^{14}~M_\odot$)} &
\colhead{($10^{14}~M_\odot$)} &
\colhead{($10^{14}~M_\odot$)} &
\colhead{($10^{14}~M_\odot$)} &
\colhead{($\Rv$)} \\ 
}
\startdata 
G000.44-41.83 & 21:04:18.603 & -41:20:39.36 & 0.165 & 4.81 & 0.47 & 4.89 & 0.32 & 5.30 & 0.33 & 1.68 \\
G002.74-56.18 & 22:18:39.822 & -38:53:58.47 & 0.141 & 5.34 & 0.51 & 4.24 & 0.30 & 4.41 & 0.30 & 1.55 \\
G003.90-59.41 & 22:34:27.334 & -37:44:07.88 & 0.151 & 9.23 & 0.89 & 7.26 & 0.24 & 7.19 & 0.26 & 1.37 \\
G006.47+50.54 & 15:10:56.117 & 05:44:40.38 & 0.077 & 7.06 & 0.64 & 7.19 & 0.17 & 7.04 & 0.20 & 1.43 \\
G006.70-35.54 & 20:34:46.912 & -35:49:24.54 & 0.089 & 5.22 & 0.50 & 4.40 & 0.21 & 4.17 & 0.23 & 1.66 \\
G006.78+30.46 & 16:15:46.073 & -06:08:54.61 & 0.203 & 17.53 & 2.01 & 16.25 & 0.26 & 16.12 & 0.29 & 1.08 \\
G008.30-64.75 & 22:58:48.095 & -34:48:04.62 & 0.312 & 7.97 & 0.75 & 7.70 & 0.40 & 7.75 & 0.40 & 1.50 \\
G008.44-56.35 & 22:17:45.701 & -35:43:32.55 & 0.149 & 3.84 & 0.39 & 4.41 & 0.27 & 4.79 & 0.30 & 1.82 \\
G008.93-81.23 & 00:14:19.305 & -30:23:29.33 & 0.307 & 10.22 & 0.93 & 9.23 & 0.36 & 9.84 & 0.39 & 1.37 \\
G018.53-25.72 & 20:03:30.848 & -23:23:37.54 & 0.317 & 9.79 & 0.93 & 8.65 & 0.45 & 8.99 & 0.47 & 1.42
\enddata
\tablecomments {Columns list cluster name (ESZ {\em Planck} cluster name with the prefix PLCKESZ omitted), right ascension, declination, redshift, $\YX$ derived {\em Chandra} mass, $\YX$ derived {\em Chandra} mass uncertainty, PSX {\em Planck} mass, PSX {\em Planck} mass uncertainty, PSZ2 {\em Planck} mass, PSZ2 {\em Planck} mass uncertainty, and maximum cluster radius where the emission integral is computed in terms of $\Rv$ 
($\Rv$ was computed using the $\YX$ derived {\em Chandra} mass).}
\label{tab:ESZmasses}
\end{deluxetable*}


\begin{deluxetable*}{lcccccccccc}[t!]
\tablecaption{Cluster masses using SZ and X-ray proxies for the clusters in
  the X-ray sample (this table is available online in its entirety
  in machine-readable format).} 
\tablewidth{0pt} 
\tablehead{ 
\colhead{Cluster} & 
\colhead{RA} &
\colhead{DEC} &
\colhead{$z$} & 
\colhead{$\MX$} &
\colhead{$\sigma_{\MX}$} &
\colhead{$M_{\rm 500}^{\rm PSX}$} &
\colhead{$\sigma_{M_{500}^{\rm PSX}}$} &
\colhead{$M_{500}^{\rm PSZ2}$} &
\colhead{$\sigma_{M_{500}^{\rm PSZ2}}$} &
\colhead{$r_{\rm max}$} \\ 
\colhead{} & 
\colhead{} &
\colhead{} &
\colhead{} & 
\colhead{($10^{14}~M_\odot$)} &
\colhead{($10^{14}~M_\odot$)} &
\colhead{($10^{14}~M_\odot$)} &
\colhead{($10^{14}~M_\odot$)} &
\colhead{($10^{14}~M_\odot$)} &
\colhead{($10^{14}~M_\odot$)} &
\colhead{($\Rv$)} \\ 
}
\startdata 
G006.47+50.54 & 15:10:56.117 & 05:44:40.38 & 0.077 & 7.06 & 0.64 & 7.19 & 0.17 & 7.04 & 0.20 & 1.43 \\
G006.70-35.54 & 20:34:46.912 & -35:49:24.54 & 0.089 & 5.22 & 0.50 & 4.40 & 0.21 & 4.17 & 0.23 & 1.66 \\
G006.78+30.46 & 16:15:46.073 & -06:08:54.61 & 0.203 & 17.53 & 2.01 & 16.25 & 0.26 & 16.12 & 0.29 & 1.08 \\
G021.09+33.25 & 16:32:46.854 & 05:34:31.61 & 0.151 & 8.47 & 0.77 & 8.07 & 0.27 & 7.79 & 0.30 & 1.42 \\
G029.00+44.56 & 16:02:14.068 & 15:58:16.23 & 0.035 & 2.86 & 0.26 & 2.64 & 0.13 & 3.53 & 0.06 & 1.01 \\
G033.78+77.16 & 13:48:52.710 & 26:35:31.20 & 0.062 & 5.67 & 0.51 & 4.72 & 0.14 & 4.47 & 0.14 & 1.53 \\
G042.82+56.61 & 15:22:29.473 & 27:42:18.76 & 0.072 & 5.63 & 0.51 & 4.44 & 0.20 & 4.08 & 0.19 & 1.47 \\
G044.22+48.68 & 15:58:21.100 & 27:13:47.87 & 0.089 & 12.36 & 1.12 & 9.44 & 0.18 & 8.77 & 0.20 & 1.25 \\
G046.50-49.43 & 22:10:19.489 & -12:10:10.03 & 0.085 & 5.09 & 0.48 & 4.51 & 0.20 & 4.39 & 0.20 & 1.60 \\
G048.05+57.17 & 15:21:12.694 & 30:38:00.59 & 0.078 & 3.92 & 0.36 & 3.66 & 0.23 & 3.59 & 0.21 & 1.77 \\
\enddata
\tablecomments{Same as Table \ref{tab:ESZmasses}, except for the X-ray sample. }
\label{tab:Xraymasses}
\end{deluxetable*}

We extracted the SZ signal using multi-frequency matched filters \citep[MMF,][]{her02,mel06}. These require information on the instrumental beam, the SZ frequency spectrum, and a cluster profile; noise auto- and cross-spectra are estimated directly from the data.  We used the MMF in a targeted mode, extracting the SZ flux from a position centered on the X-ray emission peak using the ``universal'' pressure profile of \citet{2010Arnaud} as a spatial template, in an aperture $\theta_{500}$ corresponding to the $\Rv$ determined in Sect.~\ref{sec:yxmass} above. The extraction was achieved by excising a $10^\circ \times 10^\circ$ patch with pixel size $1\farcm72$, centered on the X-ray position, from the six HFI maps, and estimating the SZ flux using the MMF. The profiles were truncated at $5\,R_{\rm 500}$ to ensure integration of the total SZ signal.  The flux and corresponding error were then scaled to $\Rv$ using the profile assumed for extraction, yielding the refined estimate of $\YSZ$. The resulting refined mass estimate $M^{\rm SZ}_{\rm 500}$~PSX was calculated as described above, using the $M^{\rm SZ}_{\rm 500}$--$\YSZ$ relation and the measured $(\YSZ, \theta_{500}$) posterior information. 



\begin{figure}[!t]
\centerline{
\includegraphics[width=0.475\textwidth]{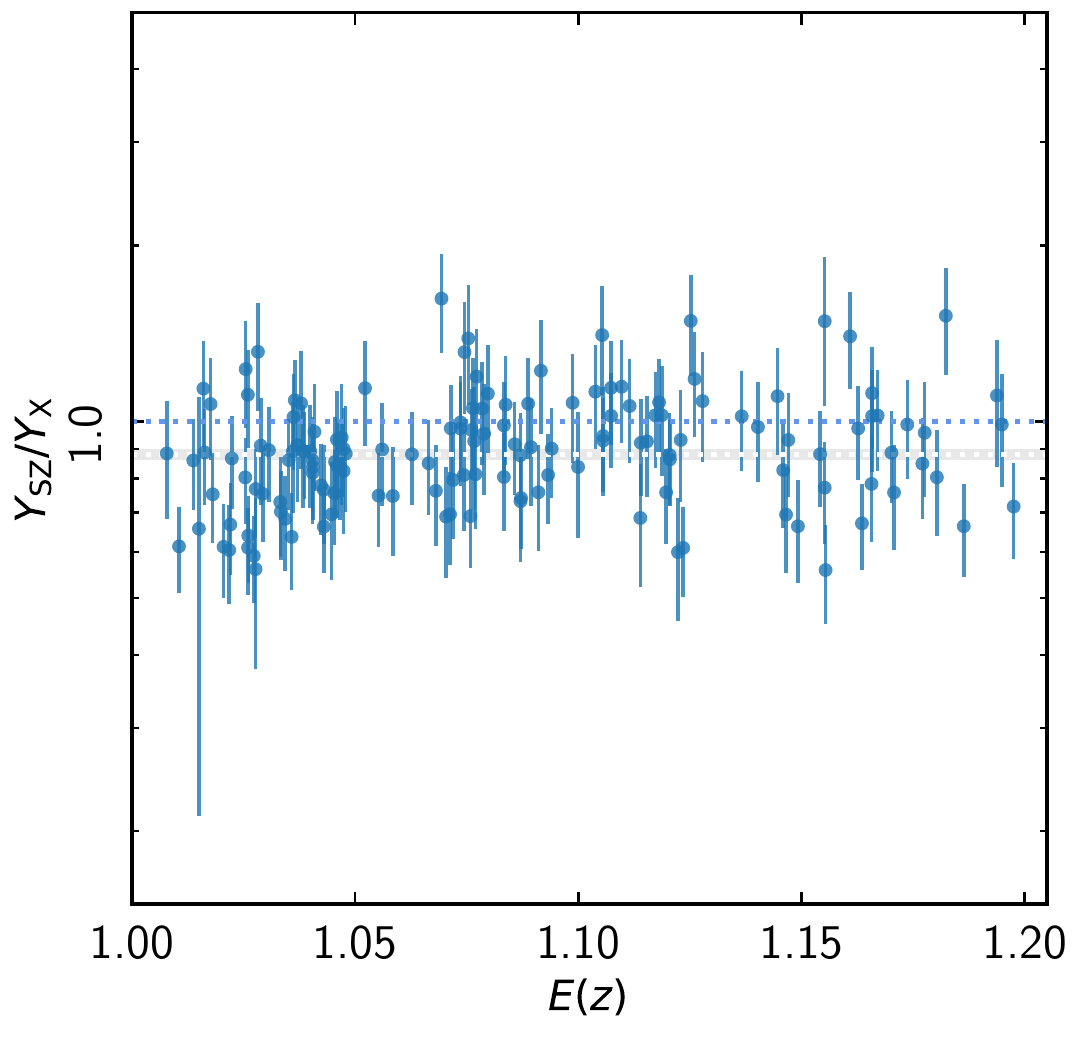} 
}
\caption{Ratio between $\YSZ$ and $\YX$, plotted as a function of $E(z)$. The white dashed line and grey envelope show the error-weighted mean and associated $1\sigma$ uncertainty, $\YSZ/\YX = 0.88\pm0.02$. }\label{fig:yrat}
\end{figure}


\subsection{Fitting Procedure}
\label{sec:fitting}

We fit each scaling relation between a pair of observables $(X,Y)$ with a power law of the form 
\begin{equation}
 Y/Y_0 = A [(X/X_0)]^B  
\label{eqn:powl}
\end{equation}
assuming no evolution, as expected in the self-similar model. 
The reason for the latter decision is that the redshift leverage is insufficient to put a strong constraint on the evolution because the maximum redshift of the sample is  $z < 0.35$. Figure~\ref{fig:yrat} shows that the data are indeed consistent with no evolution in the current sample. 
The pivot points $X_0, Y_0$ were $7 \times 10^{14}$\,\Msun,\   $6\times10^{-5}$ Mpc$^2$, and $3 \times 10^{-3}$ arcmin$^2$ for  $M$, $Y\ [{\rm Mpc}^2]$, and $Y\ [{\rm arcmin}^2]$ respectively. 

The fit was undertaken using linear regression in the log--log plane,  taking the uncertainties in both variables into account and intrinsic scatter. The X-ray and SZ measurements are independent, so there is no covariance between statistical errors for the scaling laws studied here.

We used  the orthogonal Bivariate Correlated Errors and intrinsic Scatter \citep[][\bces]{1996Akritas} linear regression method to perform the baseline fits.  \bces\ is widely used by the astronomical community, giving results that may easily be compared with other data sets fitted using the same method. 
The intrinsic vertical scatter was computed from the quadratic difference between the raw scatter and that expected from the statistical uncertainties, as described in \citet{pra09} and \citet{planck2011-5.2b}.  

We also fit the data using a Bayesian maximum likelihood estimation approach, with Markov Chain Monte Carlo (MCMC) sampling. Using the \texttt{emcee} package developed by \citet{for13},  we write the likelihood as described by  \citet{rob15}.
{\small 
\begin{eqnarray}
     \ln{L}  & =  & \sum_{i=1}^N \left[    \ln{  \frac{ B^2 + 1}{\sigma_i^2} }  - \frac{\left(  \ln{(Y_i/Y_0)}- \ln{A} -B \ln{(X_i/X_0)}   \right)^2  } {\sigma_i^2} \right] \\
  \sigma_i^2  &= & \sigma^2    + B^2\sigma_{X,i}^2  + \sigma_{Y,i} ^2 
\end{eqnarray}
}
with the intrinsic scatter, $ \sigma^2$, as a free parameter. We used flat priors on the parameters, with ranges  $[0.6,1.4]$, $[-1.0,1.0]$, and $[0.0,1.0]$, for $B$, $\ln{A}$, and $\sigma$, respectively. For comparison, we also used the \linmix\ \citep{kel07} and  \lira\ \citep{2016Sereno} Bayesian regression packages, with default priors. The latter, as well as \linmix, allows to correct for the Eddington bias due to the non-uniform distribution of the covariate  in the presence of intrinsic scatter \citep{2015aSereno, 2016Sereno}. For \lira, we used the option \texttt{sigma.XIZ.0=‘prec.dgamma’} and a single Gaussian component for the corresponding latent variable (we checked that adding an  additional component does not change the results). For \linmix, we use the default setting of three gaussians.

The fit results are detailed in Table~\ref{tab:main_results}, where it can be seen that the differences in slope and normalization between methods are always less than the $1\sigma$ statistical error on the parameters. 
We also note the excellent agreement between \lira\ and \bces, suggesting that the Eddington bias due to the non-uniform distribution in the X variable is not an issue.

Table~\ref{tab:main_results} also gives the  mean ratio between SZ and X-ray quantities. This was computed by maximizing the likelihood of the logarithm of the ratios in log-log space, taking into account the statistical errors and an intrinsic scatter. We also computed the ratio by weighting each value of the logarithm ratio by the quadratic sum of the statistical error and the intrinsic scatter. This scatter and the weighted mean were then estimated simultaneously by iteration. Both methods give consistent results. 

In \citet[]{2014Planck29}, the \YSZYX\ relation for the PSZ1 catalog was corrected for the Malmquist bias affecting the $\YSZ$ determination (see their Figure 32).  The correction was made on a cluster-by-cluster basis, following the approach proposed by \citet{2009Vik}.  The clusters in our study are from the ESZ catalog, which were detected in the \planck\ maps corresponding to the first 10 months of the survey. However, here we use the SZ signal from the PSZ2 catalog, re-extracted from the full mission maps. Although the maps are not fully independent, they correspond to significantly different  noise realizations and we thus expect the re-extracted SZ signal from the PSZ2 maps to be much less affected by  Malmquist bias than the original ESZ detection signal.  As a first approximation, in the following   we  assume that the SZ signal is independent of the detection and  neglect residual selection effects from Malmquist bias.  We discuss the impact of  this assumption extensively in Sect.~\ref{sec:mbias} and Appendix~\ref{appx:mbias}.

\begin{figure*}[!ht]
\centerline{
\includegraphics[width=0.485\textwidth]{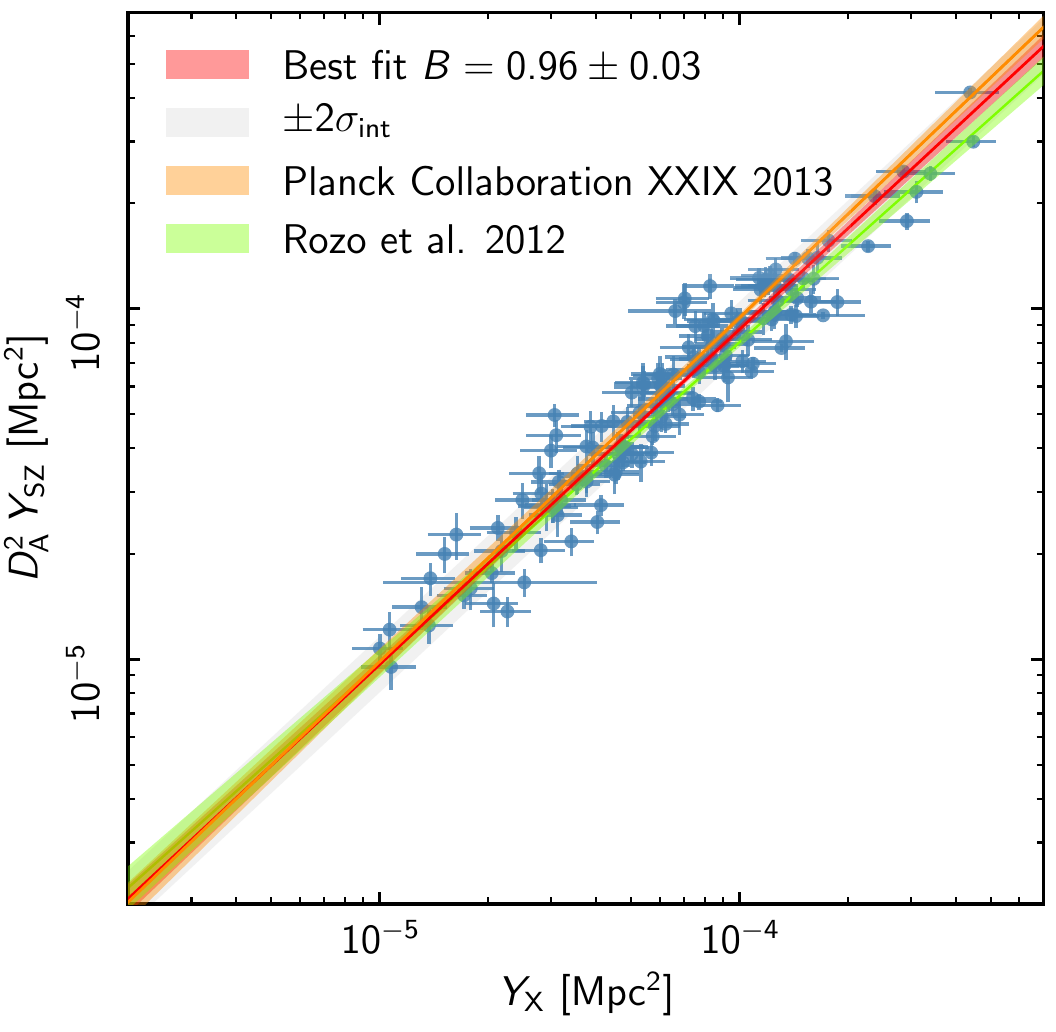} 
\hfill
\includegraphics[width=0.485\textwidth]{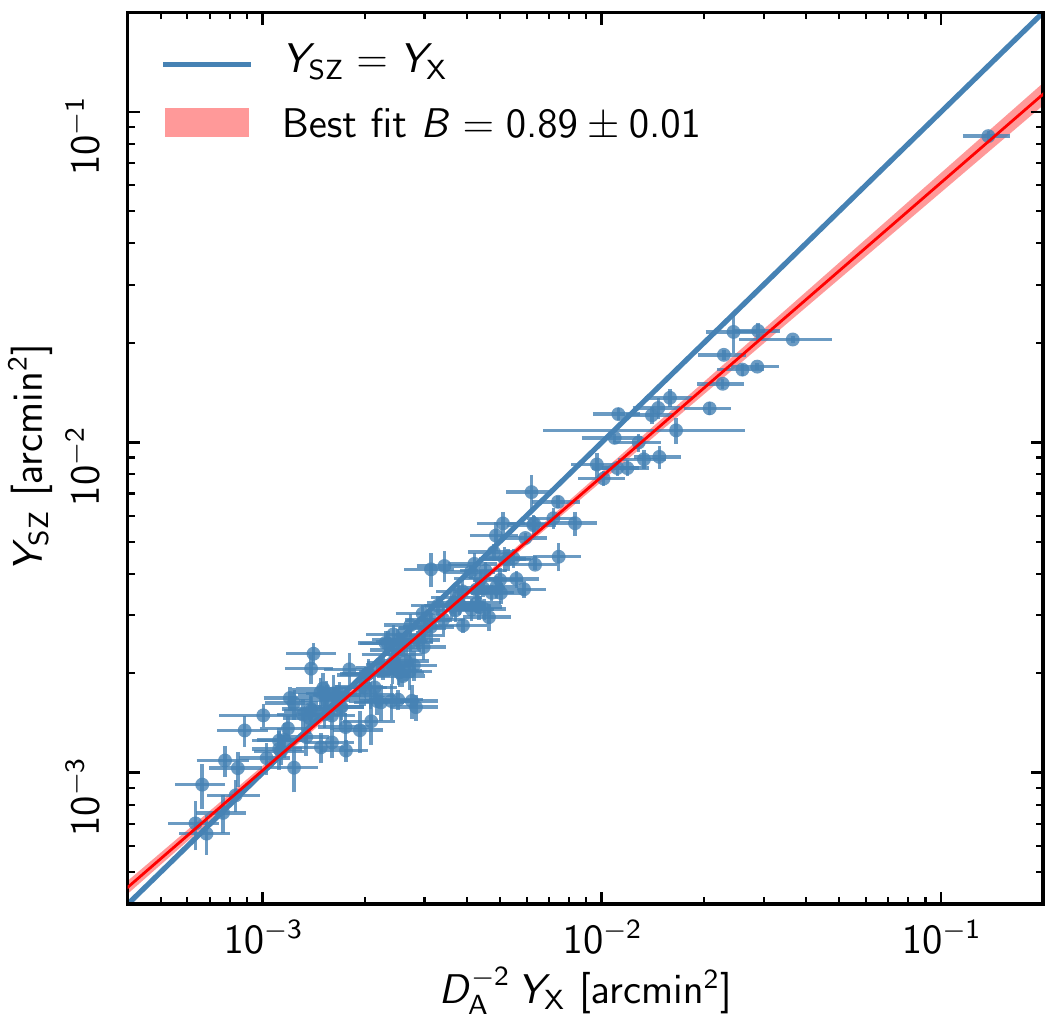} 
}
\caption{The \YSZYX\ relation for the ESZ sample. Left: Correlation between $\DA^{2}\YSZ$ and $\YX$ in units of Mpc$^2$, and comparison with previous works. Right: Correlation between $\YSZ$ and $\DA^{-2}\YX$ in units of arcmin$^2$. The SZ signal is re-extracted in the full-mission maps at the X-ray position and size.} \label{fig:yy} 
\end{figure*}
\begin{figure*}[!th]
\begin{center}
\resizebox{\textwidth}{!} {
\includegraphics[width=0.475\textwidth]{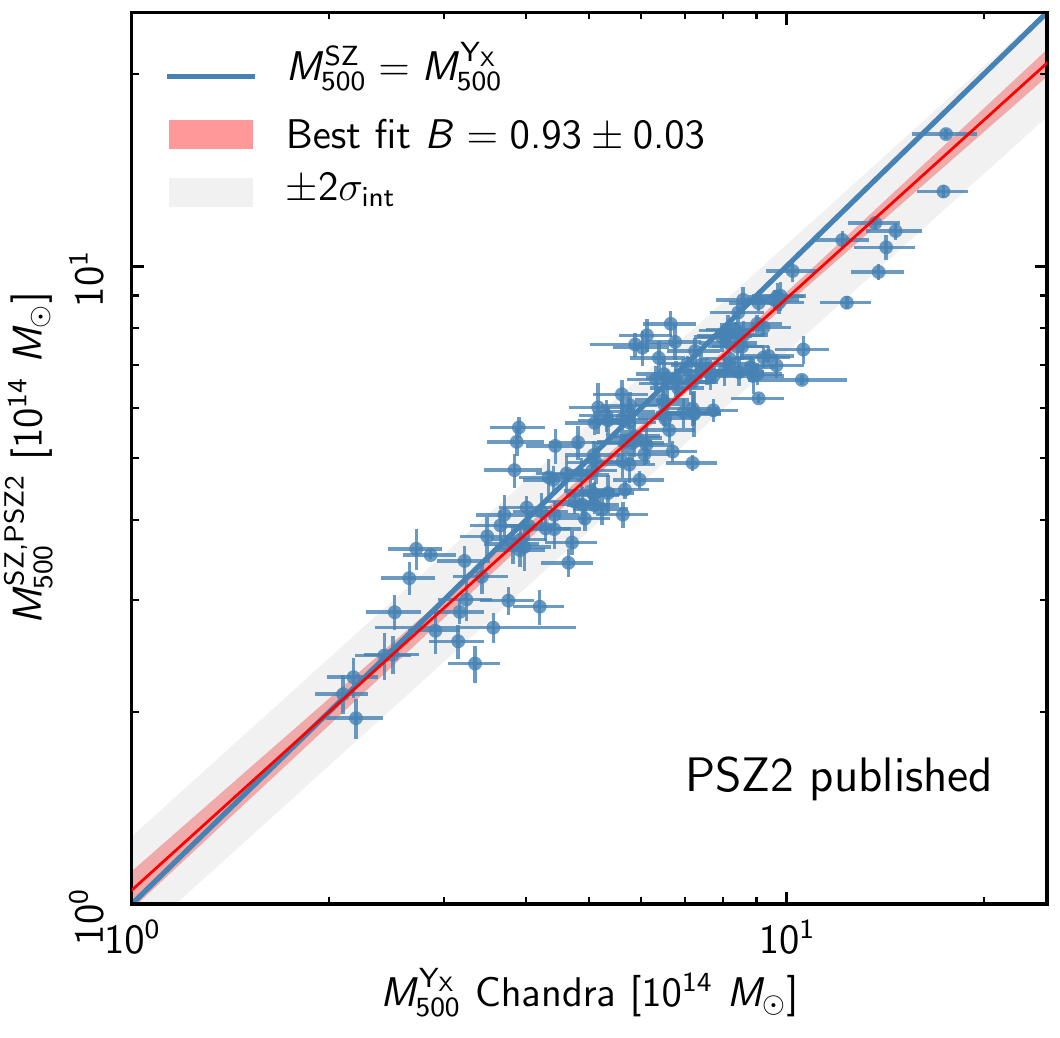} 
\hfill
\includegraphics[width=0.475\textwidth]{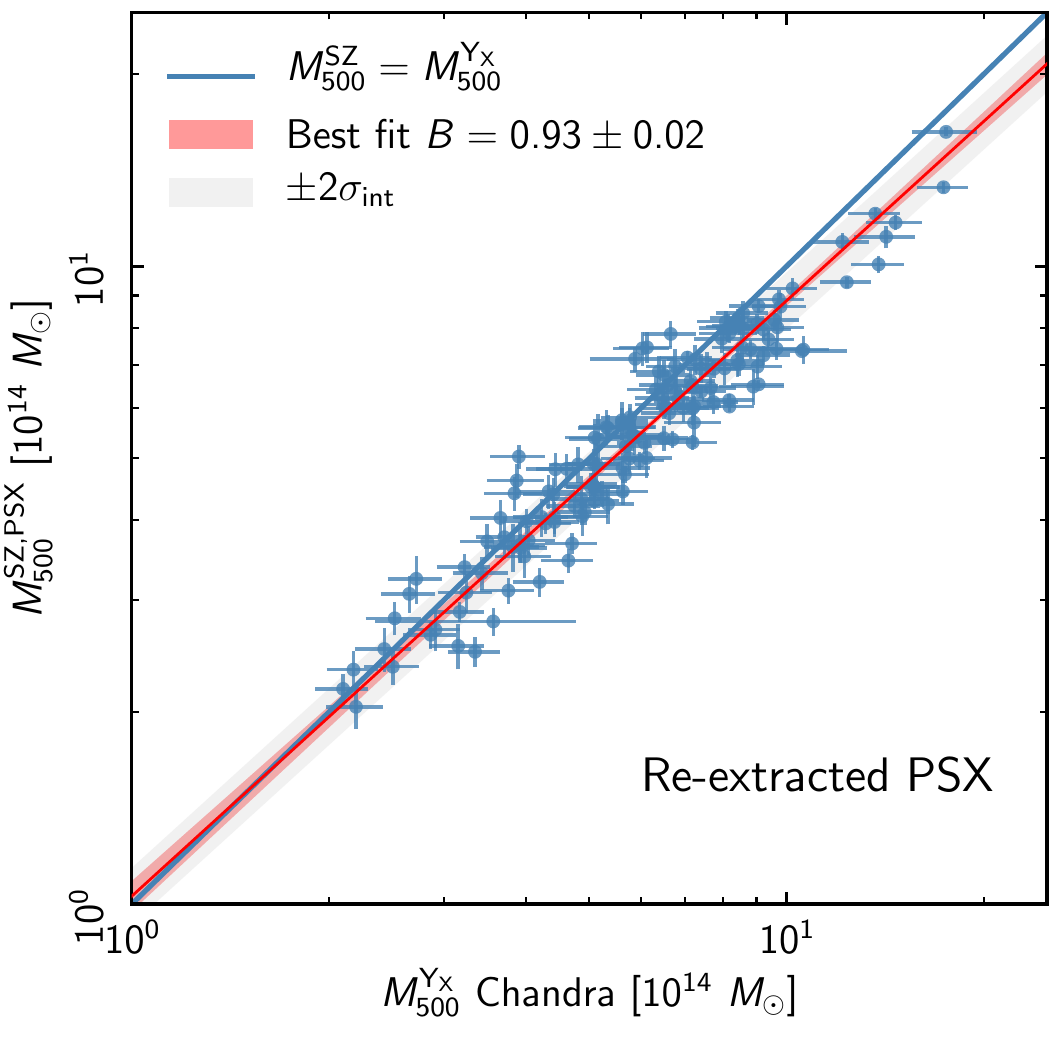}
}
\end{center}
\caption{\footnotesize{Comparison between X-ray and SZ derived masses
for clusters in the PSZ2 sample. Left: \planck\ mass
published in the PSZ2 catalog, $\MPSZ$, iteratively estimated from the $\MSZ$--$\YSZ$ relation and the $\YSZ$ -- size
degeneracy curve. The X-ray mass is derived from \chandra\  data using the $\YX$ mass
proxy.  Right: Same with SZ signal and masses re-extracted at
the X--ray position and size, $\MPSX$. The blue and red lines 
correspond to the identity and the best fit power law, respectively.  
The red regions correspond to the $\pm 1\sigma$ uncertainty and scatter
around the best fit. The gray regions correspond to the $\pm 2\sigma$
intrinsic scatter.}}
\label{fig:mm}
\end{figure*}

\begin{table}[]
\newcolumntype{L}{>{\columncolor{white}[0pt][\tabcolsep]}l}
\newcolumntype{R}{>{\columncolor{white}[\tabcolsep][0pt]}l}
\caption{{\footnotesize Best fit parameters for scaling relations, obtained with \bces, \linmix,  MCMC,  and \lira\  fitting procedures.} 
\label{tab:screl}}       
\begin{center}
\begin{tabular}{@{}Lccc}
\toprule
\toprule

\multicolumn{1}{c}{Relation} & \multicolumn{1}{c}{ $\ln{A}$ }&
\multicolumn{1}{c}{$B_{}$} & \multicolumn{1}{c}{$\sigma_{\rm ln,i}$} \\
\midrule

BCES orthogonal \\

\YSZYXmpc\ &  $-0.115\pm 0.016$ & $0.96\pm 0.03$ & $0.092\pm0.028$ \\
\YSZYXarc\ & $-0.106\pm 0.016$ & $0.89\pm 0.01$ & -- \\
$\MPSZ$--$\MX$ & $-0.089\pm 0.011$ & $0.93\pm 0.03$ & $0.096\pm0.012$ \\
$\MPSX$--$\MX$ & $-0.101\pm 0.009$ & $0.93\pm 0.02$ & $0.051\pm 0.016$ \\

\midrule

LINMIX \\

\YSZYXmpc\ &  $-0.118\pm 0.018$ & $0.95\pm 0.02$ & $0.099\pm0.071$ \\
\YSZYXarc\ & $-0.103\pm 0.015$ & $0.89\pm 0.02$ & $0.056\pm0.049$ \\
$\MPSZ$--$\MX$ & $-0.096\pm 0.012$ & $0.89\pm 0.03$ & $0.098\pm0.048$ \\
$\MPSX$--$\MX$ & $-0.105\pm 0.011$ & $0.92\pm 0.02$ & $0.055\pm 0.039$ \\

\midrule

MCMC \\

\YSZYXmpc\ &  $-0.118\pm 0.017$ & $0.94\pm 0.02$ & $0.085\pm0.028$ \\
\YSZYXarc\ & $-0.103\pm 0.015$ & $0.88\pm 0.02$ & -- \\
$\MPSZ$--$\MX$ & $-0.096\pm 0.013$ & $0.90\pm 0.03$ & $0.095\pm0.012$ \\
$\MPSX$--$\MX$ & $-0.107\pm 0.010$ & $0.91\pm 0.02$ & $0.050\pm 0.015$ \\

\midrule
LIRA & & &$\sigma_{\rm ln,i,Y\vert Z}$ \\

\YSZYXmpc\ &  $-0.118\pm 0.017$ & $0.96\pm 0.02$& $0.053 \pm 0.032$ \\ 
\YSZYXarc\ & $-0.103\pm 0.015$ & $0.89\pm 0.02$& $0.032 \pm 0.019$\\
$\MPSZ$--$\MX$ & $-0.090\pm 0.013$ & $0.93\pm 0.04$& $0.058 \pm 0.031$ \\
$\MPSX$--$\MX$ & $-0.102\pm 0.010$ & $0.93\pm 0.02$ & $0.032\pm 0.016$\\ 
\midrule
Ratio &   Mean value & $\sigma_{\rm ln,i}$\\
\midrule
$\YSZmpc/\YX$  & $0.882\pm0.015$& $0.094_{-0.027}^{+0.035}$ \\
$\MPSZ/\MX$  & $0.922\pm0.09$ &$0.105_{-0.013}^{+0.012}$\\
$\MPSX/\MX$ & $0.910\pm0.09$& $0.057_{-0.015}^{+0.019}$\\
\bottomrule
\end{tabular}
\end{center}
\tablecomments{Relations are expressed as  $Y/Y_0 = A [(X/X_0)]^B$ (Eq.~\ref{eqn:powl}). 
The pivot points $X_0, Y_0$ were $7 \times 10^{14}$\,\Msun,\ $6\times10^{-5}$ Mpc$^2$ and $3 \times 10^{-3}$ arcmin$^2$ for $M$, $(\YSZmpc,\YX) [{\rm Mpc}^2]$, and $(\YSZ,\YXarc) [{\rm arcmin}^2$], respectively. 
For all the relations $\sigma_{\rm ln,i}$ is the  intrinsic scatter in log-log space,  except  for \lira, for which we give  the intrinsic scatter with respect to the latent variable, $\sigma_{\rm ln,i,Y\vert Z}$, constrained by the method. The intrinsic scatter is not presented for the BCES orthogonal and MCMC fitting procedures of the \YSZYXarc\ relation, because for each of those, the raw and statistical scatters were compatible.}
\label{tab:main_results}
\end{table}


\section{Results}


\subsection{The \YSZYX\ relation}

The total  cylindrical Sunyaev-Zel'dovich signal is given by the integral of the Compton
parameter, $\YSZ = \int y ~d\Omega$, where $\Omega$ is the solid
angle, and the Compton parameter $y$ is given by:
\begin{eqnarray}\label{eq:ysz}
y = \frac{\sigma_{\rm T}}{m_{\rm e}c^2} \int_l n_{\rm e}~kT_{\rm e} ~dl,
\end{eqnarray}
where $\sigma_{\rm T}$ is the Thomson cross section, $m_{\rm e}c^2$ is
the electron rest mass energy, $l$ is the distance along the 
line of sight, and $k$ is the Boltzmann constant. The total
Sunyaev-Zel'dovich signal within $\Rv$  can also be expressed as (in its spherical form)\footnote{The \planck\ Collaboration performed the Compton integral within a sphere, instead of performing it along the line of sight to infinity (cylindrical integral).} :
\begin{eqnarray}\label{eq:ysz_int}
\YSZ = \frac{\sigma_{\rm T}}{D_{\rm A}^2m_{\rm e}c^2} \int_V n_{\rm e}~kT_{\rm e} ~dV,
\end{eqnarray}
where $D_{\rm A}$ is the angular size distance of the cluster and $V$ is the sphere of radius $\Rv$ .

The X-ray analog of the Sunyaev-Zel'dovich Compton parameter is
defined as $\YX = M_{\rm gas} \times kT_{\rm X}$, where 
$M_{\rm gas}$ is the gas mass given by:
\begin{eqnarray}\label{eq:mgas}
  M_{\rm gas} = \int_{V} \rho_{\rm gas} ~dV =  \mu_{\rm e}  m_{\rm p} \int_{V} n_{\rm e} ~dV,
\end{eqnarray}
where $\mu_{\rm e}$ is the mean molecular weight per electron,
$m_{\rm p}$ is the proton mass, and $T_{\rm X}$ is the X-ray measured
temperature in the (0.15--1.0) $\times ~ \Rv$ radial range.

On the other hand, the mass-weighted temperature is given by:
\begin{eqnarray}\label{eq:tspec}
T_{\rm mw} \equiv \frac{\int_{V} n_{\rm e} T_{\rm e}~dV}{\int_{V} n_{\rm e}~dV},
\end{eqnarray}
which, when combined with Equations \ref{eq:ysz} and \ref{eq:mgas},
results in:
\begin{eqnarray}\label{eq:dAsqYsz}
D_{\rm A}^2 \YSZ = c^\ast M_{\rm gas}T_{\rm mw},
\end{eqnarray}
where $c^\ast$ is given by:
\begin{eqnarray}\label{eq:C}
c^\ast = \frac{\sigma_{\rm T}}{m_{\rm e}c^2} \frac{1}{\mu_{\rm e} m_{\rm p}} 
= 1.38 \times \frac{10^{-5}~{\rm Mpc^2}}{10^{14}~{\rm keV}~ M_\odot},
\end{eqnarray}
where we used $\mu_{\rm e} = 1.1706$, which was calculated assuming 
a fully ionized plasma with metallicity of 0.3 $Z_\odot$, and 
reference values from \citet{1989AndersGrevesse}.

 In practice, when computing $M_{\rm gas}$, we used the fact that the emission measure
is proportional to $\langle n_{\rm e}^2 \rangle$, averaged
in radial annuli, to estimate $\langle n_{\rm e} \rangle^2$. For a uniform spherical gas distribution, $\langle n_{\rm e}^2 \rangle = \langle n_{\rm e} \rangle^2$, 
so one can recover the gas density. However, the reality is different,
with typically $\langle n_{\rm e}^2 \rangle > \langle n_{\rm e} \rangle^2$. Defining the clumping
factor as $Q = \langle n_{\rm e}^2 \rangle / \langle n_{\rm e} \rangle ^2$, we have 
$M_{\rm gas}^{\rm obs} = Q M_{\rm gas}^{\rm true}$. By definition
$\YX = M_{\rm gas}^{\rm obs} T_{\rm X}$, so we can write:
\begin{eqnarray}\label{eq:dAsqYsz_2}
\frac{D_{\rm A}^2 \YSZ}{c^\ast \YX} = \frac{1}{Q} \frac{T_{\rm mw}}{T_{\rm X}}.
\end{eqnarray}

\begin{figure*}[!t]
\centerline{
\includegraphics[width=0.475\textwidth]{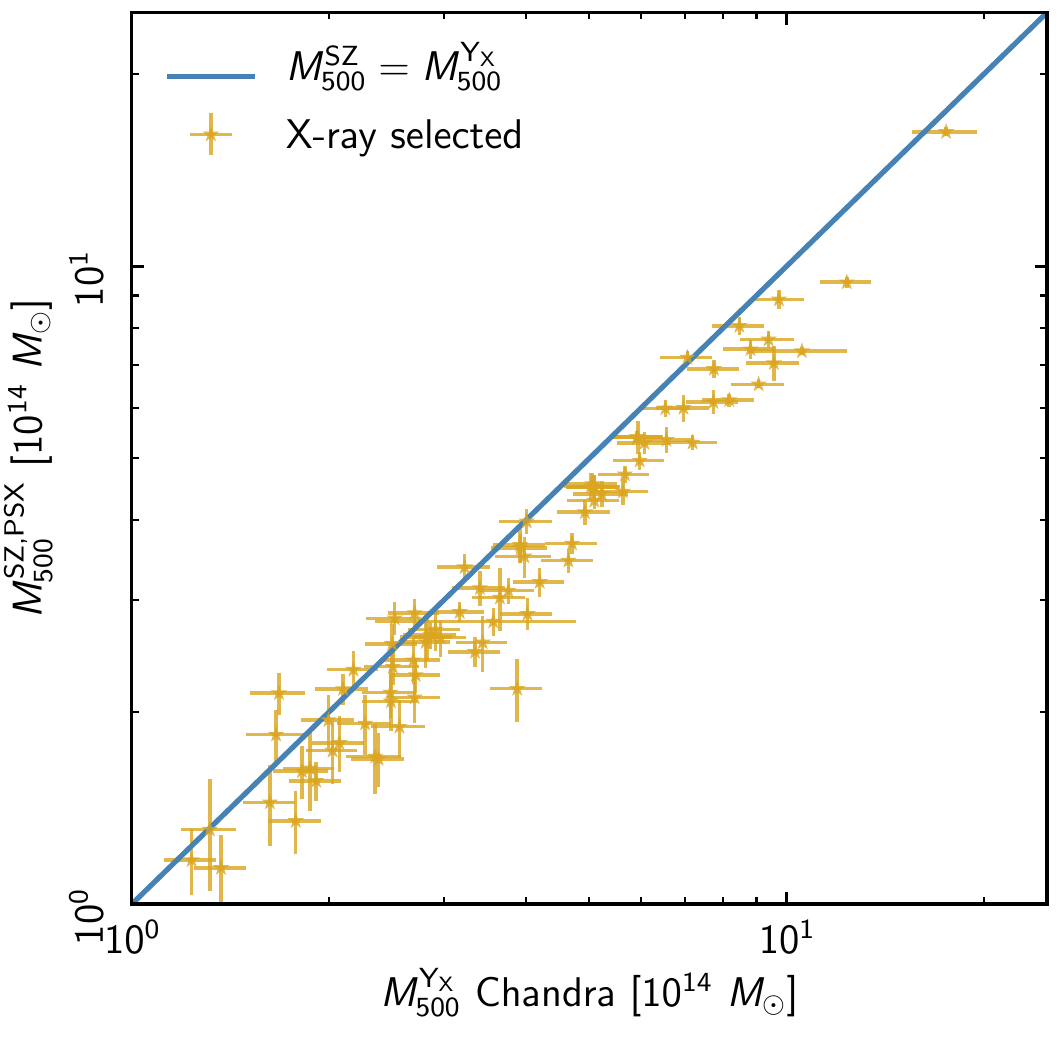}
\hfill
\includegraphics[width=0.475\textwidth]{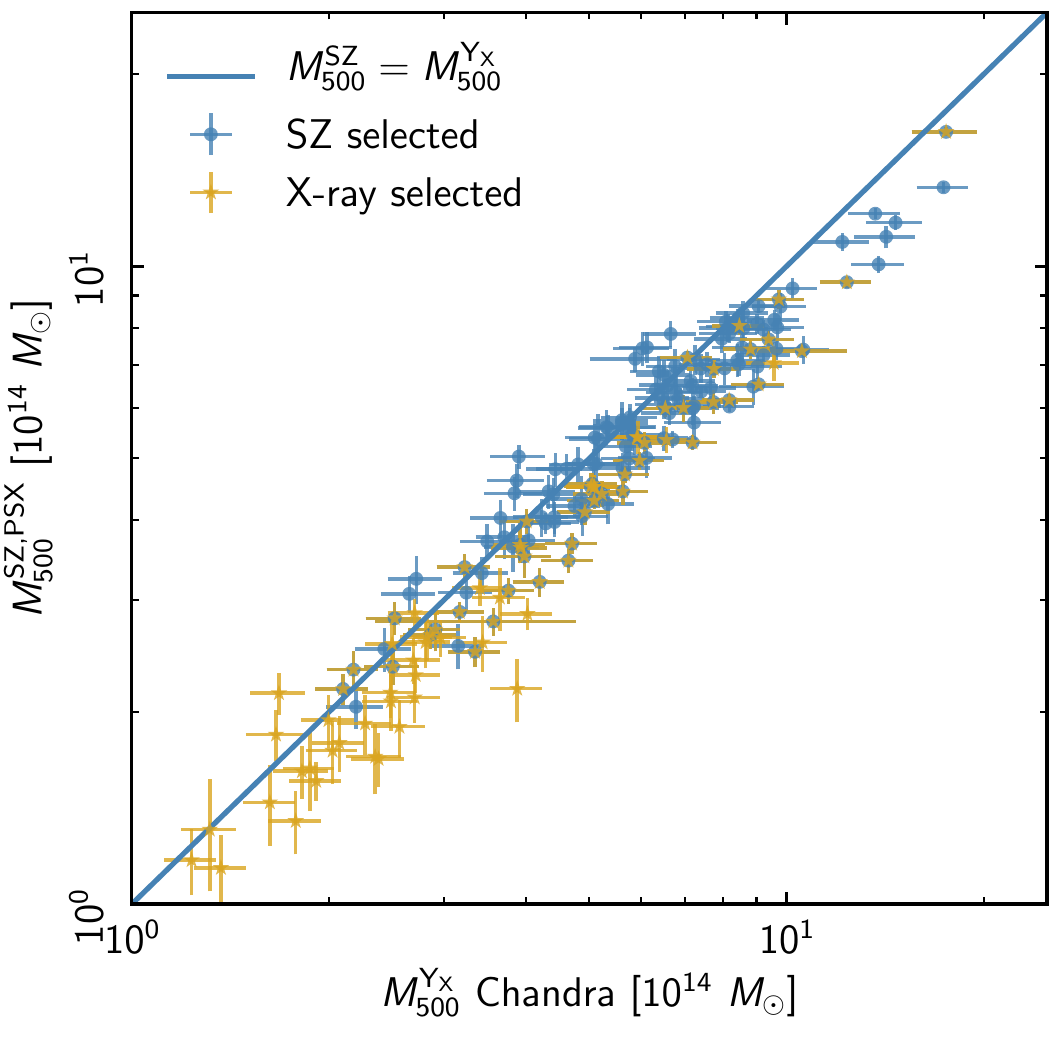}
}
\centerline{
\includegraphics[width=0.475\textwidth]{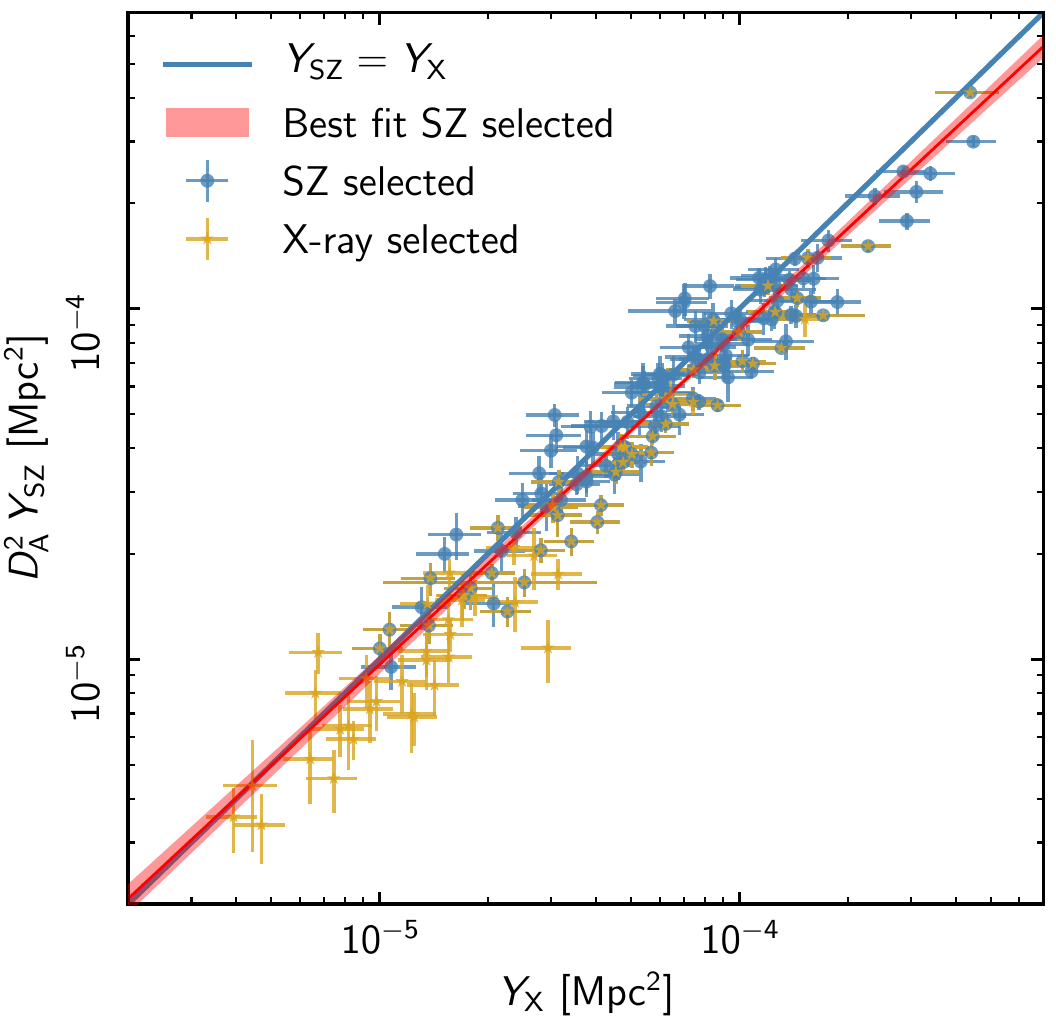} 
\hfill
\includegraphics[width=0.475\textwidth]{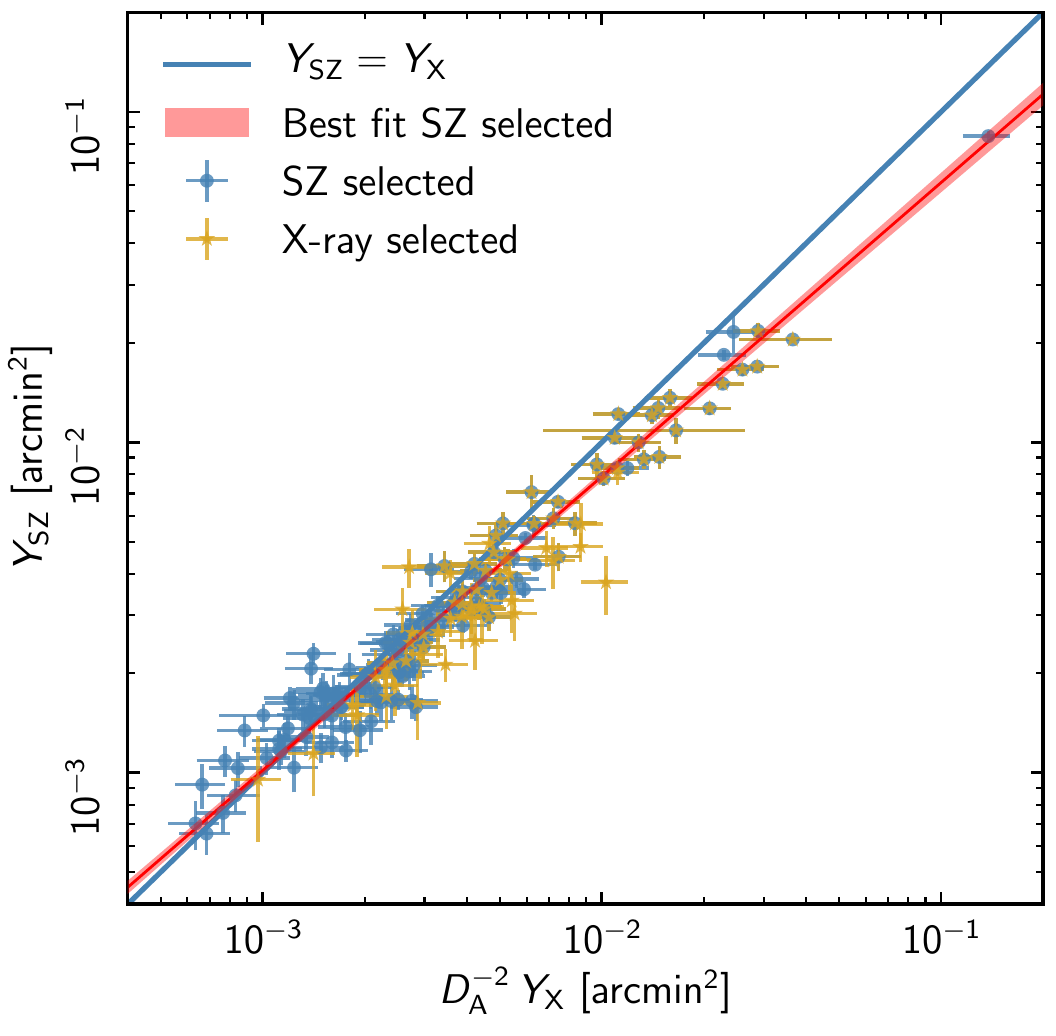} 
}
\caption{\footnotesize{Top left: the \MSZMX\ relation for X-ray selected clusters only. Top right: the \MSZMX\ relation with SZ and X-ray selected clusters indicated. Bottom left: the \YSZYXmpc relation, with SZ and X-ray selected clusters indicated. 
 Bottom right: same as bottom left panel, except for   the flux \YSZYXarc relation. The SZ signal is extracted at the X--ray position and size.}}
\label{fig:addx}
\end{figure*}

The \YSZYX\ relation is thus a fundamental probe of the consistency of the SZ and X-ray signal. When extracted within the same aperture (here
\chandra\  $\Rv$), the ratio does not depend on any mass calibration. It depends on clumpiness, internal structure (mass weighted temperature versus  spectroscopic temperature), and absolute calibration of X-ray measured temperature \citep[see e.g. the discussion in][]{ada17}. The \YSZYX\ relation is expected to have a slope of one (if clusters are self-similar in shape), and a normalization less than one due to the radial dependence of the temperature.  In the following, $\YX$ will refer to the quantity renormalised by the $\c$ factor  (Eq.~\ref{eq:C}), ie. expressed  in units of Mpc$^2$.

The left-hand panel of Figure \ref{fig:yy} shows the   the \YSZYXmpc\ relation, i.e. the correlation between the  cluster physical properties -- the intrinsic  Compton parameter and $\YX$. The slope, $B = 0.96\pm0.03$, is consistent with unity at slightly more than one sigma, while the normalization is $A = 0.89\pm0.09$.  
The slope of the relation is consistent with that found by \citet{2014Planck29}, displayed in Figure \ref{fig:yy} by the orange line,  within the statistical errors. The relation found by \citet{roz12} using \chandra\ data, shown in green in Figure \ref{fig:yy}, has a somewhat shallower slope.

However, the picture changes when the \YSZYXarc\ relation is plotted, i.e.  when considering the observed SZ flux (arcmin$^2$).
When plotted this way, the slope is clearly not consistent with unity,  $B = 0.89\pm0.01$. The effect is similar to, and in agreement with, that observed in \citet{2014Planck29}, $B= 0.91\pm 0.02$, although at higher statistical precision (see their  Section 7 and Appendix D). 


\subsection{Comparison between SZ and X-ray derived masses}

We next investigated the relation between the published \planck\ masses, $\MPSZ$, and those obtained from the \chandra\ data via the \MYX\ relation, $\MX$, as described above in Section \ref{sec:yxmass}. 
The left-hand panel of Figure \ref{fig:mm} compares the  published \planck\ PSZ2 masses to those from \chandra\ for the 147 ESZ objects. The slope of the relation, $B = 0.93\pm0.03$, is less than unity at slightly more than the $2\sigma$ level. This translates into an $\MSZ/M_{500}^{\rm X}$ ratio that decreases from $\sim 0.99$ to $\sim 0.88$ between $\Mv = 2
\times 10^{14} ~\msol$ and  $\Mv = 10^{15} ~\msol$. 

The right-hand panel of Figure \ref{fig:mm} shows the SZ mass re-extracted at the X-ray position and size, $\MPSX$, compared to the \chandra\ values. The main
effect is a decrease of the intrinsic scatter by about a factor of two (from 11\% to 6\%). The slope, $B = 0.93\pm0.02$ is  unchanged, although the uncertainty is smaller as a result of the decreased scatter. As the X-ray mass is slightly higher than the PSZ2 mass, fixing the size to the X-ray value tends to increase the SZ mass on average, while extracting the SZ signal at the X-ray position has the opposite effect (the blind SZ signal is extracted at the position maximising the signal-to-noise). As a result the median SZ to X-ray mass ratio, $\MSZ/\MX= 0.91 \pm 0.09$,  is  unchanged, within the statistical errors.   

In the following, we use the $\MPSX$  value as reference, as it  is directly  physically related to the  $\MX$ quantity, and we will refer to it as $\MSZ$.


\subsection{X-ray selected clusters}
\label{sec:xcluster}

Here we consider the effect of adding the X-ray selected clusters to the relations  discussed above. These clusters are discussed only for comparison to the SZ-selected objects.

The top-left panel of Figure~\ref{fig:addx} shows the \MSZMX\ relation for the X-ray selected clusters only. It can be seen that these objects clearly populate the low mass part of the plot, with the majority of the systems having $\Mv < 5 \times 10^{14}$ \Msun. The effect of adding the SZ selected clusters, as shown in the right-hand panel of Figure~\ref{fig:addx}, is to transform what was a relation with a slope slightly less than unity (i.e. the SZ selected clusters only) into one with a clear offset in normalization. For the full sample of SZ + X-ray selected clusters, the slope $B = 0.97\pm0.02$, and the median ratio $\MSZ / \MX = 0.90\pm0.14$. 

To shed light on the above, we can use 
the corresponding intrinsic \YSZYXmpc\ relation, which is shown 
in the bottom-left hand panel of Figure \ref{fig:addx}.
 We observe that as expected, the SZ and X-ray selected objects are distributed in the \YSZYX\ plane in a similar fashion to their distribution in the \MSZMX\ plane: the X-ray selected clusters  tend to have higher $\YX$ values for a given $\DA^2\YSZ$.

The bottom-right hand panel of Figure \ref{fig:addx} shows the $\YSZ$-$\YX$ relation  expressed in apparent flux 
 (\YSZYXarc). 
Here we notice that the X-ray selected clusters follow the distribution of SZ selected systems, but are typically distributed towards higher values. This is a consequence of the X-ray selected systems being on average at lower redshift. As a result, when the relation is plotted
in intrinsic quantities
these systems are redistributed in the \YSZYX\ plane
via the $\DA^2$ factor. As noted in \citet{2014Planck29}, the shallower slope in arcmin$^2$ results in an overestimate of the dispersion about the relation in Mpc$^2$ as a result of this redistribution. This is discussed in more detail below.


\section{Discussion}


The  slope of the  \YSZYXarc relation, $B = 0.89 \pm 0.01$, is in good agreement with that found by  \citet{2014Planck29}, and points to a very  significant departure from unity. This is not the case for the \YSZYXmpc\ relation between intrinsic properties, indicating that the $\YSZ/\YX$ ratio variations are more related to the SZ flux than to the intrinsic Compton parameter. This suggests a measurement origin for deviation of the slope from unity, rather than a physical variation of the  $\YSZ/\YX$ ratio ( e.g. with mass).  We first discuss possible measurement biases which may explain  the slope, starting with a possible residual Malmquist bias.

\subsection{Selection bias, scatter, and covariance}
\label{sec:mbias}

When studying scaling relations,  the importance of  taking into account measurement biases is well recognised \citep[e.g.][]{2012Angulo}. 
Relations between observed quantities 
result from the power-law scaling of each observable  with redshift and mass, the most fundamental cluster properties. The intrinsic scatter of each quantity  around the mean relation  may be correlated, through, for instance, a common physical origin. For a cluster sample constructed from a survey, an  important measurement bias is the so-called Malmquist bias. It is most critical for the observable used for the detection, $\mathcal{O}_{\rm det}$, when it is measured from the survey data.  At a given mass, clusters with values that are scattered to higher values of $\mathcal{O}_{\rm det}$, either by noise or by the intrinsic scatter between the observable  and the mass, will be preferentially detected. The average value of $\mathcal{O}_{\rm det}$ will be then biased high, particularly when  close to detection threshold, affecting  the apparent slope of any scaling  relation with $\mathcal{O}_{\rm det}$. This effect is amplified by the nature of the mass function, the number of clusters increasing with decreasing mass \citep[][see their Fig. 5]{2011Allen}.

One may wonder if the deviation from unity of the slope of the \YSZYXarc\  relation is not simply due to residual Malmquist bias. Such a residual bias may be due to correlation between the noise in the ESZ  and PSZ2 maps, as well as the intrinsic scatter of the Compton parameter at given mass. The intrinsic scatter of the $\YX$ proxy with the underlying mass should also be considered, as should its probable correlation with the scatter in $\YSZ$. 

Appendix~\ref{appx:mbias} describes extensive simulations  that were designed to investigate these effects on the observed \YSZYXarc\ relation. These were based on the production of mock ESZ catalogs through the injection of simulated clusters, following the Tinker mass function, into the \planck\ 10-month maps. This was followed by the application of the MMF algorithm to obtain the SZ detections above a $S/N=6$, the threshold of the ESZ catalog.  The detected clusters were then re-injected into the full mission maps and re-analysed by the MMF algorithm to extract their PSZ2 flux values. 
Ten mock catalogs were produced, with a total of 1657 detections. To investigate selection, scatter, and covariance effects, we assumed a Gaussian log-normal correlated distribution for $\DA^2\YSZ$ and $\YX$ at fixed mass $\Mv$ with a covariance matrix. The reference quantity at given mass, following exactly the scaling relation, was denoted $\YM$. The \YSZYXarc\ relation of the mock data was then analysed exactly as for the real data. Introducing  each effect successively (see Fig.~\ref{fig:rels}),  we found that:

\begin{itemize}
\item  The  residual Malmquist bias, due to the fact that the noise realisations of the ESZ and PSZ2 maps are not strictly fully independent, is negligible. The $\YSZ$ flux extracted from the PSZ2 map is unbiased as compared to the reference $\DA^{-2}\YM$ value, in the absence of scatter. This is also confirmed by the absence of correlation, in the real data,  between the $\YSZ/\YX$  ratio and the $S/N$ ratio of the ESZ detection at a given mass (Fig.\ref{fig:mb}). 
\item 
There is a small residual Malmquist bias due to the scatter in the \YSZM\ relation. This is expected, as clusters detected above the ESZ survey threshold, because they are scattered up due to the intrinsic scatter,  will still have a higher $\YSZ$ signal than the average in the PSZ2 maps.  The importance of this effect on the \YSZYM\ relation depends on the relative magnitude of the ESZ  noise and the intrinsic scatter, i.e. which effect drives the cluster detectability.  For reasonable values of  scatter derived from numerical simulations of cluster formation, the effect on the slope is modest, $\Delta\,B\sim-0.02$. 
\item  Intrinsic scatter  in the \YXM\ relation decreases the slope of the \YSZYXarc\ relation, as compared to the \YSZYM\  relation, by $\Delta\,B\sim-0.02$. We do not have  a clear explanation of this  effect, which cannot simply be due to a residual  Eddington-like bias.  However the likely correlation between the $\YSZmpc$ and $\YX$ scatter {\it redresses} the slope back towards unity ($B=0.973\pm0.005$).  This is a well-known effect:  selection bias due to intrinsic scatter  has  little impact on the  relation between observables when clusters scatter almost along the relation \citep{2012Angulo}. 
\end{itemize}

In summary, residual Malmquist bias due to \planck\ noise (statistical scatter)  is negligible, and selection effects due to typical values of intrinsic scatter as predicted by numerical simulations cannot explain the observed behaviour of the \YSZYXarc\ relation. 

An independent mass estimate (such as a lensing mass) is required to make further progress on this issue. This would enable one to constrain simultaneously the \YXM\ and \YSZM\ scaling relations and their intrinsic scatter, taking into account the \planck\  selection function. Such an approach, based on a full modelling of the maximum likelihood of the observed data ($\YSZ$, $\YX$ and mass for each cluster) is now frequently applied to survey data \citep[e.g.][]{2006Stanek,2009Vik,2010Mantz,2016Giles,2019Dietrich}. Application of such a method to the current data set would be more complicated, however, as we would need to combine the detection in the ESZ maps with the measurement of the $\YSZ$ flux in the PSZ2 maps. Such a study is outside the scope of the paper. 
%


\subsection{Other measurement systematic effects}\label{sec:slope_ysz_sx}

Following the tests performed in Section \ref{sec:mbias} and in Appendix~\ref{appx:mbias},  indicating that selection effects
cannot explain the slope of the \YSZYXarc\ relation, we next considered the possibility that 
its departure from unity is due to  systematic issues with the $\YSZ$ or $\YX$ estimates.  We therefore considered a range of possible explanations, including:

\begin{itemize}

\item Relativistic effects: the results presented in this work neglect relativistic effects, which might have an impact on the measured fluxes. In principle the relativistic effect is stronger in hotter, more massive clusters, so it should be most apparent in the slope of the correlation between intrinsic properties. This is clearly not the case (Figure~\ref{fig:yy}). However, to further test this hypothesis, we re-extracted \planck\ fluxes, assuming relativistic corrections to the SZ spectrum \citep{ito98} based on the observed {\em Chandra} temperatures. 
Re-fitting the data obtained with these relativistic corrections, the slope of the \YSZYXarc\ relation remains identical ($0.89 \pm 0.01$). The slope of the \YSZYXmpc\ relation increases only slightly, to $0.99 \pm 0.03$. We conclude that relativistic corrections cannot explain the deviation of the slope from unity for the observed \YSZYXarc\ relation.

\item Non Gaussian beams: the approximation of the \planck\ beam with a Gaussian may cause a loss of flux due to neglect of the side-lobes. This effect would be more important for the most extended objects (i.e. nearby or hot clusters). To test the magnitude of this effect, we re-extracted the $\YSZ$ values using the \planck\ effective beams available through the \planck\ Legacy Archive\footnote{\url{https://pla.esac.esa.int/}}. We found a median ratio of $\YSZ^{\rm Gauss} / \YSZ^{\rm eff.} = 1.00\pm0.02$, with essentially zero dependence on angular extent.

\item Variation in profiles: as detailed in Section \ref{sec:planckdata}, re-extraction of the \planck\ flux is undertaken assuming the `universal' pressure profile of \citet{2010Arnaud} as a fixed spatial template. It is well known that individual profiles may deviate significantly from this. Furthermore, nearer clusters are better resolved in the \planck\ maps than more distant objects, and so there may be a distance dependence in this variation. However, comparison of $\YSZ$ fluxes measured using a fixed `universal' template and one based on the individual pressure profile yielded a best-fit slope and normalisation entirely consistent with unity \citep[see Figure B.1 of][]{PIPV}.

\item X-ray temperature calibration issues: the observed good agreement in slope seen in Figure \ref{fig:yy} between the present study 
and the previous work by \citet{2014Planck29}, which used \xmm\ data,  suggest, empirically,  
that this  cannot explain the observed slope.  Furthermore, as for the  relativistic effect, any temperature-dependent systematics should be more visible  in the correlation in Mpc$^2$ than in arcmin$^2$, which is clearly not the case. The effect of \xmm\ versus \chandra\ calibration  on the scaling relations is further discussed Sec.~\ref{sec:xmmchandra}.

\end{itemize}

In summary, none of the above possibilities explains why the slope of the \YSZYX\ relation is significantly less than unity when expressed in units of apparent flux (arcmin$^2$).

\subsection{Link between the \MSZMX\  and the \YSZYX\  relations.}

The slope of the  intrinsic  \YSZYXmpc\ relation is significantly different from that of the \YSZYXarc\ relation, and is consistent with unity. The difference between the two relations is a conversion from flux to intrinsic properties, i.e. a linear rescaling of each quantity by the angular distance  $\DA^2$ of each cluster.  In Appendix \ref{appen:diff_slope}, we demonstrate analytically that this transformation results in a  \YSZYXmpc\ relation slope that is closer to unity, as observed ($B = 0.96 \pm 0.03$), and increases the dispersion about the relation. This effect is actually  caused  by re-ordering of the data from the  \YSZYXarc\ plane to the \YSZYXmpc\ plane, as already noted in Sec.~\ref{sec:xcluster} for X--ray clusters. 

The properties of the \MSZMX\ relation, derived from the mass proxies  
$\MSZ$ and $\MX$, originate in  
the properties  of the intrinsic  \YSZYXmpc\ relation. One can translate between the two 
by making use of the \MYX\ and the \MYSZ\  relations.  We can express the relations as follows:
\begin{eqnarray}\label{eq:scaling_relations}
\YSZmpc  &\propto&(\YX)^\alpha \nonumber \\
\MSZ &\propto&( \MX)^\beta \nonumber \\
\MX &\propto& (\YX)^\gamma \nonumber \\
\MSZ &\propto& (\YSZmpc)^\delta,
\end{eqnarray}
where $\alpha$, $\beta$, $\gamma$, and $\delta$ noting the slope  of the relations above (only here). The above leads to $\beta = \alpha \times \delta / \gamma$. In this work we obtained $\alpha = 0.96\pm 0.03$. From Equation \ref{eq:y_m} we have 
$\delta = 1/1.789=0.56$, and from \citet{2009Vik} we have $\gamma = 0.57$, which leads to 
$\beta = (0.96\pm 0.03) \times (0.56) / 0.57 = 0.94 \pm 0.03$, which is
basically what one gets in Figure \ref{fig:mm} from the direct fit.  
In other words, the deviation of the slope from unity is simply slightly 
amplified  when considering the mass proxies,  
rather than the measured $Y$ parameters, because the
\MYSZ\ and \MYX\ relations do not have exactly the 
same slopes (although they are consistent).
Furthermore, the logarithmic intrinsic scatter of the  \MSZMX\  is $\sigma_{\rm ln,i} = 0.051$, about $\gamma$  times the intrinsic scatter of the \YSZYXmpc\ relation, $\sigma_{\rm ln,i} =0.092$,   as expected.

\subsection{Effect of X-ray temperature calibration and of a different \MYX\ calibration}
\label{sec:xmmchandra}

It is well-known that there is a systematic difference between the X-ray  temperature derived from \chandra\ and \xmm\, the former yielding temperatures  $\sim$ 7\% larger than \xmm\ in the mass range considered here \citep{2015Schellenberger}. Consistently, there is also a slight difference of $\sim 4\%$ in the normalisation of the \MYX\ relations obtained by \citet{2009Vik}, which is used here, and that obtained by \citet{2010Arnaud}. (The slope difference between the two relations of $0.01\pm0.02$ is completely negligible.)  Here we estimate analytically the effect of using {\em XMM-Newton} data and the  \citet{2010Arnaud} relation on the \YSZYX\ and \MSZMX\ relations.

\subsubsection{Effect on measured quantities}
We consider first the effect  on $\YX$ and the corresponding mass $\MX$. We write the relation as $\MX\! ~ \propto~ \!A \YX^B$, with $\YX\!=\!\Mg(r\!<\!\Rv)\,\TX$. The quantities $\MX$ and $\YX$ are estimated iteratively, and $\YX$ does not simply scale linearly with  $\TX$ as $\Mg$ depends sensitively on the aperture. For simplicity, we assume that  the gas density is given by a $\beta$-model with $\beta\!=\!2/3$, which leads to a gas mass $\Mg(r\!<\!\Rv)\propto \Rv$.  On the other hand, we can consider that the gas mass derived from \xmm\ and \chandra\ data are the same \citep{2017Bartalucci} and the core excised gas temperature  $\TX$ is not changed by a (small) change in aperture.  Noting that $\MX \propto  \Rv^3 $ by definition, we then obtain:
\begin{eqnarray}
 \YX &\propto  &A^{1/(3-B)}\TX^{3/(3-B)}  =  A^{0.41}\TX^{1.23}  \nonumber \\
  \MX &\propto &A^{3/(3-B)}\TX^{3B/(3-B)}   = A^{1.23}\TX^{0.70} \nonumber \\
 \Rv &\propto  &A^{1/(3-B)}\TX^{B/(3-B)}  = A^{0.41}\TX^{0.23} 
\end{eqnarray}
for $B=0.57$ (Eq.~\ref{e_yx_m}). 

The SZ quantities, $\YSZ$ and the corresponding $\MSZ$ re-extracted within the X-ray aperture,  will also  be affected by the $\Rv$ aperture change.
From our analysis of \planck\ data for the present sample, the variation of $\YSZ$ with aperture $\theta_{500}$ around the nominal $\theta_{500}$ value has a logarithmic slope of $~0.8$, so that:
\begin{equation}
 \YSZ  \propto A^{0.8/(3-B)}\TX^{0.8 \times B/(3-B)}
 \end{equation}
The corresponding change in SZ mass,  $\MSZ$,  can be obtained  from  the \MYSZ\ relation (Eq.~\ref{eq:angle}). We can then insert the numerical values for the \chandra\ and \xmm\ temperature difference and for the normalisation of the  \MYX\ relation: $\TXc\!=\!1.07\,\TXx$;  $A_{\rm XMM}\!=\!0.96\,A_{\rm CXO}$ with  $B\!=\!0.57$.  We then obtain $\YXx\!\sim\!0.90\,\YXc$,   $\MXx\!\sim\!0.88\,\MXc$, $\YSZx\!\sim\! 0.97\,\YSZc$ and $\MSZx\!\sim\! 0.99\,\MSZc$. 

\subsubsection{Effect on the scaling relations}

We can now estimate how these changes in cluster parameters  would affect the  \YSZYXmpc\ and \MSZMX\ relations. 
For constant multiplying factors that change $\YX$, $\YSZ$ and corresponding masses,  the slopes would not change, however, the normalizations would. We consider  the following power-laws:  $\YSZmpc\!=\!A (\YX)^\alpha$ and $\MSZ\!=\!B (\MX)^\beta$, with   $\alpha=0.96$ and $\beta=0.93$ corresponding to the best fitting values (Table~\ref{tab:main_results}). The normalisation will change following the ratio of \xmm\ and \chandra\ quantities, given above, as 
$A_{\rm XMM} = A_{\rm CXO} \times \left[0.97/0.90\right]^{\alpha}$, and   
$B_{\rm XMM} = B_{\rm CXO} \left[0.99/0.88\right]^{\beta}$.  
Had we used {\em XMM-Newton} data and \citet{2010Arnaud} \MYX\ relation,  we estimate that the slopes for the   \YSZYX\  and the  $\MSZ$--$\MX$ relations would be the same, and the normalizations would be higher, by $\sim 7\%$ and $12\%$, respectively. 

Finally, we compare our results with those from \citet{2014Planck29}. For comparison purposes, we express the \YSZYXmpc\ relation using the pivot point of \citet{2014Planck29}, $Y_{\rm p} = 10^{-4} \rm Mpc^2$. In this work, we obtain $\YSZ/Y_{\rm p} = 0.87 \times (\YX^{\rm CXO}/Y_{\rm p})^{0.96}$. Using the estimated changes in slope and normalization for {\em XMM-Newton} data derived above, $\alpha_{\rm XMM} = \alpha_{\rm CXO}$ and $A_{\rm XMM} = 1.07 ~ A_{\rm CXO}$, we obtain:  
$\YSZ/Y_{\rm p} = 0.93 \times (\YX^{\rm XMM}/Y_{\rm p})^{0.96}$, which is in perfect agreement with the result from \citet{2014Planck29}, $\YSZ/Y_{\rm p} = (0.94 \pm 0.02) \times (\YX/Y_{\rm p})^{0.98 \pm 0.03}$.  The  \planck\ SZ masses were calibrated from the \YSZYX\ and \MYX\ relations derived from \xmm\ and the corresponding   normalisation  of the \MSZMX\  relation is one by construction. The fact that we find a  mean mass ratio  ($0.91\pm0.09$)  $10\%$ smaller than unity,  is also consistent with the calibration differences. 

In conclusion the impact of \xmm\ versus \chandra\ calibration is small, but larger than the statistical errors on mean quantities. There is a good agreement in slope between the present study and previous study by  \citet{2014Planck29} based on \xmm\ data. Differences in mean Y and mass ratio are consistent with \xmm\ versus \chandra\ calibration.

\subsection{Comparison to previous results}

\citet{2017Schellenberger} compared the \planck\ masses with
masses derived assuming hydrostatic equilibrium (HE), for 50  clusters in
common in the HIFLUCGS and PSZ2 catalogs. 
They derive  a larger discrepancy between X-ray and SZ masses, with a
very significant  mass dependence (slope of $\alpha=0.76\pm0.08$).
They suggest that the differences at low masses can be due to
Malmquist bias effect, in conjunction with and  underestimate of 
the \planck\  masses at the high mass end.  Our study does not confirm
this trend. We emphasize that we showed that our study is free from Malmquist bias.
On the other hand, the selection function of the
sub-sample of HIFLUCGS clusters discussed in \citet{2017Schellenberger}
is a complex combination of X-ray and SZ selections.  
Moreover, their masses are estimated from the Hydrostatic Equilibrium (HE) equation and 
all the most massive clusters in their sample are highly unrelaxed objects.  
While generally we expect the HE mass to be lower than the true mass, 
the HE mass may actually  overestimate the true mass when it is obtained from  
extrapolation of NFW profile fitted to data  in the central region 
\citep[see Figure 5 of][]{2006Rasia}. \citet{2020Lovisari} also compared hydrostatic masses to the {\em Planck} masses. They compared the hydrostatic masses of 117 clusters from the ESZ sample with their {\em Planck} PSZ2 derived masses, finding a small offset (i.e. 4\%) when the enclosed hydrostatic masses were computed at $R_{500}$ as determined by {\em Planck}.

\section{Conclusions}

Using {\em Chandra} observations, we derived the $\YX$ proxy and associated  mass for of 147 clusters with $z < 0.35$ from the {\em Planck} Early Sunyaev-Zel'dovich catalog and for 80 clusters with $z < 0.22$ from an X-ray flux-limited sample. The \chandra\ ESZ follow-up is complete within the \planck\ cosmological mask region. We re-extracted the \planck\ $\YSZ$ measurements centered on the more precise X-ray position and within the X-ray characteristic size ($\Rv$) from the full \planck\ mission maps. This re-extraction of the SZ signal for clusters originally detected in the 10-month maps minimizes the Malmquist bias, with the residual bias due to statistical scatter being negligible.

We discuss the results in terms of two $\YSZ$ measurements: the apparent flux $\YSZ$ in units of arcmin$^2$, and the intrinsic Compton parameter $\YSZmpc$ in units of Mpc$^2$. Our conclusions are as follows:

\begin{itemize}
\item 
The \YSZYXmpc\  relation is consistent in slope with that derived by the \planck\ Collaboration from \xmm\ data, with $B=0.96\pm0.03$.  There is a slight  offset in normalisation of $(6.5 \pm 0.1)\%$ compared to the \planck\ results, which we showed is consistent with known calibration systematic uncertainties between \xmm\ and \chandra.    The resulting $\YSZ/\YX$ ratio is $0.88 \pm 0.02$, in good agreement with X-ray expectations given radially-decreasing temperature profiles, and with previous determinations. It also suggests that there is a low level of gas clumping within $R_{500}$. This result is independent of any mass calibration and scaling relations.
\item 
We compared the \chandra\ X-ray  masses, derived from the $\YX$ proxy,  to the  \planck\  SZ masses, derived from the $\YSZmpc$ proxy. The  \planck\  masses are  about $10\%$ smaller. The median ratio of $\MSZ/\MX= 0.91\pm0.01$, is consistent with the aforementioned calibration differences.  The slope of the relation, $0.93\pm0.03$, differs from unity at slightly more than the $2\sigma$ level. The use of the re-extracted masses does not change the \MSZMX\ relation, but results in a significant reduction in the scatter (by a factor of two).
\item
The slope of the \YSZYXarc\ flux relation, at $B=0.89\pm0.01$, is significantly less than unity. We showed that this effect is not due to measurement issues. We performed extensive simulations, involving injection of simulated clusters into \planck\ maps, and subsequent detection and re-extraction using the \planck\ SZ detection algorithm MMF3. Using these simulations, we showed that this result is not due to selection effects, intrinsic scatter, or covariance between quantities.  
The X-ray selected sample follows the general trend exhibited by the SZ-selected sample, and extends it down to lower SZ fluxes.
\item
We showed analytically that  changing the \YSZYX\ relation from apparent flux (in units of arcmin$^2$) to intrinsic properties (in units of Mpc$^2$) results in a best-fit slope that is closer to unity, as observed ($0.96\pm0.03$), and increases the dispersion about the relation. The redistribution resulting from  this transformation implies that the best-fit parameters and dispersion of the intrinsic  \YSZYXmpc\   relation, and by extension the \MSZMX\ relation derived from these quantities, will be sample-dependent.
\end{itemize}

By itself, our study cannot estimate the absolute value of any bias between  the \planck\  mass  and the  true mass, or exclude any
mass dependence of this bias. Such a mass dependence has been suggested by
several lensing studies \citep{2014Linden,2015Hoekstra} 
although the latest compilation study by \citet{2017Sereno} 
suggests that the apparent mass dependent bias is actually due to an underlying 
redshift dependence. All these works are based on sub-samples of 
\planck\ clusters, with no simple selection criteria, so that selection 
effects are potentially complex. To make further progress in this
field, we require a complete follow-up of a well-characterised \planck\ cluster sample with mass estimation using various techniques (X-ray proxies, HE mass, lensing masses). 
Such data will be available through the CHEX-MATE programme \citep{CHEXMATE}.

\acknowledgments

\noindent
F.A-S. acknowledges support from {\em Chandra} grant GO3-14131X. G.W.P., M.A., and J.-B.M. acknowledge funding from the European Research Council under the European Union's Seventh Framework Programme (FP7/2007-2013)/ERC grant agreement No. 340519, and from the French Space Agency, CNES.
W.F., C.J., A.V., R.K., and L.D. acknowledge support from the Smithsonian Institution and the {\em Chandra} High Resolution Camera Project through NASA contract NAS8-03060.  R.J.v.W. acknowledges support from the VIDI research programme with project number 639.042.729, which is financed by the Netherlands Organization for Scientific Research (NWO). L.L. acknowledges support from the contracts ASI-INAF Athena 2019-27-HH.0, “Attivit\`adi Studio per la comunit\`a scientifica di Astrofisica delle AlteEnergie e Fisica Astroparticellare” (Accordo Attuativo ASI-INAFn. 2017-14-H.0) and INAF “Call per interventi aggiuntivia sostegno della ricerca di mainstream di INAF”, and financial contribution from NASA through contracts 80NSSCK0582 and 80NSSC19K0116. Basic research in radio astronomy at the Naval Research Laboratory is supported by 6.1 Base funding.
S.B. acknowledges financial support from the agreement ASI-INAF n.2017-14-H.0, the INFN INDARK grant.

\bibliographystyle{apj}
\bibliography{main}

\appendix
\section{Sample selection bias, scatter, and covariance tests}
\label{appx:mbias}

\subsection{Simulations}

The present clusters were selected from the ESZ catalog,  produced from detections on the \planck\ 10-month maps, while their PSZ2 $\YSZ$  flux values were extracted from the full mission maps.  To quantify any possible residual  bias due to the sample  selection, we undertook an extensive series of simulations by generating mock ESZ and PSZ2 catalogs. 

Starting from a Tinker mass function \citep{tin08}, simulated clusters were injected into the \planck\ SZ `cosmological' mask region (excluding the Galactic plane and the Magellanic clouds) of the 10-month maps used for the original ESZ catalog construction. Each simulated object was modeled with the \citet{2010Arnaud} pressure profile. The $\YSZ$ value corresponding to the cluster redshift and mass was drawn from the \YSZM\  relation, with a bias between the X--ray calibrated mass and the true mass, $(1-b)=0.65$,  tuned to recover the cluster number counts for the \planck\ cosmology.  

 For the $\YX$ value, we considered  $\YX$ physical quantities  normalised by the $\c$ factor given by Eq.~\ref{eq:C}, i.e. expressed in same units as the intrinsic Compton parameter. We note that  any selection effect will not depend on the mean ratio between the true $\YX$ and $\YSZmpc$ value. The slope of the \YSZYXmpc\ relation is expected to be close to unity, and we are interested in  possible slope biases  induced by selection and measurements. For simplicity,  we thus assume that the normalisation and slope of the \YXM\ and  \YSZM\  scaling relations are the same.

However, we allow for different scatters and a possible covariance between $\YX$ and $\YSZmpc$ variations from the relation.  We can thus write: 
\begin{eqnarray} 
\YM & = & A E(z)^{2/3} M^{B} \nonumber \\
P(\YSZmpc, \YX \vert Y_{M}) &=& \mathcal{N}(\YM,V_{\sigma})
\end{eqnarray}
where  $\YM$ is the latent $Y$ value at a given mass (i.e. the values of $\YX$ and $\YSZmpc$ obtained  if there were no scatter), and  $\YSZmpc$ and $\YX$ are the true values.  
We assume a Gaussian log-normal correlated distribution for $\YSZ$ and $\YX$ at fixed mass,  $\mathcal{N}(\YM,V_{\sigma})$, with a covariance matrix, $V_{\sigma}$:
\begin{equation*}
\begin{pmatrix}
 \sigsz^2 & r \, \sigsz\, \sigx \\
r \, \sigsz\, \sigx &    \sigx^2 
\end{pmatrix}\label{eqn:pmatrix}
\end{equation*}
where $\sigsz$ and $\sigx$ are the scatter in the log--log plane  of the Compton parameter and $\YX$ at fixed mass, respectively. The intrinsic $\YX$ and $\YSZmpc$ quantities were converted to flux using the angular distance to each cluster.  In the absence  of bias, the {\it  observed} \YSZYXarc\ relation should be the identity, i.e. have a slope and normalisation equal to unity. In the above, all quantities refer to values computed within a sphere of radius $\Rv$. 

SZ detections were then obtained by applying the MMF algorithm \citep{mel06} to the 10-month  maps. The mock ESZ catalogs were constructed from application of a $S/N>6$ threshold to these detections. These mock ESZ clusters were then re-injected into the full mission maps. The MMF algorithm was then run on these full mission maps, and the mock PSZ2 value  at the `true' position was extracted  for the mock ESZ clusters. We extracted both the mock PSZ2 value at the `true' size and  at the X--ray size. 
The measurement errors are estimated from the maps as described in  \citet{mel06}. 
The simulations are therefore fully representative of the observations. We ran 10 ESZ simulations, yielding a total number of 1657 clusters, allowing a  very precise estimate of any bias on the slope (the precision is better than $\sim0.5\%$). For simplicity we assume that the statistical errors on $\YX$ are negligible. 

\begin{figure*}[!t]
\centerline{
\includegraphics[width=\textwidth]{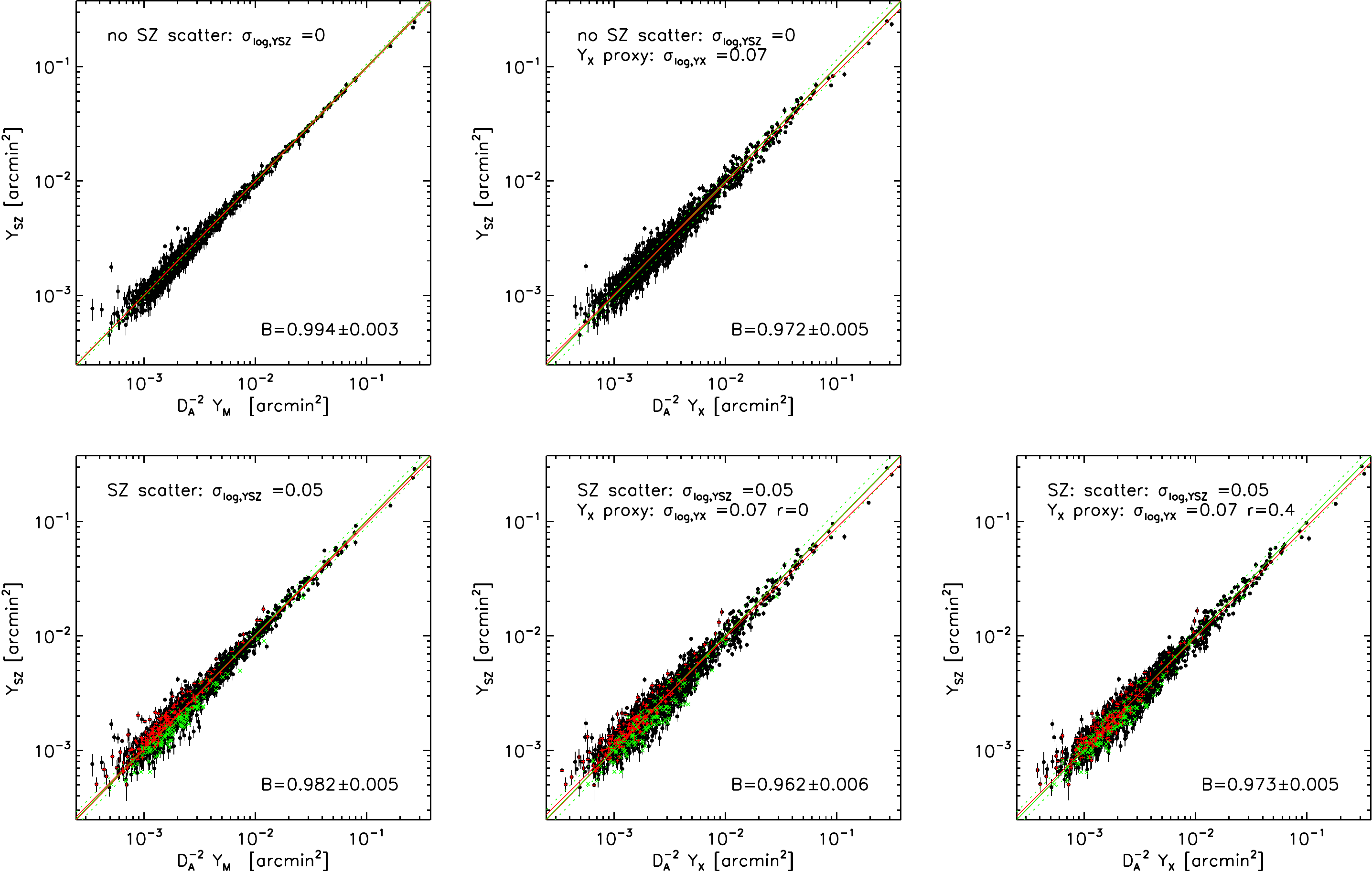}
}
\caption{\footnotesize{Simulated $\YSZ$ re-extracted from PSZ2 maps, as a function of $\YMarc$ (left columns) and $\YXarc$ (other columns),  for various assumptions on scatter and co-variance. $\YM$ is the  Compton parameter corresponding to the cluster mass for the mean \YSZM\ relation, i.e. the value that would be obtained if there was no scatter in the relation. The scatter in $\YSZmpc$ and $\YX$ at fixed mass, $\sigsz$ and $\sigx$, respectively,  are indicated in each panel. 
 {\it Middle panels:}  There is no correlation between  $\YSZ$ and $\YX$ deviations {\it Bottom right:} There is covariance  with $r=0.4$ (Eq.~\ref{eqn:pmatrix}). The different cases are discussed  in the text. }}
\label{fig:rels}
\end{figure*}

In the following, we detail a step-by-step analysis of the simulated relations and the inter-dependence of Malmquist bias, intrinsic scatter, and covariance between the quantities.  We focus on the \YSZYXarc\ relation between the SZ flux and its X-ray equivalent. This is the most fundamental observed relation, on which selection effects are expected to be most visible (particularly at the low end).  The left-hand panel of Fig.~\ref{fig:rels} shows the  effect of Malmquist bias on the observed $\YSZ$ values  as compared to the latent $\YMarc$ value, 
while the other panels show the corresponding \YSZYXarc\ relation and the effect of $\YX$  intrinsic scatter and covariance with $\YSZmpc$.  Unless otherwise stated the fit results are obtained with the \bces\ method. 

\subsubsection{Malmquist bias due to Planck noise and intrinsic scatter in the \YSZM\   relation} 

We first assume that there is no intrinsic scatter in the \YSZM\ relation. The corresponding observed  $\YSZ$ value, extracted at the true size, is shown in the  top-left panel of Fig.~\ref{fig:rels}  as a function of  $\YMarc$.  The scatter in the y-axis direction is entirely due the observational uncertainties in the SZ measurements.  Object selection was performed on the 10-month (ESZ) map and the $\YSZ$ signal is extracted from the full (29-month) mission maps.  The noise between the ESZ and PSZ2 maps is slightly correlated, so  we estimate here the residual effect of the ESZ selection on the flux estimation in the PSZ2 map. The effect on the slope is very small (slope $0.994 \pm 0.003$), implying a negligible impact.

We now include intrinsic scatter in the \YSZM\  relation, assuming $\sigma_{\log \YSZ}=0.05$ \citep{kay12, leb17}. Addition of intrinsic scatter increases  the dispersion in the Y-direction, and has the effect of changing the sample selection. The resulting relation is shown in the bottom-left panel of Fig.~\ref{fig:rels}. Objects that are newly-detected as compared to the previous case (about $10\%$) are plotted in red, and objects that are lost (no longer detected) are plotted in green ($7\%$). 
As expected, lost objects are on average less intrinsically bright than newly-detected objects, specially at low flux, as more up-scattered systems   pass the threshold than down-scattered systems. The intrinsic scatter then pushes the slope slightly away from unity, to $0.98 \pm 0.005$ (a $2\%$ effect). This is however still far from the observed value $0.89 \pm 0.01$.

\subsubsection{Using the $\YX$ proxy and the Eddington bias}

We now consider $\YXarc$ as the covariate, the observable we consider in this study. We assumed  $\sigx=0.07$ following numerical simulations  \citep{pla14, leb17, tru18}.  The corresponding \YSZYXarc\ relations  are shown in the middle panel of Fig.~\ref{fig:rels}, for the two cases described in the previous Section.
$\YX$  is a scattered estimate of  $\YM$, so the effect is to introduce scatter in the X-axis direction. The extraction at the X-ray size rather than at the `true' size changes slightly the $\YSZ$ value, but the effect is very small.  We also note that the slope is decreased,  as compared to that of the $\YSZ$--$\YM$ relation, by about $\Delta B=-0.02$, with a slope of $0.972 \pm 0.005$, for  $\sigsz=0$, and $0.962 \pm 0.006$ for $\sigsz=0.05$, respectively. 

This slope decrease  is reminiscent of the Eddington-like bias, as described extensively by   \citet{2015aSereno}. This bias is due to the the intrinsic scatter of the covariate,  (here $\YX$) with respect to the latent quantity ($\YM$)  when it has a non-uniform distribution. However this does not fully explain the effect, which we fail to fully understand. Indeed, the same slope is found using the  \lira\ regression code, designed to correct for this effect: $0.95\pm 0.02$, for the nominal case with intrinsic scatter in $\YSZ$. The same agreement between \lira\ and \bces\ results was noted for the observations ((Table \ref{tab:screl})

We next added covariance between the $\YX$ and $\YSZ$  deviations, assuming a correlation coefficient $r=0.4$ \citep[e.g.][]{far19,nag19}.  The bottom-right panel of Figure~\ref{fig:rels} shows the resulting relation, indicating that addition of covariance pushes the slope back towards unity, to $0.973 \pm 0.005$. This is due to the correlation between the deviations of the quantities  in the two axes.   Increasing the correlation coefficient from $r=0.4$ to $r=0.6$ increases the push towards unity with the slope changing to $0.982 \pm 0.005$. For a perfect correlation,  $\YX$ and $\YSZ$ would simply move along the line, canceling the Malmquist bias due to intrinsic scatter.

\begin{figure*}[t]
\begin{center}
\resizebox{\textwidth}{!} {
\includegraphics[]{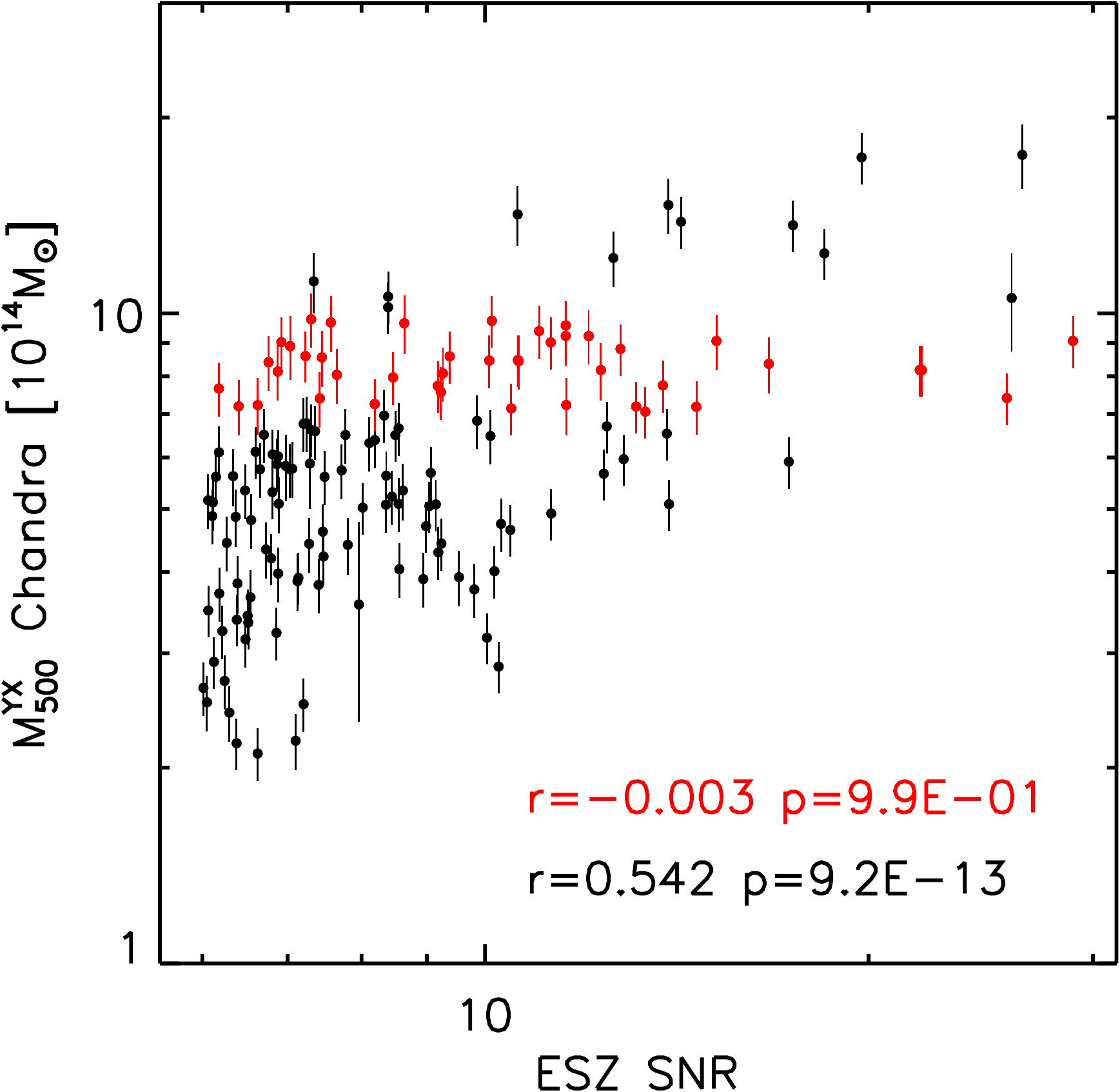} 
\hspace{0.1cm}
\includegraphics[]{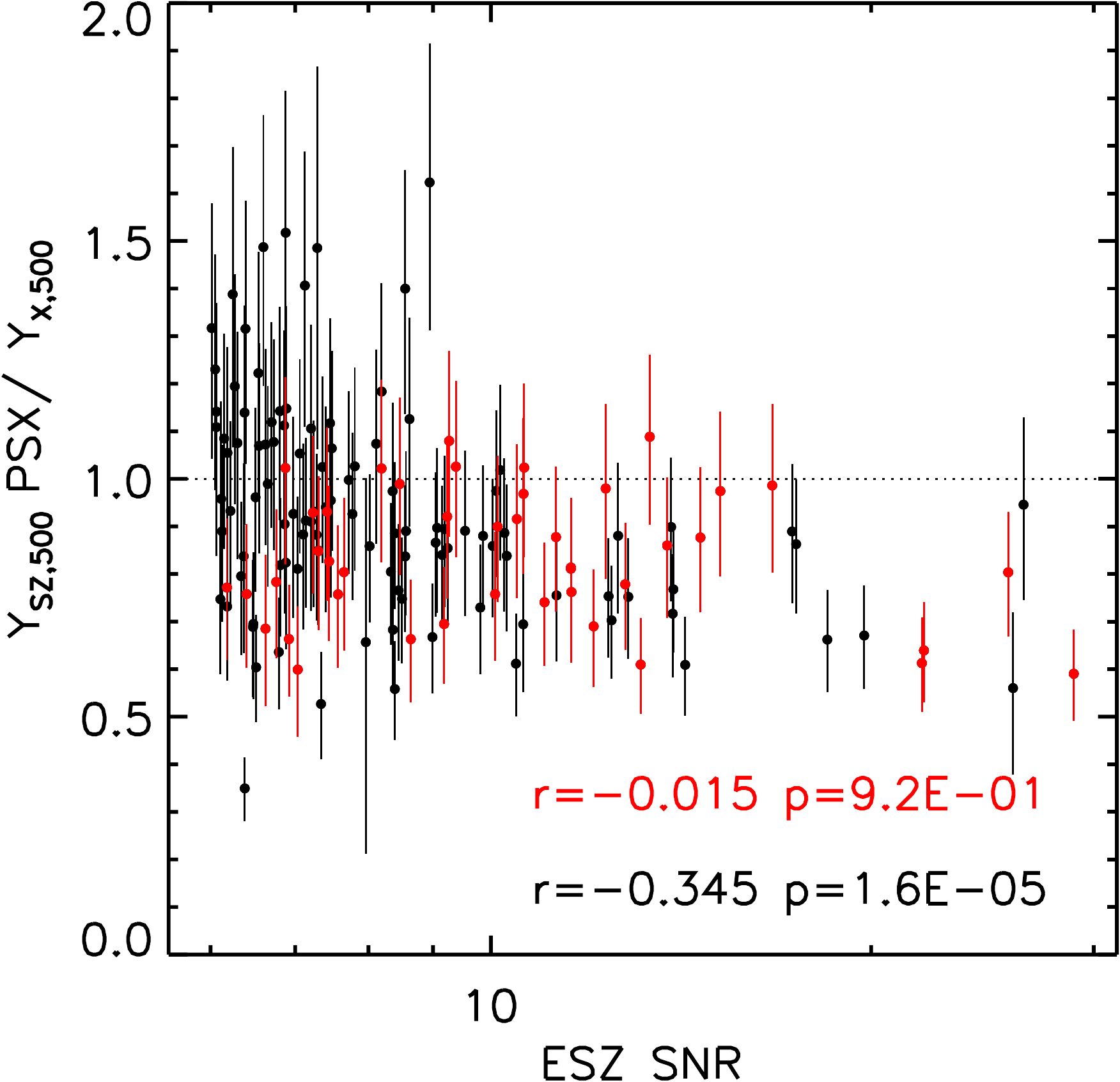} 
\hspace{0.1cm}
\includegraphics[]{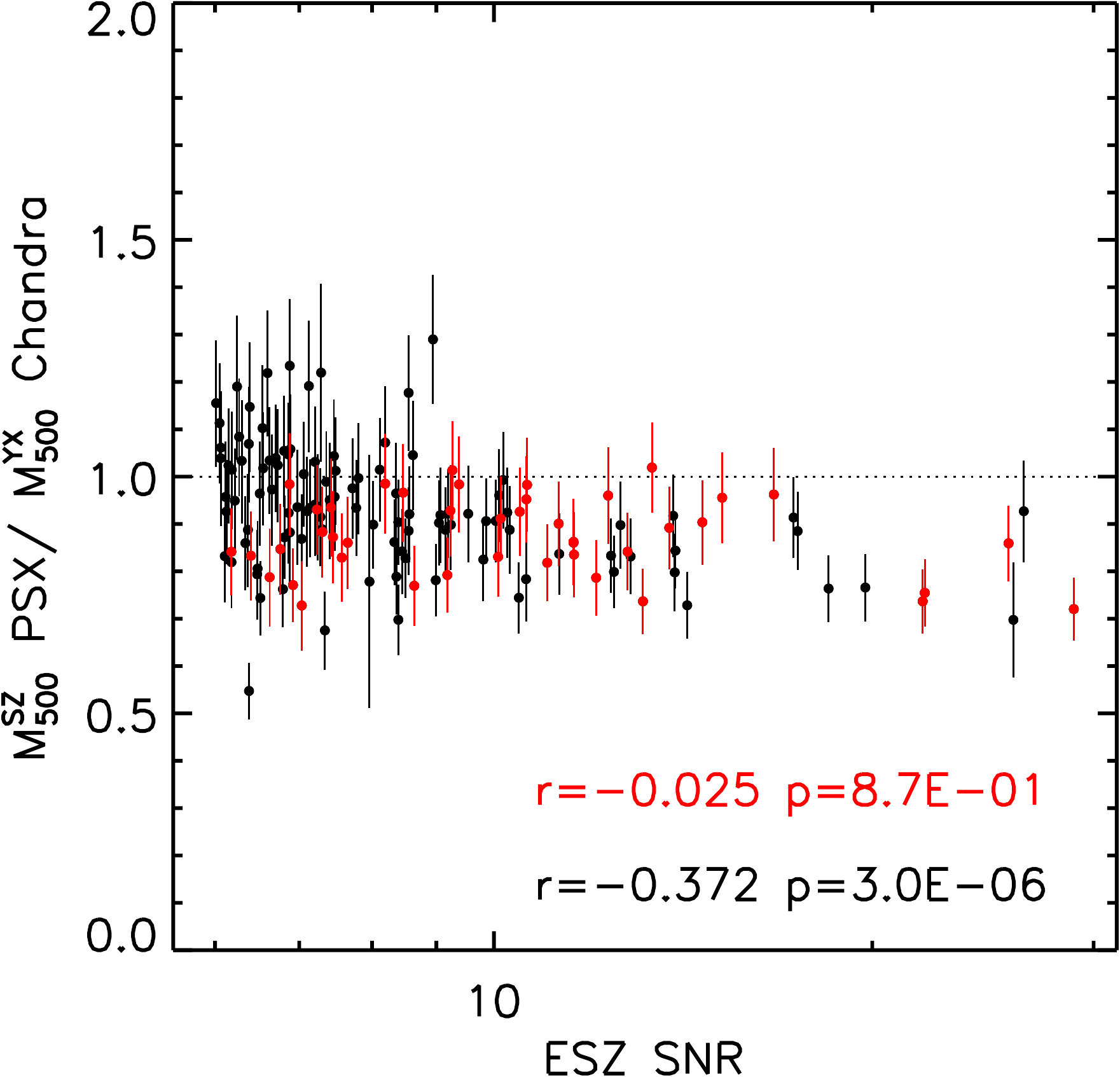} 
}

\end{center}
\caption{\footnotesize{Verification of the importance of the Malmquist
    bias. Left panel: Correlation between  $\MSZ$ and the $S/N$ of the
    ESZ detection.  Red points corresponds to the sub-sample used to
    check for the Malquist bias effect, effectively removing the
    intrinsic mass dependence from the S/N of the detection (see
    text).  Center and Right panel: correlation between the SZ and
    mass ratio as a function of $S/N$. The correlations coefficients
    for the full sample and sub-sample are given in the plots and
    Table \ref{tab:mb}.}}
\label{fig:mb}
\end{figure*}

\subsubsection{Extreme case: large value of an intrinsic scatter in the $\Mv$--$\YSZ$ relation}

As a final test, we doubled the intrinsic scatter in the $M_{500}$--$\YSZ$ relation, so that $\sigma_{\log \YSZ}=0.10$. If $r=0$, the increased dispersion in the Y-direction again affects the sample selection, but more strongly than in the bottom-left panel of  Fig.~\ref{fig:mb}, driving the slope to $0.930\pm0.008$. However, as in the nominal case, inclusion of covariance again redresses the slope towards unity: for $r=0.4$, the slope becomes $0.955\pm0.007$, and for $r=0.6$, the slope is $0.967\pm0.006$. 

The general conclusion is that biases due to selection effects and intrinsic scatter cannot explain the observed slope of the \YSZYXarc\ flux relation. As we used PSZ2 SZ values, largely independent of the detection values, the residual Malmquist bias is dominated by the intrinsic scatter between the $\YSZmpc$ and the mass.  The intrinsic scatter in $\YX$  is found to slightly further reduce the slope, but   covariance between $\YX$ and $\YSZmpc$ redresses the slope towards unity. For typical scatter and covariances derived from simulations, the net effect is an decrease of the slope to $B\sim0.97$, far from the observed value of $0.89 \pm 0.01$. The combination of factors that gives the most deviant slope from unity requires unrealistically large intrinsic scatter in the \YSZM\ relation and zero covariance between $\YX$ and $\YSZmpc$, and even when  doubling the $\YSZmpc$ scatter the slope still does not agree with the data.


\subsection{Further tests}

We further checked that the Malmquist bias due to \planck\  noise fluctuations is indeed negligible in the
data when using PSZ2 values for the ESZ clusters, by examining the correlation between the PSZ2 $\YSZ/\YX$ ratio and the $S/N$ of the ESZ detection.  
The $\YSZ$ measured in the ESZ maps is affected by Malmquist bias, as clusters with a signal that is boosted by positive noise have a higher probability to be detected, particularly close to detection threshold.  The signature of this bias is a correlation between  the  ratio of  the observed signal to the  true signal, $\YSZ/Y_{\rm SZ, true}$, with the   $S/N$ of the ESZ detection.  If the PSZ2 measurements were fully independent, this correlation would  disappear. We do not have access to $Y_{\rm YSZ, true}$, but we can use  $\YX$ as a proxy as it is not affected by the \planck\ noise in the SZ detection. A negligible dependence of the $\YSZ/\YX$ ratio on $S/N$ would imply negligible Malmquist bias due to \planck\ noise. 

However, a difficulty is that more massive clusters are easier to
detect, so that the $S/N$ is correlated to mass as shown on the left
panel of Figure \ref{fig:mb}. 
To disentangle a possible intrinsic mass dependence of the $\YSZ/\YX$
ratio from a  dependence on the $S/N$ due to the Malmquist bias, we
selected a sub-sample of clusters of nearly the same mass, 
in a very restricted mass range of $\MX= (8.5 \pm 1.5) \times 10^{14}~M_\odot$.  
The mean mass of the sub-sample was chosen to maximize the $S/N$
leverage, while $\Delta M$ is small enough to ensure  that there is 
no residual correlation between mass and $S/N$. 
The clusters in this sub-sample are identified by red points in the
left panel of Figure \ref{fig:mb}.  
The $\YSZ/\YX$  for the full sample is significantly correlated with
the $S/N$ and increases at low $S/N$ (central panel of Figure
\ref{fig:mb}). However the correlation with $S/N$ disappears for the  
sub-sample at a given mass range ($r=-0.015$, $p=0.92$). 
The same is found for the mass ratio ($r=-0.025$, $p=0.87$).  

This test shows that the Malmquist bias due to \planck\ noise fluctuations  is indeed negligible when
considering PSZ2 $\YSZ$ values. Note that this does imply a negligible Malmquist bias due to intrinsic scatter of $\YSZ$ with mass.  Clusters detected because their $\YSZ$ are scattered up at a given mass may have also a higher $\YX$  value due to  covariance between  $\YSZ$ and $\YX$ deviations at given mass, so that the ratio  $\YSZ/\YX$ remains weakly dependent on $S/N$.

\begin{figure}[]
\centerline{
\includegraphics[width=0.36\textwidth]{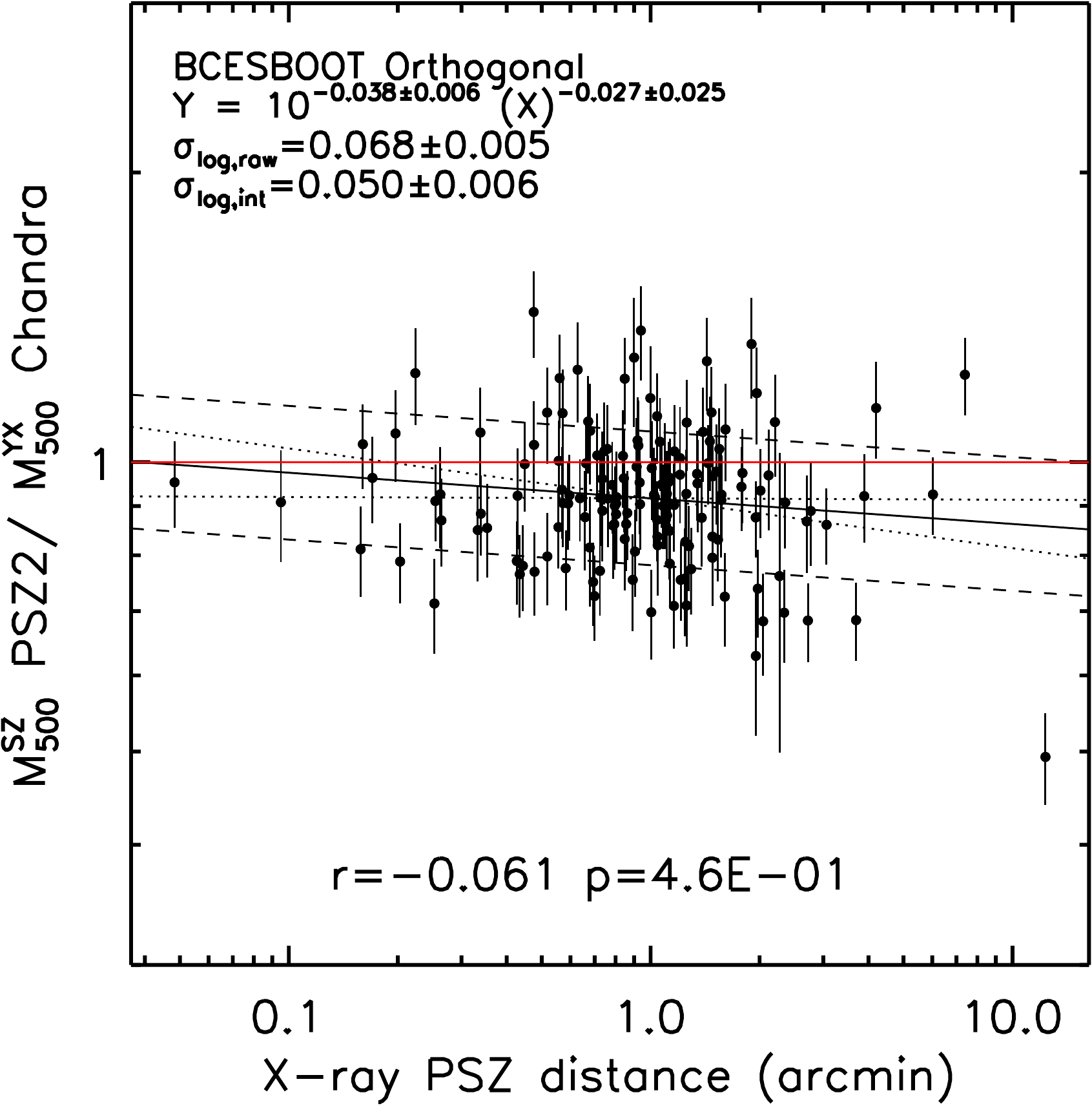}
}
\caption{\small{Ratio between SZ and X-ray derived masses
plotted against the projected distances between the cluster centers as measured
using {\em Planck} and {\em Chandra} data. The solid black line corresponds to
the power-law best fit, while the solid red line corresponds to equal 
SZ and X-ray derived masses. The dashed line corresponds to the 1$\sigma$ confidence range. The dotted line corresponds to the 1$\sigma$ envelope on the best-fitting line.
}}\label{fig:dist_mass_ratio}
\end{figure}

\begin{deluxetable}{lcccc}[]
\tablecaption{Spearman rank coefficient and significance of Malmquist bias tests.} 
\tablewidth{0.5\columnwidth} 
\tablehead{ 
\colhead{Relation}&
\colhead{$r$} &
\colhead{$p$} &
\colhead{$r$ } &
\colhead{$p$ } \\
\colhead{}&
\colhead{} &
\colhead{} &
\colhead{sub-sample} &
\colhead{sub-sample} 
}
\startdata 
$M_{\rm Y_{X}}$--$S/N$ & 0.542 & $9.2 \times 10^{-13}$ & -0.003 & 0.99 \\
$\YSZ / \YX$--$S/N$ & -0.345 & $1.6 \times 10^{-5}$ & -0.015 & 0.92 \\
$\MSZ / M_{500}^{\rm X}$--$S/N$ & -0.372 & $3.0 \times 10^{-6}$ & -0.025 & 0.87 
\enddata
\tablecomments{Columns list the rank coefficient and significance of
  Malmquist bias tests for the full and sub-samples. The null hypothesis $p$-value is the probability that the observed coefficient is obtained by chance if the two parameters are completely independent.}
\label{tab:mb}
\end{deluxetable}

\begin{figure*}[!t]
\centerline{
\includegraphics[width=0.5\textwidth]{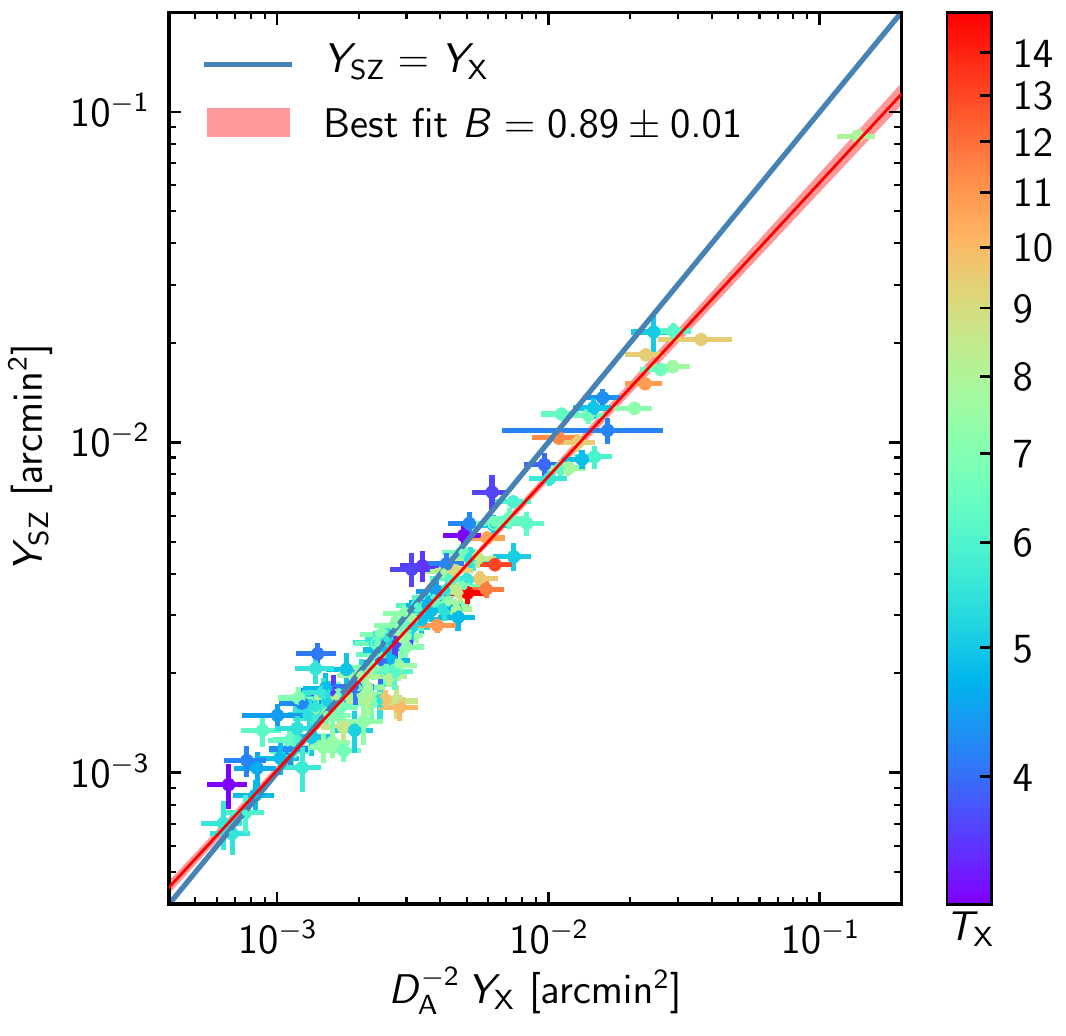}
\includegraphics[width=0.5\textwidth]{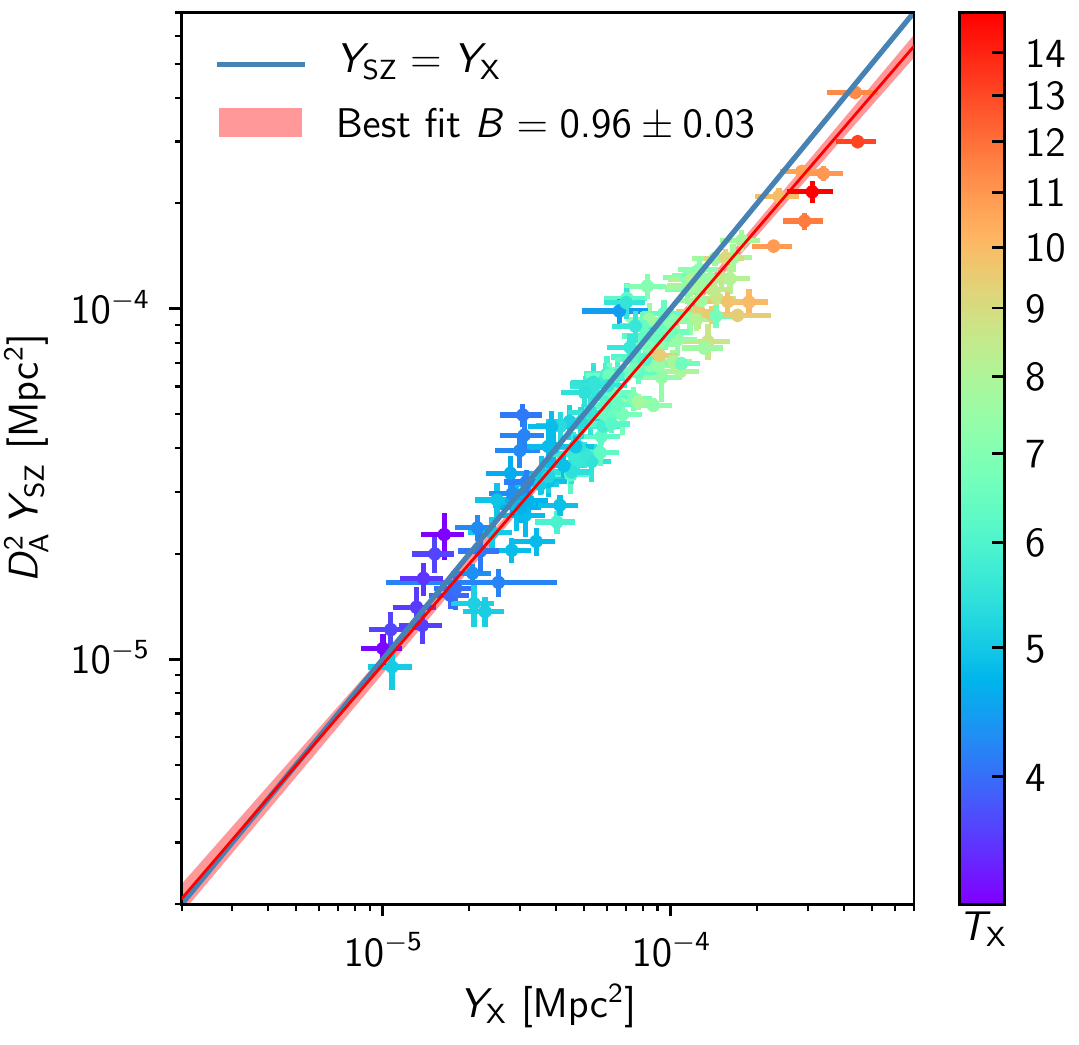}
}
\caption{\small{Observed change in slope in the \YSZYX\ relation caused by the change  from apparent flux to intrinsic properties. The points are color-coded according to the values of the gas temperature. As we can see, despite being separated in the \YSZYX\ relation in  intrinsic properties (units of Mpc$^2$), the colors are mixed when the same relation is  expressed in apparent flux (units of arcmin$^2$), illustrating the redistribution that happens when the  \YSZYX\ relation is expressed different physical quantities. This redistribution ultimately changes the slope of the \YSZYX\ relation. Left: the best fit relation is given by $\YSZ \propto (\YX)^{0.89}$ in  apparent flux (units of arcmin$^2$). Right: same as top left panel, except for $\YSZ \propto (\YX)^{0.96}$ in  intrinsic properties (units of Mpc$^2$).
}}\label{fig:observed_change_in_slope}
\end{figure*}

\section{Testing a Mass Dependence on {\em Chandra}--{\em Planck} Center Offset}

The uncertainty on the {\em Planck} cluster center is determined by the
spatial resolution of its instruments. Here, we compare the cluster centers
determined by the {\em Planck} detection algorithm ($\sim$ arcmins) to the precise
($\sim$ arcsec) {\em Chandra} centroids. The mean offset between the {\em Chandra} and 
{\em Planck} centers is $1.'7$, similar to the result obtained by
\citet{2013PlanckXMMvalid} of $1.'5$, and in agreement with expectations
from the {\em Planck} sky simulations \citep[][see their Section 6.1]{2011PlanckCol}.

We investigate a possible dependence of the SZ to X-ray derived mass ratio (see Figure
\ref{fig:dist_mass_ratio}) on the offset between 
the {\em Planck} and {\em Chandra} centers. The relation between the SZ to
X-ray derived mass ratio and the offset between 
the {\em Planck} and {\em Chandra} positions ($\Delta R$) is given by:
\begin{equation}
\MSZ/\MX= 10^A \times \Delta R^B.\label{eq:power_law_ratio_deltaR}
\end{equation}

Our results are presented in Table \ref{tab:power-law_ratio}. We find
no evidence of a correlation between the two quantities (under the
null hypothesis of no correlation, we obtained a $p$-value of 0.46), 
suggesting that the large {\em Planck}   
position uncertainty ($\sim 2'$) is not driving the
differences in cluster mass determinations.

\begin{deluxetable}{lcccc}[h!]
\tablecaption{Best Fit Parameters for the Ratio Between SZ and X-ray Masses} 
\tablewidth{0pt} 
\tablehead{ 
\colhead{Relation}&
\colhead{$A$} &
\colhead{$\sigma_{\rm A}$} &
\colhead{$B$} &
\colhead{$\sigma_{\rm B}$}
}
\startdata 
Ratio versus $\Delta R$ &              -0.038 & 0.006 & -0.027 & 0.025 \\
\enddata
\tablecomments{Columns list best fit parameters and their
  uncertainties for the power-law given by
  Equation (\ref{eq:power_law_ratio_deltaR}).} 
\label{tab:power-law_ratio}
\end{deluxetable}


\section{Difference in slope of the \YSZYX\ relation when  expressed in apparent flux (units of arcmin$^2$) and intrinsic properties (units of Mpc$^2$)}\label{appen:diff_slope}

As noted throughout the text, we observe a statistically significant difference in the slope of the \YSZYX\ relation when  expressed in apparent flux (units of arcmin$^2$) compared to intrinsic properties (units of Mpc$^2$), as clearly seen also in Figure \ref{fig:observed_change_in_slope}. When clusters are color-coded according to their gas temperatures, we clearly see on the right panel of Figure \ref{fig:observed_change_in_slope} that clusters are ordered by color, as expected when the \YSZYX\ relation is expressed in  intrinsic properties (units of Mpc$^2$). On the other hand, on the left panel of the Figure \ref{fig:observed_change_in_slope} the colors are more evenly spread throughout the \YSZYX\ relation. A re-ordering of data points occurs when the  \YSZYX\ relation is expressed in intrinsic properties instead of apparent flux because this change depends on the underlying cosmology and cluster redshift, the result being a different multiplying factor for each cluster. In this Appendix we show that this change
 from apparent flux to intrinsic properties results in a best-fit $\YSZ$ --$\YX$ relation slope that is closer to unity when expressed in  intrinsic properties (units of Mpc$^2$).

\subsection{Analytical demonstration of the change in slope and dispersion}\label{sec:app_analytical}

For simplicity, the linear regression presented here is undertaken using the least squares method.
For $n$ pairs of data $(x,y)$, the method of least squares may be used to write a linear relationship between x and y. 
The least squares regression line is the locus that minimizes the sum of the squares of the vertical deviation from each data point from the best fit linear relation.

The least square regression line for a set of $n$ data points is given by the equation of a line:

\begin{equation}
y = a + b x,
\end{equation}

where $a$, $b$, and the variance, $\sigma^2$, about the best fit are given by ($\sum_{i=1}^{n} \rightarrow \sum$):

\begin{equation}
b = \frac{n \sum x_{\rm i} y_{\rm i} - \left(\sum x_{\rm i}\right)\left( \sum y_{\rm i}\right)}{n \sum x_{\rm i}^2 - \left(\sum x_{\rm i}\right)^2}\label{eq:b}
\end{equation}

\begin{equation}
a = \frac{\sum y_{\rm i} - b \sum x_{\rm i}}{n}\label{eq:a}
\end{equation}

\begin{equation}
\sigma^2 = \frac{\sum (y_{\rm i} - a - b x_{\rm i})^2}{n-1}\label{eq:disp}.
\end{equation}
\begin{widetext}

Assuming that the data are now best described by a power-law relation given by $y = a x^b$, one can linearize this relation by applying a logarithmic function to both sides of this equality:
\begin{eqnarray}
{\rm log} (y)  = {\rm log} (a) + b  ~ {\rm log} (x) 
                \rightarrow  z = a_0 + a_1 ~ w.
\end{eqnarray} 

Now, let us perform the following linear transformation on the data $(x,y)$:
\begin{eqnarray}
y^\prime = \zeta ~ y: y_{\rm i}^\prime = \zeta_{\rm i} ~ y_{\rm i} {\rm ~and~}
x^\prime = \zeta ~ x: x_{\rm i}^\prime = \zeta_{\rm i} ~ x_{\rm i},                     
\end{eqnarray}
where the coefficient $\zeta_{\rm i}$ is unique for each pair of data ($x_{\rm i},y_{\rm i}$). If each data point ($x_{\rm i},y_{\rm i}$) has a dispersion about the best fit relation given by $1+\epsilon_{\rm i}$, such that $y_{\rm i} = a x_{\rm i}^b (1+\epsilon_{\rm i})$, the linear transformation leads to $y_{\rm i}^\prime = a \zeta_{\rm i} x_{\rm i}^b (1+\epsilon_{\rm i}) = a \zeta_{\rm i}^{1-b} x^{\prime b}_{\rm i} (1+\epsilon_{\rm i})$, which can be linearized as:
\begin{eqnarray}
{\rm log} (y^{\prime}_{\rm i})  = {\rm log} (a) + (1-b) {\rm log} (\zeta_{\rm i}) +
{\rm log} (1+\epsilon_{\rm i}) + b  ~ {\rm log} (x_{\rm i}^{\prime}) \rightarrow
                  z_{\rm i}^{\prime} = a_0 + (1-b) a_{1,\rm i} + a_{2,\rm i} + b w_{\rm i}^{\prime}.\label{eq:linear_powerlaw_2}
\end{eqnarray}

\begin{figure*}[!b]
\centerline{
\includegraphics[width=0.5\textwidth]{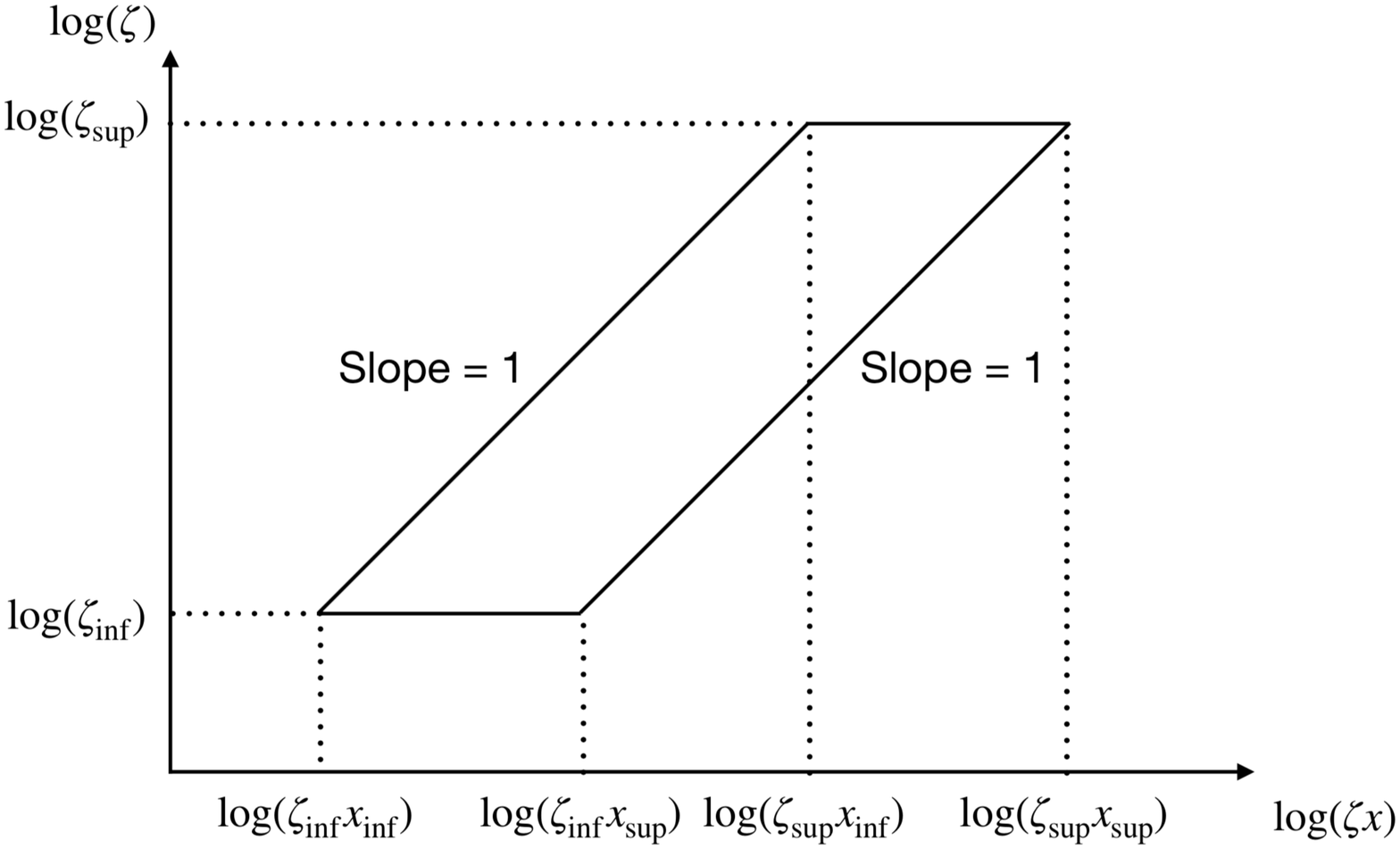}
\includegraphics[width=0.5\textwidth]{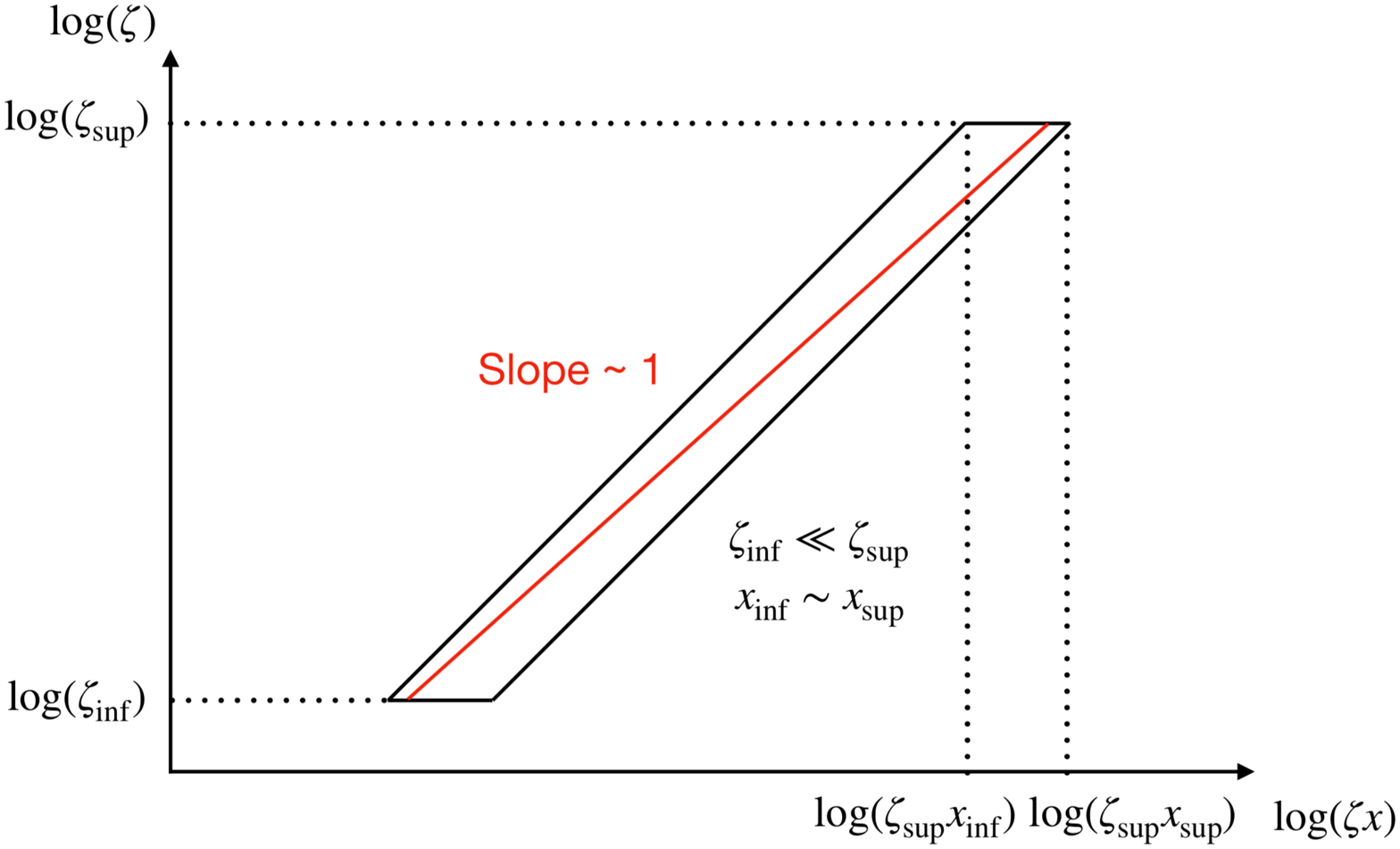}
}
\vspace{-0.5cm}
\centerline{
\includegraphics[width=0.5\textwidth]{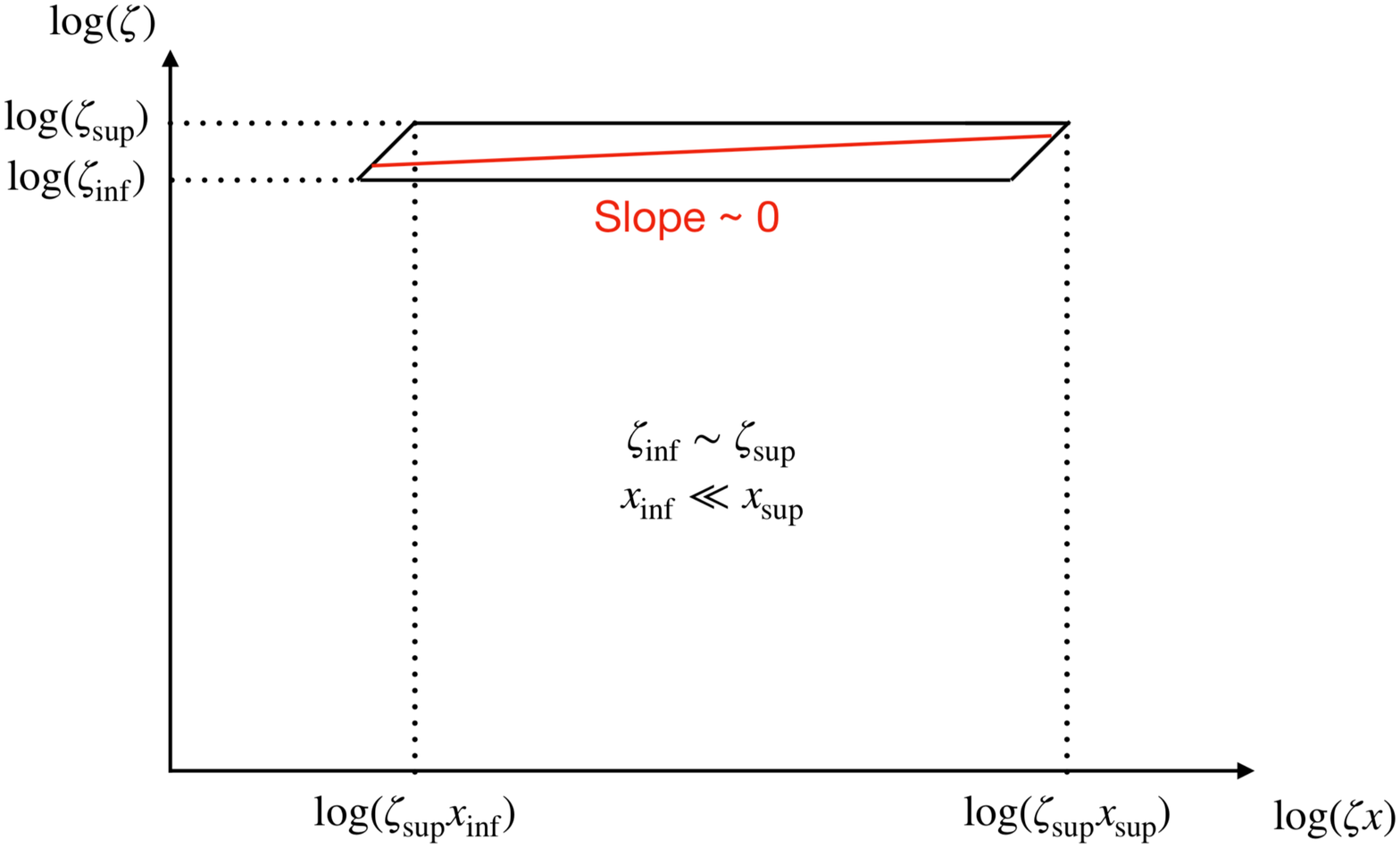}
\includegraphics[width=0.5\textwidth]{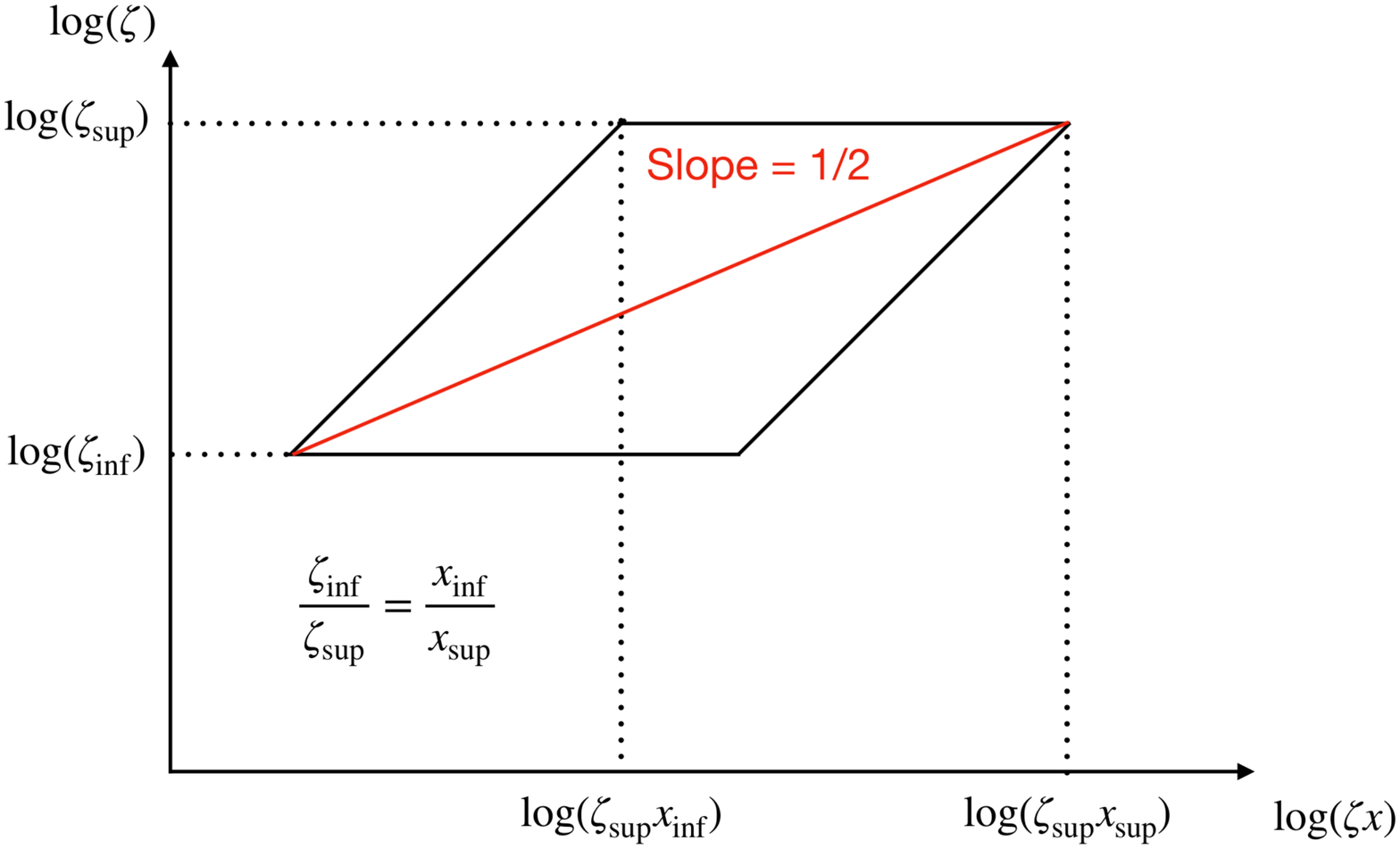}
}
\caption{\small{Visualization of the $\zeta$--$\zeta x$ relation in logarithmic space. Top left: illustration of boundaries for the pairs ($\zeta_{\rm i}$, $\zeta_{\rm i} x_{\rm i})$ when $\zeta_{\rm i} \in [\zeta_{\rm inf}, \zeta_{\rm sup}]$ and   $x_{\rm i} \in [x_{\rm inf}, x_{\rm sup}]$. Top right: illustration of the slope of the best power-law fit when $x_{\rm inf} \sim x_{\rm sup}$ and $\zeta_{\rm inf} \ll \zeta_{\rm sup}$. 
Bottom left: same as top right panel, except for $x_{\rm inf} \ll x_{\rm sup}$ and $\zeta_{\rm inf} \sim \zeta_{\rm sup}$. Bottom right: same as top right panel, except for
$x_{\rm inf}/x_{\rm sup}$ = $\zeta_{\rm inf}/\zeta_{\rm sup}$.
}}\label{fig:zeta_vs_zetax}
\end{figure*}


\begin{figure*}[t]
\centerline{
\includegraphics[width=0.5\textwidth]{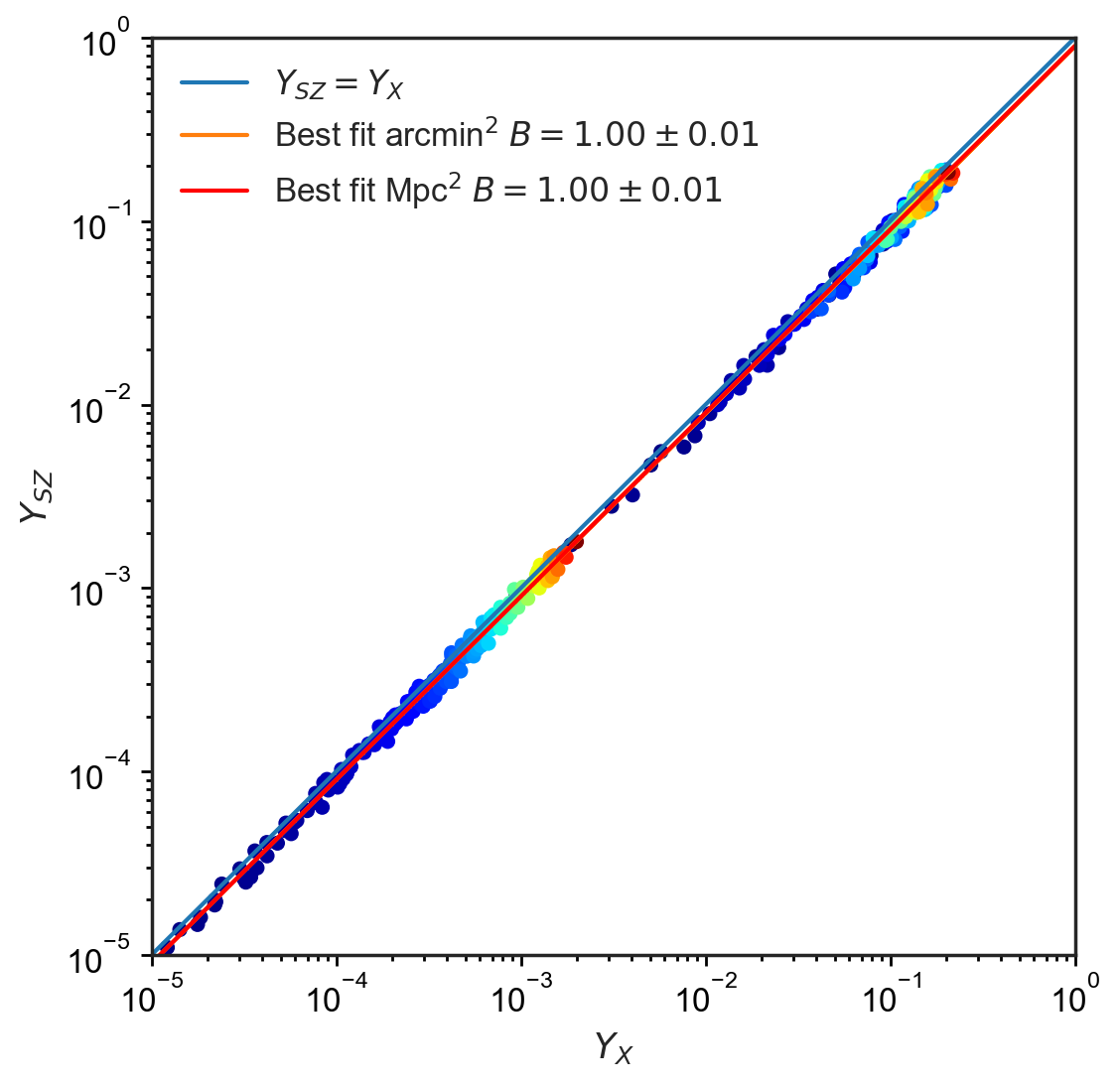}
\includegraphics[width=0.5\textwidth]{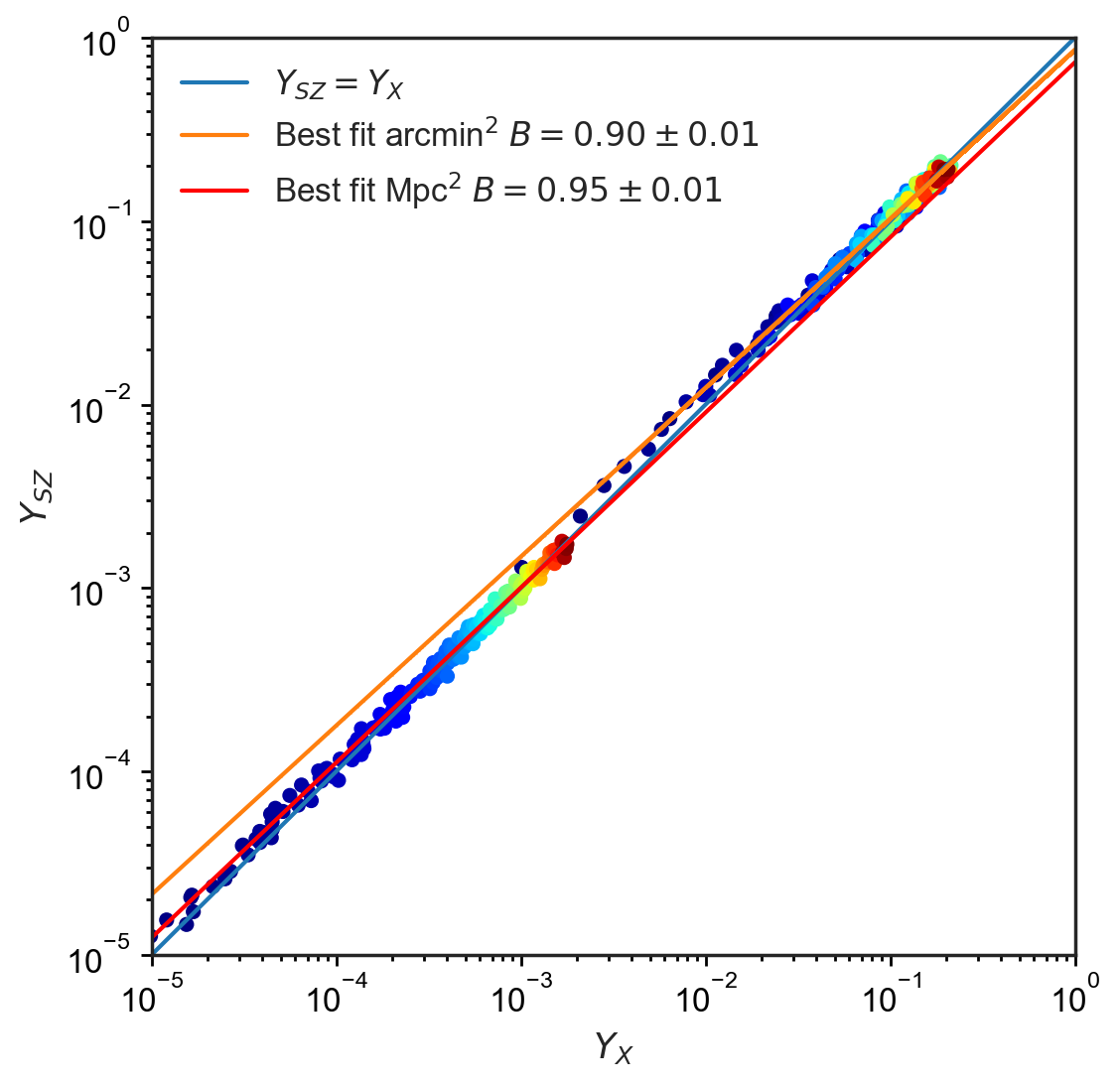}
}
\centerline{
\includegraphics[width=0.5\textwidth]{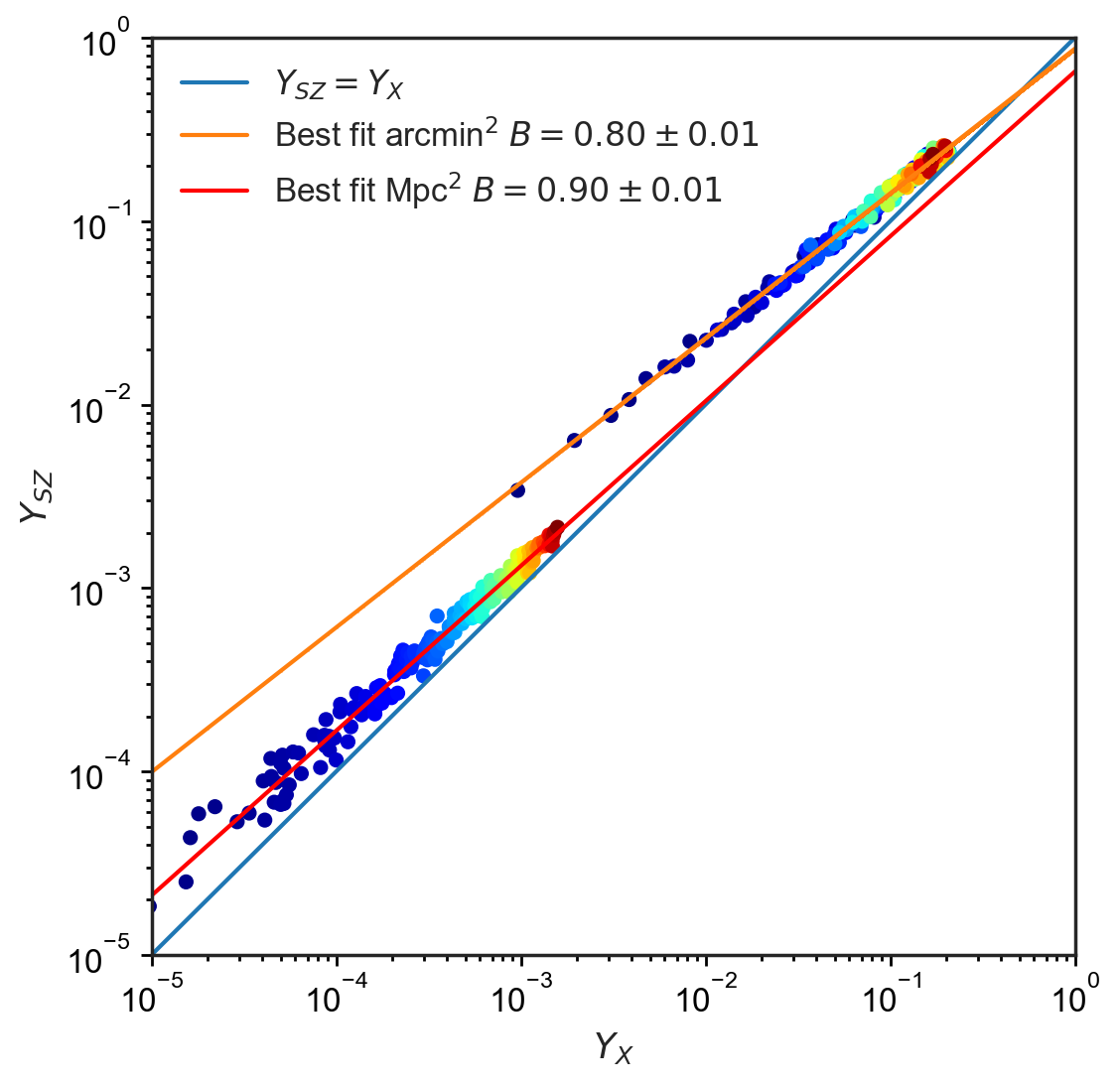}
\includegraphics[width=0.5\textwidth]{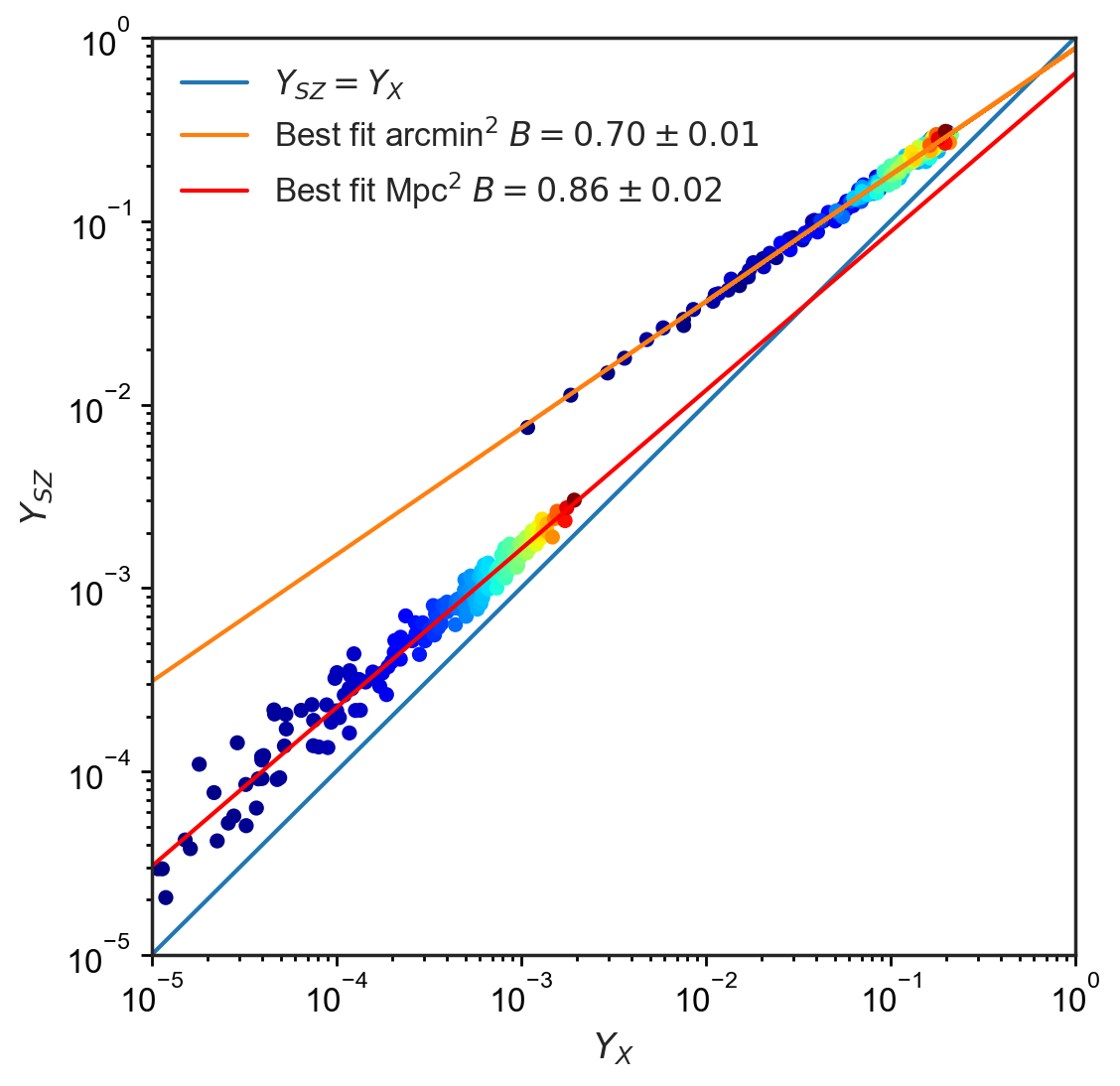}
}
\caption{\small{Toy model of the change in slope caused by the change of units from arcmin$^2$ to Mpc$^2$. The points are color-coded according to the mock values of $\YX$ in units of Mpc$^2$, which should be closely related to a temperature dependence. As we can see, despite being separated in the \YSZYX\ relation in units of Mpc$^2$, the colors are mixed when the same relation is plotted in units of arcmin$^2$, illustrating the re-distribution that happens when the values are presented in other units. This re-distribution ultimately changes the slope of the best fit, in this particular set of examples as given by $b^\prime = b + (1-b)/2$. Top left: creating a distribution that follows $\YSZ \propto (\YX)^1$ in units of arcmin$^2$, one gets the same slope in units of Mpc$^2$, as expected from $b^\prime = b + (1-b)/2$. Top right: same as top left panel, except for $\YSZ \propto (\YX)^{0.90}$ in units of arcmin$^2$ leading to $\YSZ \propto (\YX)^{0.95}$ in units of Mpc$^2$. Bottom left: same as top left panel, except for $\YSZ \propto (\YX)^{0.80}$ in units of arcmin$^2$ leading to $\YSZ \propto (\YX)^{0.90}$ in units of Mpc$^2$. Bottom right: same as top left panel, except for $\YSZ \propto (\YX)^{0.70}$ in units of arcmin$^2$ leading to $\YSZ \propto (\YX)^{0.86}$ in units of Mpc$^2$.
}}\label{fig:mock}
\end{figure*}

Using the linearized equation presented in (\ref{eq:linear_powerlaw_2}), equation (\ref{eq:a}) takes the following form:
\begin{eqnarray}
b^\prime &=& \frac{n \sum w^{\prime}_{\rm i} z^{\prime}_{\rm i} - \left(\sum w^{\prime}_{\rm i}\right)\left( \sum z^{\prime}_{\rm i}\right)}{n \sum w^{\prime 2}_{\rm i} - \left(\sum w^{\prime}_{\rm i}\right)^2} \nonumber \\
&=& \frac{n \sum (a_0 w^{\prime}_{\rm i} + (1-b) a_{1,\rm i} w^{\prime}_{\rm i} + a_{2,\rm i} w^{\prime}_{\rm i} + b w_{\rm i}^{\prime 2}) - \left(\sum w^{\prime}_{\rm i}\right)\left( \sum (a_0 + (1-b) a_{1,\rm i} + a_{2,\rm i} + b w_{\rm i}^{\prime})\right)}{n \sum w^{\prime 2}_{\rm i} - \left(\sum w^{\prime}_{\rm i}\right)^2} \nonumber \\ 
&=& \frac{n \sum a_0 w^{\prime}_{\rm i} + n \sum (1-b) a_{1,\rm i} w^{\prime}_{\rm i} +  n \sum a_{2,\rm i} w^{\prime}_{\rm i} + n \sum b w_{\rm i}^{\prime 2} - n \sum a_0 w^{\prime}_{\rm i} - \sum (1-b) a_{1,\rm i} \sum w_{\rm i}^{\prime} - \sum a_{2,\rm i} \sum w_{\rm i}^{\prime} - b \left(\sum w^{\prime}_{\rm i}\right)^2}{n \sum w^{\prime 2}_{\rm i} - \left(\sum w^{\prime}_{\rm i}\right)^2} \nonumber \\
&=& b + (1-b) \frac{ n \sum a_{1,\rm i} w^{\prime}_{\rm i} - \sum a_{1,\rm i} \sum w_{\rm i}^{\prime}}{n \sum w^{\prime 2}_{\rm i} - \left(\sum w^{\prime}_{\rm i}\right)^2} 
+ \frac{ n \sum a_{2,\rm i} w^{\prime}_{\rm i} - \sum a_{2,\rm i} \sum w_{\rm i}^{\prime}}{n \sum w^{\prime 2}_{\rm i} - \left(\sum w^{\prime}_{\rm i}\right)^2}
\label{eq:a2},
\end{eqnarray}
which becomes
\begin{eqnarray}
b^\prime = b 
+ 
(1-b) \frac{ n \sum {\rm log} (\zeta_{\rm i}) {\rm log} (\zeta_{\rm i}x_{\rm i})- \sum {\rm log} (\zeta_{\rm i}) \sum {\rm log} (\zeta_{\rm i}x_{\rm i})}{n \sum {\rm log} (\zeta_{\rm i}x_{\rm i})^2 - \left(\sum {\rm log} (\zeta_{\rm i}x_{\rm i})\right)^2} 
+ 
\frac{ n \sum {\rm log} (1+\epsilon_{\rm i}) {\rm log} (\zeta_{\rm i}x_{\rm i})- \sum {\rm log} (1+\epsilon_{\rm i}) \sum {\rm log} (\zeta_{\rm i}x_{\rm i})}{n \sum {\rm log} (\zeta_{\rm i}x_{\rm i})^2 - \left(\sum {\rm log} (\zeta_{\rm i}x_{\rm i})\right)^2} 
\label{eq:a2_f},
\end{eqnarray}
where the term multiplying $(1-b)$ is the slope of the best fit of $\zeta(x^\prime) \equiv \zeta(\zeta x)$ in logarithmic space (compare Equation (\ref{eq:a2_final}) with (\ref{eq:b})), which will be named $\beta$. The last term is the slope of the best fit of the ($1+\epsilon$) -- $\zeta x$ relation in logarithmic space. Assuming
there is no correlation between the dispersion about the original best fit relation and the transformed x-coordinate ($\zeta x$), the last term of Equation (\ref{eq:a2_f}) vanishes, so that it finally becomes:
\begin{eqnarray}
b^\prime = b 
+ 
(1-b) \frac{ n \sum {\rm log} (\zeta_{\rm i}) {\rm log} (\zeta_{\rm i}x_{\rm i})- \sum {\rm log} (\zeta_{\rm i}) \sum {\rm log} (\zeta_{\rm i}x_{\rm i})}{n \sum {\rm log} (\zeta_{\rm i}x_{\rm i})^2 - \left(\sum {\rm log} (\zeta_{\rm i}x_{\rm i})\right)^2}.
\label{eq:a2_final}
\end{eqnarray}

Following the same arguments,
we can also express (for completeness) the new linear coefficient $a^{\prime}$ as:
\begin{eqnarray}
a^\prime = {\rm log} (a) + (1-b) \frac{\sum {\rm log} (\zeta_{\rm i}) - \beta \sum {\rm log} (\zeta_{\rm i}x_{\rm i})}{n} 
\label{eq:aa2_final},
\end{eqnarray}
where the term multiplying $(1-b)$ is the linear coefficient of the best fit of $\zeta(x^\prime) \equiv \zeta(\zeta x)$ in logarithmic space (compare Equation (\ref{eq:aa2_final}) with (\ref{eq:a})), which will be named $\alpha$. 

We can finally express the new dispersion $\sigma^\prime$, following again steps similar to those leading to Equation (\ref{eq:a2_final}), by:
\begin{eqnarray}
\sigma^{\prime2} = \frac{\sum {\rm log} (1 + \epsilon_{\rm i})^2}{n-1} + (1-b)^2 \frac{\sum ({\rm log} (\zeta_{\rm i}) - \alpha - \beta ~{\rm log} (\zeta_{\rm i}x_{\rm i}))^2}{n-1} 
\label{eq:disp_final},
\end{eqnarray}
where the first term is the variance of the starting relation, $\sigma^2$, the term multiplying $(1-b)^2$ is the variance of $\zeta(x^\prime) \equiv \zeta(\zeta x)$ with respect to its best fit in logarithmic space (compare Equation (\ref{eq:disp_final}) with (\ref{eq:disp})), which will be named $\gamma^2$.

\end{widetext}

We can now express Equations (\ref{eq:a2_final}), (\ref{eq:aa2_final}), and (\ref{eq:disp_final}) simply as:
\begin{eqnarray}
&&a^\prime = {\rm log} (a) + \alpha~ (1-b), \nonumber \\
&&b^\prime = b + \beta ~(1-b), \nonumber \\ 
&&\sigma^{\prime2} = \sigma^2 + \gamma^2 ~(1-b)^2 
\label{eq:a2_final_alt}.
\end{eqnarray}

From Figure \ref{fig:zeta_vs_zetax}, we can easily see that $0 \leq \beta \leq 1$. This inequality,  
in conjunction with the set of Equations (\ref{eq:a2_final_alt}), leads to the immediate conclusion that for $b \neq 1$ the linear transformation we have just presented will result in a best-fit relation that is closer to unity, while increasing the dispersion of the data around this best-fit.

Now let us assume that both $\zeta$ and $x$ are random variables uniformly distributed in the intervals $[\zeta_{\rm inf}, \zeta_{\rm sup}]$ and  $[x_{\rm inf}, x_{\rm sup}]$ (see top left panel of Figure \ref{fig:zeta_vs_zetax}).
If $x_{\rm inf} \sim x_{\rm sup}$ and $\zeta_{\rm inf} \ll \zeta_{\rm sup}$, 
the slope of the best fit of $\zeta(\zeta x)$ is $\sim$ 1 (see top right panel of Figure \ref{fig:zeta_vs_zetax}), and Equation (\ref{eq:a2_final}) becomes $b^\prime \cong 1$. 
If $x_{\rm inf} \ll x_{\rm sup}$ and $\zeta_{\rm inf} \sim \zeta_{\rm sup}$, 
the slope of the best fit of $\zeta(\zeta x)$ is $\sim$ 0 (see bottom left panel of Figure \ref{fig:zeta_vs_zetax}), and Equation (\ref{eq:a2_final}) becomes $b^\prime \cong b$. If $x_{\rm inf}/x_{\rm sup}$ = $\zeta_{\rm inf}/\zeta_{\rm sup}$, 
the slope of the best fit of $\zeta(\zeta x)$ is 1/2 (see bottom right panel of Figure \ref{fig:zeta_vs_zetax}) , and Equation (\ref{eq:a2_final}) becomes $b^\prime = b + (1-b)/2$, which means that after random distributions of $x$ and $y$ (with the same factor, $\zeta_{\rm i}$, multiplying both coordinates for each data point) the best fit slope is closer to unity, decreasing the difference by half. To test this analytical prediction we created mock representations of the \YSZYX\ relation, which are presented in the next section.

\subsection{Mock representations of the \YSZYX\ relation}

To test the analytical prediction presented in the previous section, we created mock representations of the \YSZYX\ relation, using 200 data points. Initially, we populated the \YSZYX\ relation with values in a similar range as presented in Figure \ref{fig:observed_change_in_slope} to mimic the \YSZYX\ relation  when expressed in apparent flux (units of arcmin$^2$). We then randomly re-distributed the data points using a random variable uniformly distributed in the $[0,0.1]$ range to mimic the values presented in the 
\YSZYX\ relation when  expressed in intrinsic properties (units of Mpc$^2$). The mock representations of the \YSZYX\ relation are presented in Figure \ref{fig:mock}. As we can see, the best-fit results agree perfectly with the analytical prediction ($b^\prime = b + (1-b)/2$) presented in Section \ref{sec:app_analytical} (see Table \ref{tab:measured-prediction}). 

\begin{deluxetable}{ccccc}[h!]
\tablecaption{Best Fit Slope for the \YSZYX  expressed in apparent flux (units of arcmin$^2$) and intrinsic properties (Mpc$^2$) for our Mock Relations} 
\tablewidth{0pt} 
\tablehead{ 
\colhead{$B_{\rm arcmin^2}$} &
\colhead{$\sigma_{B_{\rm arcmin^2}}$} &
\colhead{$B_{\rm Mpc^2}$} &
\colhead{$\sigma_{B_{\rm Mpc^2}}$} &
\colhead{Predicted}
}
\startdata 
1.00 & 0.01 & 1.00 & 0.01 & 1.00 \\
0.90 & 0.01 & 0.95 & 0.01 & 0.95 \\
0.80 & 0.01 & 0.90 & 0.01 & 0.90 \\
0.70 & 0.01 & 0.86 & 0.02 & 0.85 \\
\enddata
\tablecomments{Columns list best fit slopes and their
  uncertainties for the power-law fit to the \YSZYX  relation expressed in apparent flux (units of arcmin$^2$), intrinsic properties (units of Mpc$^2$), and the predicted value given by 
  $B_{\rm Mpc^2}$ = $B_{\rm arcmin^2}$ + (1-$B_{\rm arcmin^2})/2$.} 
\label{tab:measured-prediction}
\vspace{-1.5cm}
\end{deluxetable}

%

\subsection{Predicted vs. observed change in slope for our sample}

Finally, we investigate the predicted change in slope for our sample, using the 
factors ($\zeta_{\rm i}$) that were used for the conversion from  apparent flux (units of arcmin$^2$) to intrinsic properties (units of Mpc$^2$). Using the notation of Section \ref{sec:app_analytical} and starting from the \YSZYX\ relation expressed in apparent flux (units of arcmin$^2$), we have:

\begin{itemize}
	
\item $x_{\rm i} = Y_{\rm X,arcmin^2}^{\rm i}$ and $y_{\rm i} = Y_{\rm SZ,arcmin^2}^{\rm i}$;
\item $\zeta_{\rm i}=Y_{\rm SZ,Mpc^2}^{\rm i}/Y_{\rm SZ,arcmin^2}^{\rm i} = Y_{\rm X,Mpc^2}^{\rm i}/Y_{\rm X,arcmin^2}^{\rm i}$ and $\zeta_{\rm i}x_i = Y_{\rm X,Mpc^2}^{\rm i}$;
\item $b$ = slope of the best fit of the \YSZYX\ relation in  apparent flux (units of arcmin$^2$) and $b^\prime$ = slope of the best fit of the \YSZYX\ relation in  intrinsic properties (units of Mpc$^2$).

\end{itemize}

\begin{figure*}[!t]
\centerline{
\includegraphics[width=0.495\textwidth]{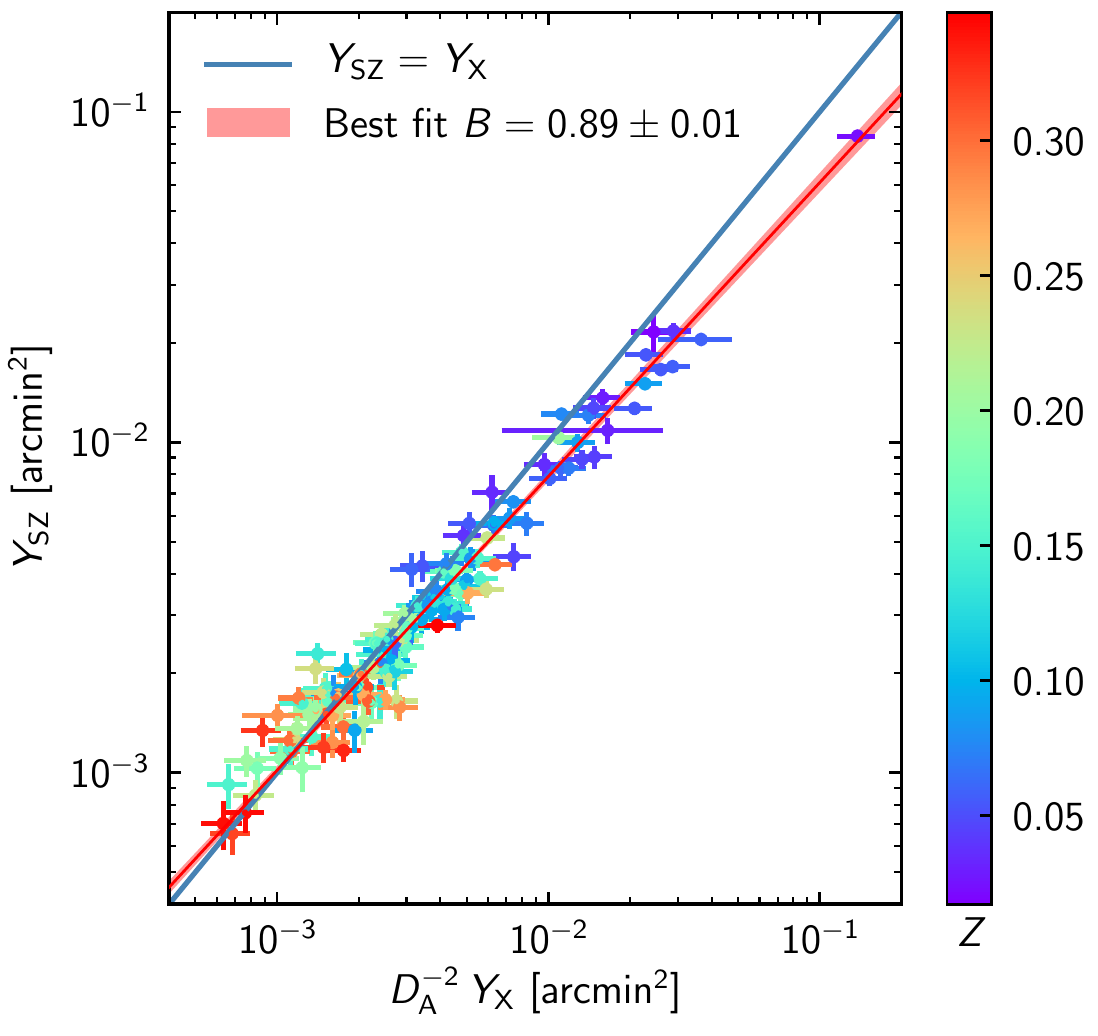}
\hfill
\includegraphics[width=0.495\textwidth]{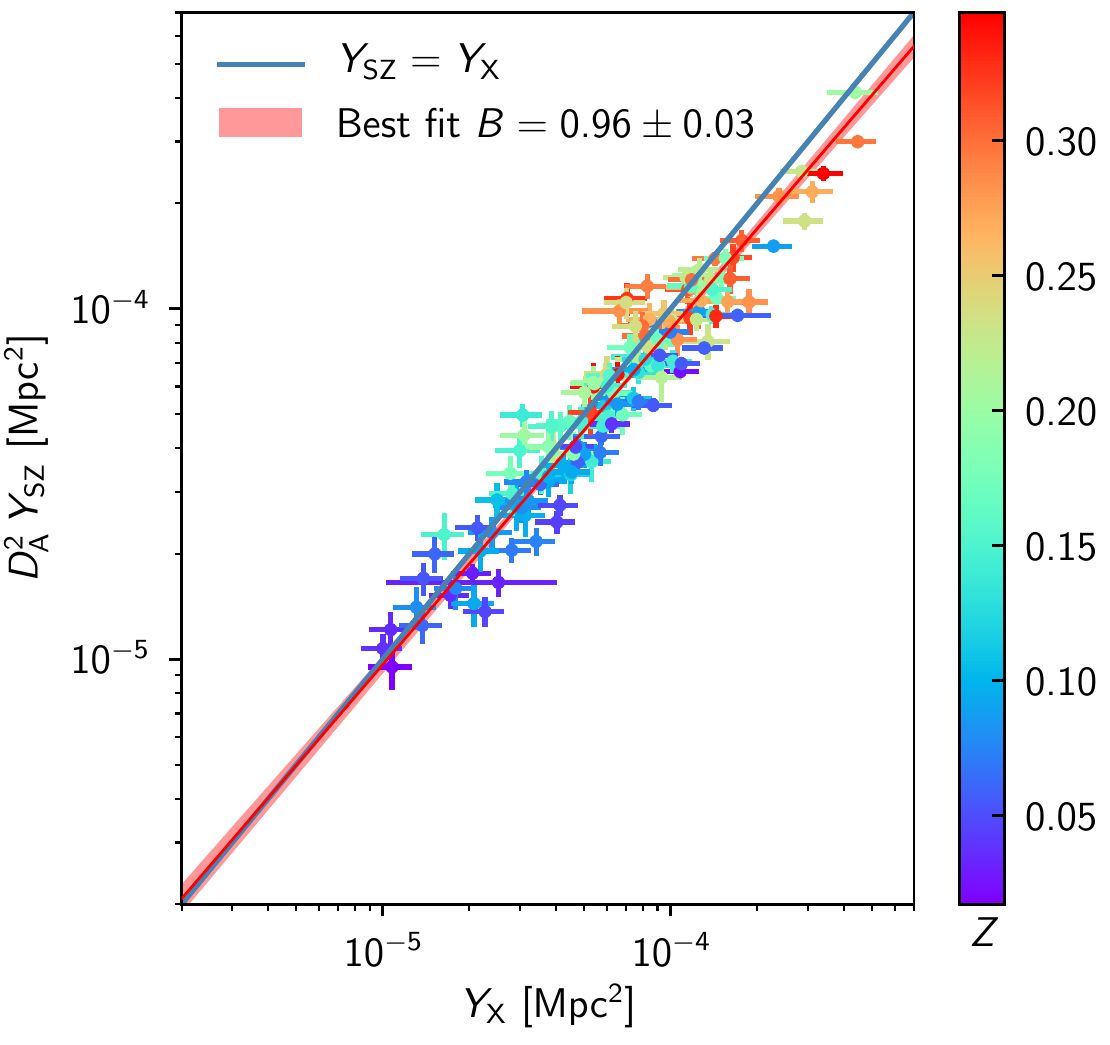}
}
\centerline{
\hspace{0.15cm}
\includegraphics[width=0.488\textwidth]{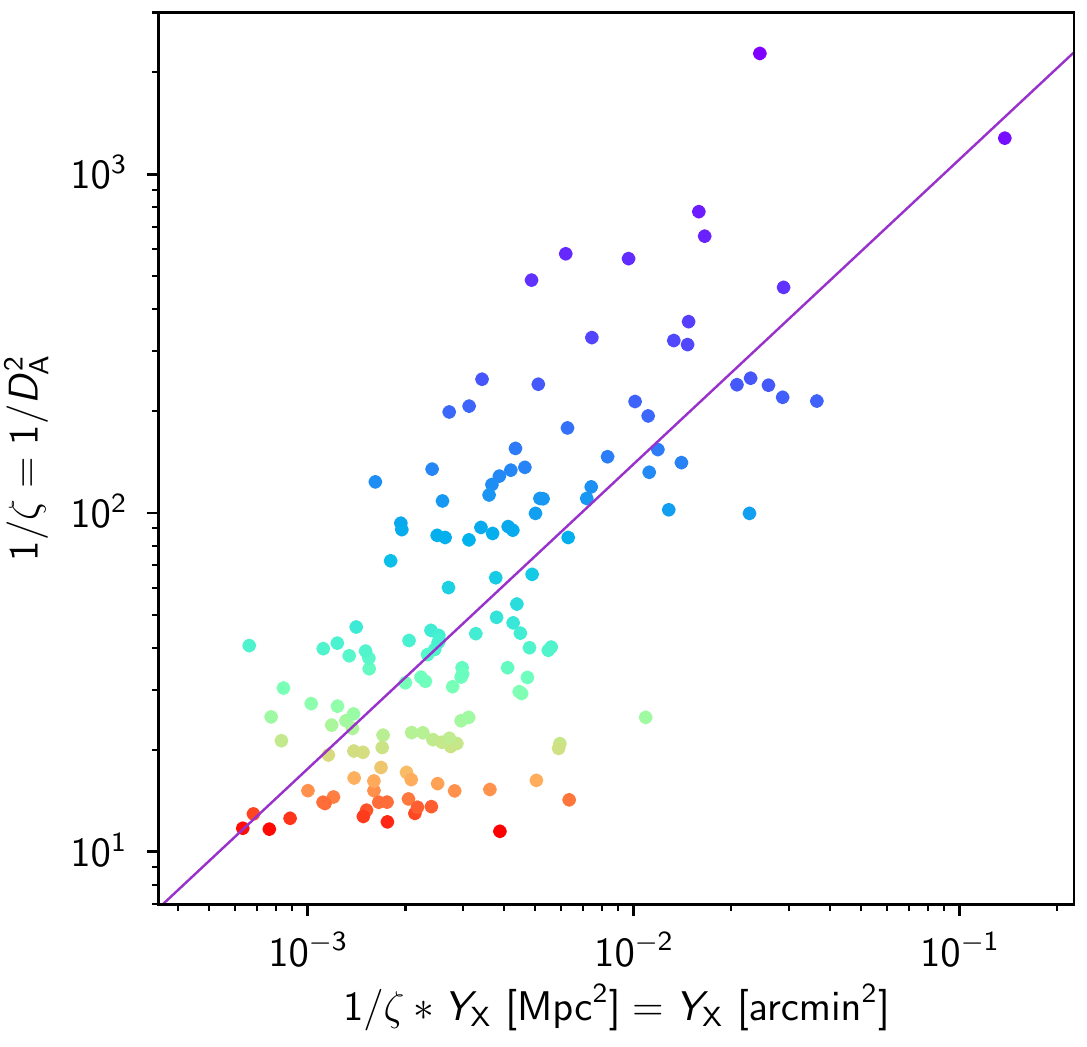}
\hfill
\includegraphics[width=0.5\textwidth]{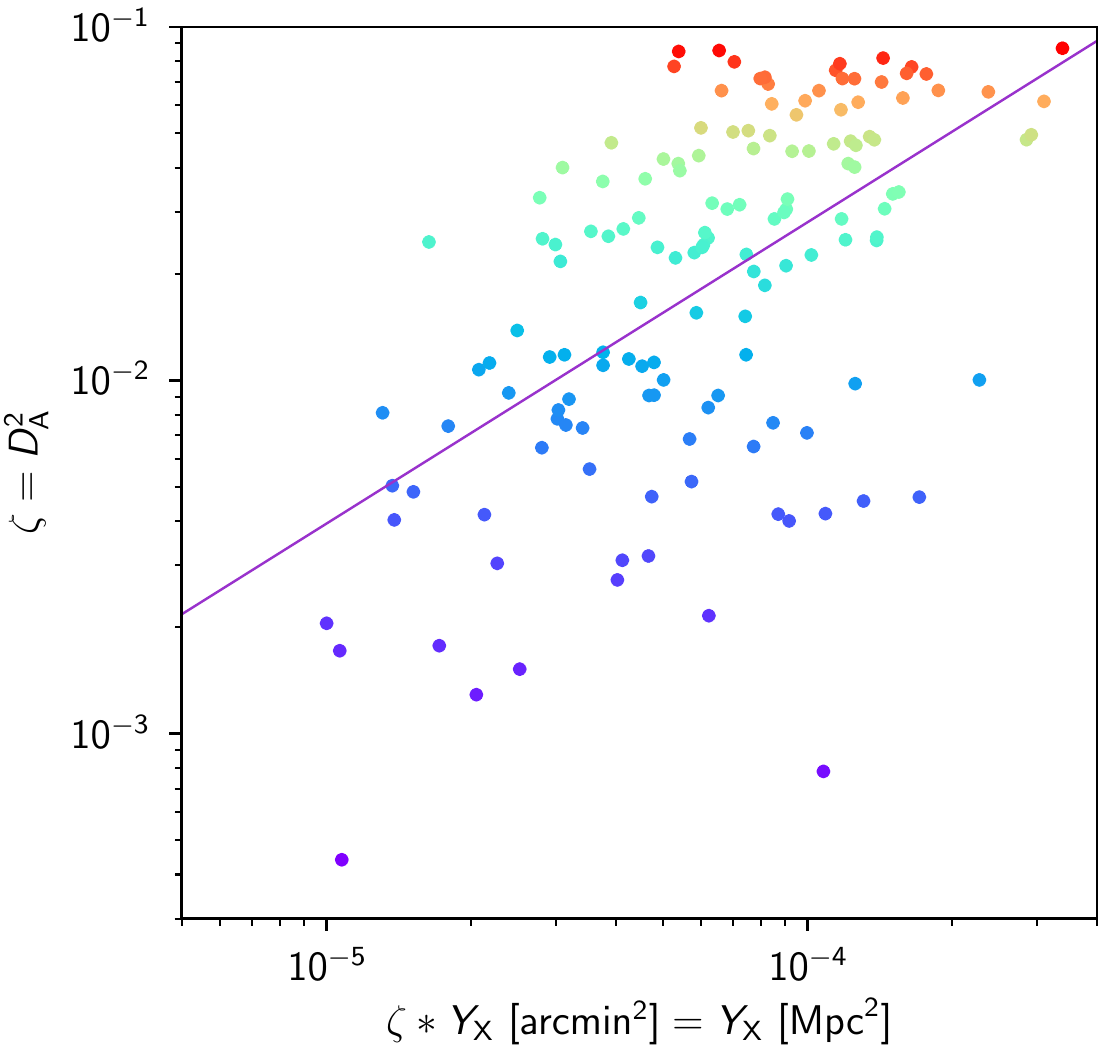}
}
\caption{\small{Top: Observed change in slope in the \YSZYX\ relation caused by the change of units from arcmin$^2$ to Mpc$^2$. The points are color-coded according to the cluster redshift. As we can see, despite being separated in the \YSZYX\ relation in units of arcmin$^2$, the colors are mixed when the same relation is plotted in units of Mpc$^2$, showing a contrary effect compared to when the colors are coded according to the cluster gas temperature (when color-coded according to the gas temperature, colors are separated in units of Mpc$^2$ -- Figure \ref{fig:observed_change_in_slope}). This illustrates the redistribution that happens when the values are presented in different units. This redistribution ultimately changes the slope of the \YSZYX\ relation. Top left: the best fit relation is given by $\YSZ \propto (\YX)^{0.89}$ in units of arcmin$^2$. Top right: same as top left panel, except for $\YSZ \propto (\YX)^{0.96}$ in units of Mpc$^2$.
Bottom: Observed $\zeta$--$\zeta x$ relation in logarithmic space for our ESZ sample. As for the top panels, the points are color-coded according to the cluster redshift. 
Bottom left: Observed $\zeta$--$\zeta x$ relation used for predicting the change of slope of the \YSZYX\ relation when units are changed from Mpc$^2$ to arcmin$^2$. The best fit slope is given by $\beta = 0.90 \pm 0.07$. 
Bottom right: Same as bottom left panel, except for changing from units of arcmin$^2$ to Mpc$^2$. The best fit slope is given by $\beta = 0.85 \pm 0.10$.
}}\label{fig:observed_change_in_slope_prediction}
\end{figure*}

Using the formalism presented in Section \ref{sec:app_analytical} and the variables as defined above, the predicted slope of the best fit of the \YSZYX\ relation when  changed from apparent flux (units of arcmin$^2$) to intrinsic properties (units of Mpc$^2$) is given by Equation (\ref{eq:a2_final}), where the term multiplying $(1-b)$ is the slope of the best fit of the ($Y_{\rm SZ,Mpc^2}/Y_{\rm SZ,arcmin^2}$) -- $Y_{\rm X,Mpc^2}$ relation ($\zeta$--$\zeta x$) in logarithmic space, which turns out to be 0.85 (see bottom panel of Figure \ref{fig:observed_change_in_slope_prediction}). This leads to a predicted slope of the best fit of the \YSZYX\ relation in  intrinsic properties of $b^\prime = 0.89 + (1 - 0.89) \times 0.85$ = 0.98, which is in agreement with the the observed value of $0.96 \pm 0.03$. Now, if we perform the same calculation to compute the predicted value of the slope of the \YSZYX in  apparent flux starting from this relation in  intrinsic properties (the slope of the best fit of $\zeta$--$\zeta x$ is 0.90 -- see bottom panel of Figure \ref{fig:observed_change_in_slope_prediction}), we obtain: $b^\prime = 0.96 + (1 - 0.96) \times 0.90$ = 1.00, which is in total disagreement with the observed value of $0.89 \pm 0.01$.

The question that then arises naturally is why  going from apparent flux to intrinsic properties results in a slope that is closer to unity, and yet the reverse is not the case.

Based on the analytical formalism presented in Section \ref{sec:app_analytical}, we know that any re-ordering of the data will result in a slope that is closer to unity. However, the analytical formalism does not indicate 
any mathematical property that allows us to know the direction  that this transformation results in a slope that is closer to unity.

The answer to this question seems to be given by understanding what is first observed, therefore what is the starting point. The observed integrated Compton parameter ($\YSZ$) is measured from the apparent flux (arcmin$^2$) from the {\em Planck} data. Similarly, the X-ray equivalent of the integrated Compton parameter, $\YX$, which is given by the product of a gas mass and a gas temperature, has a cosmological dependence on the gas mass, which is originally measured from the observed emission measure profile, and which is therefore also dependent on the apparent flux. It is only after calculations involving the underlying cosmology and cluster redshift that $\YX$ is presented in intrinsic properties ($\rm M_\odot ~keV$ or Mpc$^2$). Based on this reasoning, we see that the starting point is given by measurements that involve the apparent flux. Therefore,  going from 
apparent flux (units of arcmin$^2$) to intrinsic properties (units of Mpc$^2$) should result in a best-fit slope that is closer to unity
according to our analytical formalism presented in Section \ref{sec:app_analytical}. This is indeed what we observe.

\end{document}